\documentclass[onecolumn,notitlepage,floats,floatfix,amssymb,aps,prd,showpacs,superscriptaddress,groupedaddress,nofootinbib]{revtex4-2}  

\usepackage[T1]{fontenc}
\usepackage{lmodern} 
\usepackage{graphicx}  
\usepackage{dcolumn}   
\usepackage{bm}        
\usepackage{amssymb}   
\usepackage{amsmath}
\usepackage{bbold}
\usepackage{array}
\usepackage{makecell}
\usepackage{braket}
\usepackage{longtable}
\usepackage{supertabular,booktabs}

\usepackage{titlesec} 
\usepackage[colorlinks,citecolor=blue,urlcolor=blue,hypertexnames=true]{hyperref}
\usepackage{subfigure}

\usepackage[usenames,dvipsnames,svgnames,table]{xcolor} 

\renewcommand{\arraystretch}{2}

\newcommand{\be}{\begin{equation}}
\newcommand{\ee}{\end{equation}}
\newcommand{\bea}{\begin{eqnarray}}
\newcommand{\eea}{\end{eqnarray}}
\newcommand{\bml}{\begin{subequations}}
\newcommand{\eml}{\end{subequations}}
\newcommand{\bfig}{\begin{figure}}
\newcommand{\efig}{\end{figure}}

\newcommand{\mbf}{\mathbf}

\newcommand{\slashed}{\hspace{-1.1ex}/}

\newcommand{\bmat}{\begin{pmatrix}}
\newcommand{\emat}{\end{pmatrix}}
\usepackage{graphicx,slashed,booktabs,xcolor,multirow,float,
amsfonts,bbold,mathtools,sidecap,tikz,bm,enumitem}
\usepackage{multirow}
\usepackage{bbding}
\usepackage{titlesec}
\usepackage{hyperref}
\usepackage{wasysym}
\usepackage{amssymb}
\usepackage{pifont}
\newcommand{\grad}{\nabla}

\renewcommand{\leq}{\leqslant}
\renewcommand{\geq}{\geqslant}

\usepackage[dvipsnames, usenames]{xcolor}

\definecolor{linkcolor}{rgb}{0.55, 0.13, .32}

\definecolor{oucrimsonred}{rgb}{0.6, 0.0, 0.0}
\definecolor{persianblue}{rgb}{0.11, 0.22, 0.73}
\definecolor{forestgreen}{rgb}{0.13,0.35,0.13}
\definecolor{lightgray}{rgb}{0.83, 0.83, 0.83}
 \hypersetup{colorlinks, citecolor=oucrimsonred, linkcolor=persianblue, urlcolor=oucrimsonred}
\definecolor{cornellred}{rgb}{0.7, 0.11, 0.11}
\definecolor{navyblue}{rgb}{0.0, 0.0, 0.5}
\definecolor{amethyst}{rgb}{0.6, 0.4, 0.8}
\definecolor{yellow}{rgb}{1.0, 1.0, 0.0}
\definecolor{firebrick}{rgb}{0.7, 0.13, 0.13}
\definecolor{tangerineyellow}{rgb}{1.0, 0.8, 0.0}
\definecolor{deepfuchsia}{rgb}{0.76, 0.33, 0.76}
\definecolor{amber}{rgb}{1.0, 0.75, 0.0}
\definecolor{VioletRed4}{rgb}{0.55, 0.13, .32}
\definecolor{indiagreen}{rgb}{0.07, 0.53, 0.03}
\definecolor{VioletRed4}{rgb}{0.55, 0.13, .32}

\usepackage{hyperref}

\usepackage{graphics,appendix,afterpage,makecell}

\usepackage{bbold}
\usepackage{tikz}
\usepackage{adjustbox}
\usepackage{tikz-feynman}
\usepackage{tcolorbox}

\definecolor{oucrimsonred}{rgb}{0.6, 0.0, 0.0}
\definecolor{persianblue}{rgb}{0.11, 0.22, 0.73}
\definecolor{forestgreen}{rgb}{0.13,0.35,0.13}
\definecolor{lightgray}{rgb}{0.83, 0.83, 0.83}
 \hypersetup{colorlinks, citecolor=oucrimsonred, linkcolor=persianblue, urlcolor=oucrimsonred}
\definecolor{cornellred}{rgb}{0.7, 0.11, 0.11}
\definecolor{navyblue}{rgb}{0.0, 0.0, 0.5}
\definecolor{amethyst}{rgb}{0.6, 0.4, 0.8}
\definecolor{yellow}{rgb}{1.0, 1.0, 0.0}
\definecolor{firebrick}{rgb}{0.7, 0.13, 0.13}
\definecolor{tangerineyellow}{rgb}{1.0, 0.8, 0.0}
\definecolor{deepfuchsia}{rgb}{0.76, 0.33, 0.76}
\definecolor{amber}{rgb}{1.0, 0.75, 0.0}
\definecolor{VioletRed4}{rgb}{0.55, 0.13, .32}
\definecolor{indiagreen}{rgb}{0.07, 0.53, 0.03}
\definecolor{VioletRed4}{rgb}{0.55, 0.13, .32}

\definecolor{oucrimsonred}{rgb}{0.6, 0.0, 0.0}
\newcommand\vertarrowbox[3][6ex]{%
  \begin{array}[t]{@{}c@{}} #2 \\
  \left\uparrow\vcenter{\hrule height #1}\right.\kern-\nulldelimiterspace\\
  \makebox[0pt]{\scriptsize#3}
  \end{array}%
}

\definecolor{mtcolor}{rgb}{.8,.3,.1}

\definecolor{violachiaro}{rgb}{1,0.6,1}

\definecolor{gbcolor}{rgb}{.43,.22,.12}
 
\definecolor{gbcolor2}{rgb}{.9,.2,.6}
\definecolor{gbcolor3}{rgb}{.3,.2,.6}

\definecolor{verdechiaro}{rgb}{0.6,1,0.6}
\definecolor{giallochiaro}{rgb}{1,1,0.6}
\definecolor{bluscuro}{rgb}{0.15, 0.2, 0.9}
\definecolor{verdes}{rgb}{0.1, 0.5, 0.1}%
\definecolor{tangerineyellow}{rgb}{1.0, 0.8, 0.0}
\definecolor{smokyblack}{rgb}{0.06, 0.05, 0.03}

\definecolor{americanrose}{rgb}{1.0, 0.01, 0.24}
\definecolor{cobalt}{rgb}{0.0, 0.28, 0.67}
\definecolor{brandeisblue}{rgb}{0.0, 0.44, 1.0}
\definecolor{mycolor}{rgb}{0.0, 0.0, 0.5}
\definecolor{oxfordblue}{rgb}{0.0, 0.13, 0.28}
\definecolor{azure}{rgb}{0.0, 0.5, 1.0}
\definecolor{turquoiseblue}{rgb}{0.0, 1.0, 0.94}
\newtcolorbox{mynewbox}[1]{colback=white!5!white,colframe=azure!75!black,fonttitle=\bfseries,title=#1}
\newtcolorbox{mybox}{colback=mycolor!5!white,colframe=azure!75!black}
\newtcolorbox{mynamedbox}[1]{colback=mycolor!5!white,colframe=azure!75!black,title=#1}
\definecolor{venetianred}{rgb}{0.78, 0.03, 0.08}
\newtcolorbox{mynamedbox1}[1]{colback=venetianred!5!white,colframe=venetianred!80!black,title=#1}
\newtcolorbox{mynamedbox2}[1]{colback=azure!5!white,colframe=azure!80!black,title=#1}

\definecolor{rossocorsa}{rgb}{0.83, 0.0, 0.0}

\tikzset{->-/.style={decoration={
  markings,
  mark=at position #1 with {\arrow{>}}},postaction={decorate}}}
\tikzset{-<-/.style={decoration={
  markings,
  mark=at position #1 with {\arrow{<}}},postaction={decorate}}} 

\def\be{\begin{equation}}
\def\ee{\end{equation}}
\def\ba{\begin{eqnarray}}
\def\ea{\end{eqnarray}}

\def\L*{{\cal L}_*}
\def\L{\mathcal{L}}
\def\({\left(}
\def\){\right)}

\def\<{\langle}
\def\>{\rangle}

\def\cs2{c_{s}^{2}}

 \def\be   {\begin{equation}}   \def\ee   {\end{equation}}
 \def\ba   {\begin{array}}      \def\ea   {\end{array}}
 \def\bea  {\begin{eqnarray}}   \def\eea  {\end{eqnarray}}
 \def\bean {\begin{eqnarray*}}  \def\eean {\end{eqnarray*}}

\titleclass{\subsubsubsection}{straight}[\subsection]

\newcounter{subsubsubsection}[subsubsection]
\renewcommand\thesubsubsubsection{\thesubsubsection.\arabic{subsubsubsection}}

\titleformat{\subsubsubsection}
  {\normalfont\normalsize\bfseries}{\thesubsubsubsection}{1em}{}
\titlespacing*{\subsubsubsection}
{0pt}{3.25ex plus 1ex minus .2ex}{1.5ex plus .2ex}

\makeatletter
\renewcommand\paragraph{\@startsection{paragraph}{5}{\z@}%
  {3.25ex \@plus1ex \@minus.2ex}%
  {-1em}%
  {\normalfont\normalsize\bfseries}}
\renewcommand\subparagraph{\@startsection{subparagraph}{6}{\parindent}%
  {3.25ex \@plus1ex \@minus .2ex}%
  {-1em}%
  {\normalfont\normalsize\bfseries}}
\def\toclevel@subsubsubsection{4}
\def\toclevel@paragraph{5}
\def\toclevel@paragraph{6}
\def\l@subsubsubsection{\@dottedtocline{4}{7em}{4em}}
\def\l@paragraph{\@dottedtocline{5}{10em}{5em}}
\def\l@subparagraph{\@dottedtocline{6}{14em}{6em}}
\makeatother

\begin{document}


\definecolor{lime}{HTML}{A6CE39}
\DeclareRobustCommand{\orcidicon}{\hspace{-2.1mm}
\begin{tikzpicture}
\draw[lime,fill=lime] (0,0.0) circle [radius=0.13] node[white] {{\fontfamily{qag}\selectfont \tiny \,ID}}; \draw[white, fill=white] (-0.0525,0.095) circle [radius=0.007]; 
\end{tikzpicture} \hspace{-3.7mm} }
\foreach \x in {A, ..., Z}{\expandafter\xdef\csname orcid\x\endcsname{\noexpand\href{https://orcid.org/\csname orcidauthor\x\endcsname} {\noexpand\orcidicon}}}
\newcommand{\orcidauthorA}{0000-0002-0459-3873}
\newcommand{\orcidauthorD}{0009-0003-9227-8615}
\newcommand{\orcidauthorB}{0000-0001-9434-0505}
\newcommand{\orcidauthorC}{0000-0003-1081-0632}


\title{\textcolor{Sepia}{\textbf \huge\Large 
Scalar 
induced gravity waves from ultra slow-roll Galileon inflation 
}}


\author{{\large  Sayantan Choudhury\orcidA{}${}^{1}$}}
\email{sayantan\_ccsp@sgtuniversity.org,  \\ sayanphysicsisi@gmail.com} 
\author{{\large  Ahaskar Karde\orcidD{}${}^{1}$}}
\email{kardeahaskar@gmail.com}
\author{\large Sudhakar~Panda\orcidB{}${}^{1}$}
\email{panda@niser.ac.in}
\author{ \large M.~Sami\orcidC{}${}^{1,2,3}$}
\email{ sami\_ccsp@sgtuniversity.org,  samijamia@gmail.com}

\affiliation{ ${}^{1}$Centre For Cosmology and Science Popularization (CCSP),\\
        SGT University, Gurugram, Delhi- NCR, Haryana- 122505, India.}
\affiliation{${}^{2}$Center for Theoretical Physics, Eurasian National University, Astana 010008, Kazakhstan.}
	\affiliation{${}^{3}$Chinese Academy of Sciences,52 Sanlihe Rd, Xicheng District, Beijing.}

\begin{abstract}

We consider the production of secondary gravity waves in Galileon inflation with an ultra-slow roll (USR) phase and show that the spectrum of scalar-induced gravitational waves (SIGWs) in this case is consistent with the recent NANOGrav 15-year data and with sensitivities of other ground and space-based missions, LISA, BBO, DECIGO, CE, ET, HLVK (consists of aLIGO, aVirgo, and KAGRA), and HLV(03). 
Thanks to the non-renormalization property of Galileon theory, the amplitude of the large fluctuation is controllable at the sharp transitions between SR and USR regions. 
We show that the behaviour of the GW spectrum, when one-loop effects are included in the scalar power spectrum, is preserved under a shift of the sharp transition scale with peak amplitude $\Omega_{\rm GW}h^2\sim {\cal O}(10^{-6})$, and hence it can cover a wide range of frequencies within ${\cal O}(10^{-9}{\rm Hz} - 10^{7}{\rm Hz})$. An analysis of the allowed mass range for primordial black holes (PBHs) is also performed, where we find that mass values ranging from ${\cal O}(1M_{\odot} - 10^{-18}M_{\odot})$ can be generated  over the corresponding allowed range of low and high frequencies.

\end{abstract}

\maketitle
\tableofcontents
\newpage

\section{Introduction}

The first observational results on Gravitational Waves (GWs) generated from binary black hole mergers \cite{LIGOScientific:2016aoc} provided us with an opportunity to study the physics of the early universe. Indeed, the GWs are a unique probe that directly brings information from the early universe before recombination. Their possible sources could include phase transitions in the early universe \cite{Zu:2023olm, Abe:2023yrw, Gouttenoire:2023bqy, NANOGrav:2021flc, Xue:2021gyq, Nakai:2020oit, Athron:2023mer, Madge:2023cak}, domain walls \cite{Kitajima:2023cek, Babichev:2023pbf, Zhang:2023nrs, Zeng:2023jut, Ferreira:2022zzo, An:2023idh, Li:2023tdx}, cosmic strings \cite{Ellis:2020ena, Blasi:2020mfx, Buchmuller:2020lbh, Blanco-Pillado:2021ygr, Buchmuller:2021mbb, Madge:2023cak}, and most popularly, inflation \cite{Inomata:2023zup, Zhu:2023faa,HosseiniMansoori:2023mqh,Das:2023nmm, Balaji:2023ehk, Cai:2023dls, Wang:2023ost, Yi:2023mbm, Choudhury:2013woa, Choudhury:2023kam, Choudhury:2023tcn,Bhattacharya:2023ysp,Vagnozzi:2023lwo, Franciolini:2023pbf, Gorji:2023sil, DeLuca:2023tun, Frosina:2023nxu, Chen:2019xse, Cai:2023uhc, Huang:2023chx, Cang:2023ysz,Domenech:2021ztg,Madge:2023cak,Choudhury:2023fjs,Choudhury:2023fwk}, where GWs (tensor perturbations) are generated naturally 
\cite{Baumann:2009ds,Baumann:2022mni,Baumann:2018muz,Senatore:2016aui,Martin:2013tda,Martin:2013nzq,Mazumdar:2010sa,Lyth:1998xn}, enter the horizon and travel unimpeded  to become visible to us today using the current and proposed observational experiments \cite{Planck:2018jri, Planck:2018vyg, CMB-S4:2016ple}. These GWs would appear to us in the form of random signals forming an often-called stochastic GW background (SGWB) when looking in all the possible directions in the universe for their sources located at large redshifts. The latest announcement from various PTA collaborations, NANOGrav \cite{NANOGrav:2023gor, NANOGrav:2023hde, NANOGrav:2023ctt, NANOGrav:2023hvm, NANOGrav:2023hfp, NANOGrav:2023tcn, NANOGrav:2023pdq, NANOGrav:2023icp}, EPTA \cite{EPTA:2023fyk, EPTA:2023sfo, EPTA:2023akd, EPTA:2023gyr, EPTA:2023xxk, EPTA:2023xiy}, PPTA \cite{Reardon:2023gzh, Reardon:2023zen, Zic:2023gta}, and CPTA \cite{Xu:2023wog}, have confirmed the existence of an SGWB, which has attracted considerable work where the variety of cosmological models mentioned above are examined for being a possible source for the observed data. In this work, we will be concerned with the scalar-induced GWs scenario to explain the PTA signal.

The concept of GWs being induced by primordial density fluctuations was studied initially in the respective refs.\cite{Matarrese:1992rp, Matarrese:1993zf, Matarrese:1997ay}; however, it was confirmed that the magnitude of the produced GW spectrum was insufficient in terms of any meaningful observations. Later, the works of the authors in \cite{Ananda:2006af, Baumann:2007zm} showed that the generation of induced GWs considered in the radiation and matter-dominated eras, including the radiation-matter equality phase, would be able to produce an enhanced spectrum amplitude by examining constraints on the spectral tilt, and details of the transfer function for production of the second-order GWs between the large scales observed today to the smallest scales, in the respective works. This led to the final induced GW spectrum having an observable amplitude but with the condition of having to consider only the very low-frequency regimes. 
Another interesting observation regarding this phenomenon was made by the authors in \cite{Saito:2008jc, saito2010gravitational}, where they investigated the case of large enough primordial fluctuations which can lead to the collapse and formation of primordial black holes (PBHs)\cite{Hawking:1974rv,Carr:1974nx,Carr:1975qj,Chapline:1975ojl,Carr:1993aq,Yokoyama:1998pt,Rubin:2001yw,Khlopov:2002yi,Khlopov:2004sc,Saito:2008em,Khlopov:2008qy,Carr:2009jm,Choudhury:2011jt,Choudhury:2013woa,Lyth:2011kj,Drees:2011yz,Drees:2011hb,Hertzberg:2017dkh,Cicoli:2018asa,Ozsoy:2018flq,Byrnes:2018txb,Martin:2019nuw,Ezquiaga:2019ftu,Motohashi:2019rhu,Ashoorioon:2019xqc,Auclair:2020csm,Vennin:2020kng,Inomata:2021uqj,Ng:2021hll,Wang:2021kbh,Kawai:2021edk,Solbi:2021rse,Ballesteros:2021fsp,Rigopoulos:2021nhv,Animali:2022otk,Frolovsky:2022ewg,Escriva:2022duf,Kristiano:2022maq,Kristiano:2023scm,Karam:2022nym,Riotto:2023hoz,Riotto:2023gpm,Ozsoy:2023ryl,Choudhury:2023vuj,Choudhury:2023jlt,Choudhury:2023rks,Choudhury:2023hvf,Choudhury:2023kdb,Bhattacharya:2023ysp,Choudhury:2024ybk,Choudhury:2024jlz,Choudhury:2024dei, Choudhury:2024aji,Choudhury:2024dzw,Banerjee:2021lqu,Firouzjahi:2023ahg,Firouzjahi:2023aum,Franciolini:2023lgy,Tasinato:2023ukp,Motohashi:2023syh,Afshordi:2003zb,Frampton:2010sw,Carr:2016drx,Kawasaki:2016pql,Inomata:2017okj,Espinosa:2017sgp,Ballesteros:2017fsr,Sasaki:2018dmp,Ballesteros:2019hus,Dalianis:2019asr,Cheong:2019vzl,Green:2020jor,Carr:2020xqk,Ballesteros:2020qam,Carr:2020gox,Ozsoy:2020kat,Baumann:2007zm,Saito:2008jc,Saito:2009jt,Choudhury:2013woa,Sasaki:2016jop,Raidal:2017mfl,Ali-Haimoud:2017rtz,Di:2017ndc,Cheng:2018yyr,Vaskonen:2019jpv,Drees:2019xpp,Hall:2020daa,Ballesteros:2020qam,Ragavendra:2020sop,Carr:2020gox,Ozsoy:2020kat,Ashoorioon:2020hln,Ragavendra:2020vud,Papanikolaou:2020qtd,Teimoori:2021pte,Cicoli:2022sih,Ashoorioon:2022raz,Papanikolaou:2022chm,Papanikolaou:2023crz,Papanikolaou:2022did,Wang:2022nml, Ahmed:2021ucx,Yi:2023npi,Yuan:2021qgz,
Aghaie:2023lan}. 
This work focuses on scalar-induced GWs production from Galileon inflation \cite{Burrage:2010cu} in the presence of an ultra-slow-roll (USR) phase, which triggers the generation of large amplitude scalar perturbation. In particular, we want to discuss the observational status of the induced GWs spectrum and also examine the allowed mass range for the produced PBHs from Galileon inflation. 

Recently, there has been an active pursuit to settle the arguments concerning the formation of PBHs and the effects on their masses from the one-loop corrections to the scalar power spectrum. Related discussions are present in refs. \cite{Kristiano:2022maq,Riotto:2023hoz,Choudhury:2023vuj,Choudhury:2023rks,Choudhury:2023hvf,Choudhury:2023kdb,Kristiano:2023scm,Riotto:2023gpm,Firouzjahi:2023ahg,Firouzjahi:2023aum,Franciolini:2023lgy,Tasinato:2023ukp,Choudhury:2023hvf,Choudhury:2023kdb,Choudhury:2024ybk,Choudhury:2024jlz,Motohashi:2023syh} which concern the single-field canonical models of inflation and the EFT treatment of inflation. Amidst this, the use of the Galileon theory in \cite{Choudhury:2023hvf} has shown interesting results stemming from some important features forming the basis of our discussions in this work. See refs.\cite{Jain:2010ka,Gannouji:2010au,Ali:2010gr,deRham:2011by,Burrage:2010rs,DeFelice:2010jn,DeFelice:2010gb,Babichev:2010jd,DeFelice:2010pv,DeFelice:2010nf,Hinterbichler:2010xn,Kobayashi:2010cm,Deffayet:2010qz,Burrage:2010cu,Mizuno:2010ag,Khoury:2010xi,DeFelice:2010as,Kamada:2010qe,Kobayashi:2011pc,DeFelice:2011zh,Khoury:2011da,Trodden:2011xh,Burrage:2011bt,Kobayashi:2011nu,PerreaultLevasseur:2011wto,deRham:2011by,Brax:2011sv,DeFelice:2011uc,Gao:2011qe,Babichev:2011iz,DeFelice:2011hq,Khoury:2011ay,Qiu:2011cy,Renaux-Petel:2011rmu,DeFelice:2011bh,DeFelice:2011th,DeFelice:2011aa,Zhou:2011ix,Goon:2012mu,Shirai:2012iw,Goon:2012dy,deRham:2012az,Ali:2012cv,Liu:2012ww,Choudhury:2012yh,Choudhury:2012whm,Barreira:2012kk,deFromont:2013iwa,Arroja:2013dya,Sami:2013ssa,Khoury:2013tda,Burrage:2015lla,Koyama:2015vza,Brax:2015dma,Saltas:2016nkg} to know about the underlying Galieon framework. It includes the unique Non-Renormalization theorem, which states that the theory is stable against radiative corrections to any of the calculated correlation functions. This further limits our work to only performing the regularization procedure while neglecting the important, but redundant in this case, procedures of renormalization and resummation which is a remarkable feature to consider. Using these properties of this theory, the one-loop corrected version of the scalar power spectrum is shown to have a controllable behaviour in terms of maintaining the perturbativity argument within the theory. When working with perturbation theory up to the second order, the observed spectrum of the scalar perturbations will act as a source for generating second-order tensor modes and hence SIGWs. The inclusion of quantum loop effects will then not greatly alter the behaviour of the spectrum and only result in an introduction of oscillatory features near the tail and peak regions. We are considering that the entire inflationary phase in this theory consists of three regions, namely the first slow-roll (SRI), ultra-slow roll (USR), and the second slow-roll (SRII), such that there exists a sharp transition when passing from SRI to USR and USR to SRII phases. The behavior of the scalar power spectrum at the transition scales, and similarly for the tensor power spectrum, is controllable due to the properties of this theory mentioned at the beginning. This fact has significant implications when controlling the non-Gaussianities \cite{Choudhury:2023kdb,Choudhury:2023tcn}, and due to the phase corresponding to large primordial fluctuations having constraints on its duration, the perturbativity approximation does not break near the sharp transitions. This controlling feature results from properties intrinsic to the Galileon theory, which is not true for other single-field inflation models.  

The nature of a transition, whether sharp or smooth, reflects significantly in the allowed PBH mass and its abundance. When quantum loop effects turn out to be necessary for the overall analysis of an inflationary paradigm to consider PBHs, then the need for performing renormalization and resummation is required to reach meaningful conclusions for the completion of inflation and on the mass of PBH \cite{Choudhury:2023vuj, Choudhury:2023jlt, Choudhury:2023rks}.
An important consequence of the analysis done in the Galileon theory concerns the \textit{no-go theorem} for the masses of PBHs \cite{Choudhury:2023hvf, Choudhury:2023kdb} such that the theory is able to evade the said theorem by being able to produce solar mass PBHs along with controlling the enhancement of perturbations with successful inflation. The behaviour of the induced GW spectrum investigated when having frequencies for the sharp transitions set in both the low and high-frequency regimes provides us with an opportunity to see whether Galileon theory is able to produce a large enough GW signal that can lie within the existing observational results. Since shifting of the sharp transition scales does not affect the qualitative features of the scalar power spectrum as long as the perturbative arguments are maintained, we find that the induced GW spectrum can also show its presence where the high-frequency GW probes operate which includes LISA \cite{LISA:2017pwj}, BBO \cite{Crowder:2005nr}, DECIGO \cite{Kawamura:2011zz}, Cosmic Explorer(CE) \cite{Reitze:2019iox}, Einstein Telescope(ET) \cite{Punturo:2010zz}, the HLVK network which consists of aLIGO in Hanford and Livingstone \cite{LIGOScientific:2014pky}, aVirgo \cite{VIRGO:2014yos}, and KAGRA \cite{KAGRA:2018plz}, and the HLV network during the third observation run (O3).

This paper is outlined as follows: In Sec. \ref{s2}, we provide a short overview of the Galileon theory \textcolor{black}{which begins by analyzing the setup of interest in detail with proper realisation of each phase, including the evolution of the slow-roll parameters and the Galileon scalar field throughout, followed by}
analyzing its second-order perturbed action to get the mode solutions for the comoving curvature perturbation. Then we briefly discuss the significance of the non-renormalization theorem and introduce the third-order action responsible for the calculation of the one-loop effects. Lastly, we talk about the allowed mass range for producing PBH and compare the recent studies done in this regard. In Sec. \ref{s3}, we present a concise introduction to the theory of SIGWs and the radiation-dominated era contribution to the GW abundance formula which is going to be used by us. In Sec. \ref{s4}, we use the results of the previous sections to present our key result for the induced GW spectrum from Galileon theory. There we analyze its qualitative features in detail and comment on their observational status and their relation with the masses of PBH. In Sec. \ref{s5}, we state our conclusions.

\section{ Galileon Inflation }
\label{s2}

In this section, we present a brief overview of our findings on the Galileon framework. \textcolor{black}{ We start with a discussion on the general action of the theory followed by the realisation of an ultra-slow roll (USR) phase in the present context. After this we provide an explanation for obtaining the mode solutions for the comoving curvature perturbation using the second-order perturbed action.} Then we discuss the effects of mildly breaking the Galilean symmetry and the impact of the powerful non-renormalization theorem in the presence of a sharp transition scenario. We then discuss the third-order action in the theory responsible for one-loop effects to be used 
in the later sections. Finally, we  discuss  the case of allowed masses of PBH produced from the Galileon theory and comment on the recent studies related to this issue.

\subsection{The Set up with ultra-slow roll realisation } 
\label{s2a}

The Galileon theory is a framework where the equations of motion are of second order despite the higher-derivative terms in the action ref.\cite{Nicolis:2008in, Deffayet:2009wt},



\bea
\label{gaction}
S = \int d^{4}x\sqrt{-g}\left[\frac{M^{2}_{pl}}{2}R - V_{0} +\sum^{5}_{i=1}c_{i}{\cal L}_{i}\right]
\eea
which include the dimensionless coefficients $c_{i}$ and the remaining Lagrangians are explicitly written as follows:
\begin{eqnarray}
{\cal L}_1 & = & \phi, \quad
 {\cal L}_2=-\frac{1}{2} (\grad \phi)^2 ,\quad
	{\cal L}_3=\frac{1}{\Lambda^3} (\grad \phi)^2 \Box \phi ,\quad\nonumber\\
	{\cal L}_4&=& -\frac{1}{\Lambda^6} (\grad \phi)^2 \Big\{
					(\Box \phi)^2 - (\grad_\mu \grad_\nu \phi)
					(\grad^\mu \grad^\nu \phi)
					- \frac{1}{4} R (\grad \phi)^2
				\Big\},\\
	{\cal L}_5&=& \frac{1}{\Lambda^9} (\grad \phi)^2 \Big\{
					(\Box \phi)^3 - 3 (\Box \phi)( \grad_\mu \grad_\nu \phi)
					(\grad^\mu	 \grad^\nu \phi)
     + 2 ( \grad_\mu  \grad_\nu \phi)
					(\grad^\nu	 \grad^\alpha \phi)
					(\grad_\alpha \grad^\mu \phi)
					- 6 G_{\mu \nu} \grad^\mu \grad^\alpha \phi
					\grad^\nu \phi \grad_\alpha \phi
				\Big\}.	\nonumber	
	\end{eqnarray}
The action (\ref{gaction})  has the Galilean shift symmetry as its defining property: 
\bea \label{s1a}
\phi \rightarrow \phi+ a_{\mu}x^{\mu} + b,
\eea
where $\phi$ is the Galileon scalar field, $a_{\mu}$ is a constant vector and $b$ a constant scalar defined in the $3+1$ dimensional space-time. The framework based upon the covariant action (\ref{gaction}) is referred to as the Covariantized Galileon Theory (CGT).

\textcolor{black}{Now we move towards studying the dynamics of this Galileon scalar field in a cosmological setting where the Galileon model lives on a quasi de Sitter background spacetime such that the background field, $\bar{\phi}(t)$, is homogeneous and time-dependent. Next, to study the inflationary scenario on such a spacetime we require that variation in
the effective potential of our theory must satisfy the condition $|\Delta V/V|\ll 1$. This gives us a value of the scale factor within our CGT framework as $a(t)=\exp{(Ht)}$, where the Hubble parameter $H$ also defines the deviation from exact de sitter in form of the slow-roll parameter $\epsilon = -\dot{H}/H^{2}$. We now mention the action for the field $\phi(t)$ after collecting the first-order terms from integration by parts and discarding any boundary terms:}
\bea
\label{bcgaction}  S^{(0)}= \int dt\;a^{3}{\cal L} = \int d^4x\,a^3 \,\Bigg\{\dot{\bar{\phi}}^2\Bigg(\frac{c_2}{2}+2c_3Z+\frac{9c_4}{2}Z^2+6c_5Z^3\Bigg)+\lambda^3\bar{\phi}\Bigg\}
\eea
\textcolor{black}{where the coupling parameter is defined as $Z \equiv H\dot{\bar{\phi}}/\Lambda^3$ with $\Lambda$ as a physical cut-off scale of the theory.
In the regime where the coupling satisfies $Z \simeq 1$, we can incorporate the non-linearities within the galileon sector and also neglect any non-minimal couplings to gravity. This is the favourable regime in which we choose to work. }

\textcolor{black}{Let us briefly understand the behaviour of the scalar field in the de Sitter background through its evolution. To conduct inflation requires mildly breaking the shift symmetry, and this gets implemented starting with a linear term in the field $\bar{\phi}$ along the constant potential $V_{0}$ to give, $V = V_{0}-c_{1}\bar{\phi}$, where $c_{1}$ is a small symmetry-breaking parameter. To get the equation of motion for $\bar{\phi}(t)$ is made simple by the fact that under the condition $M_{p}\rightarrow \infty$ while maintaining the relation $3H^{2}M_{p}^{2}=V_{0}$, gives us a scalar field on exact de Sitter space. There, we have an exact shift symmetry, which allows the conservation of the associated current to provide for the equation of motion for the field $\bar{\phi}$. }

\begin{figure*}[htb!]
    	\centering
    \subfigure[]{
      	\includegraphics[width=8.5cm,height=7.5cm] {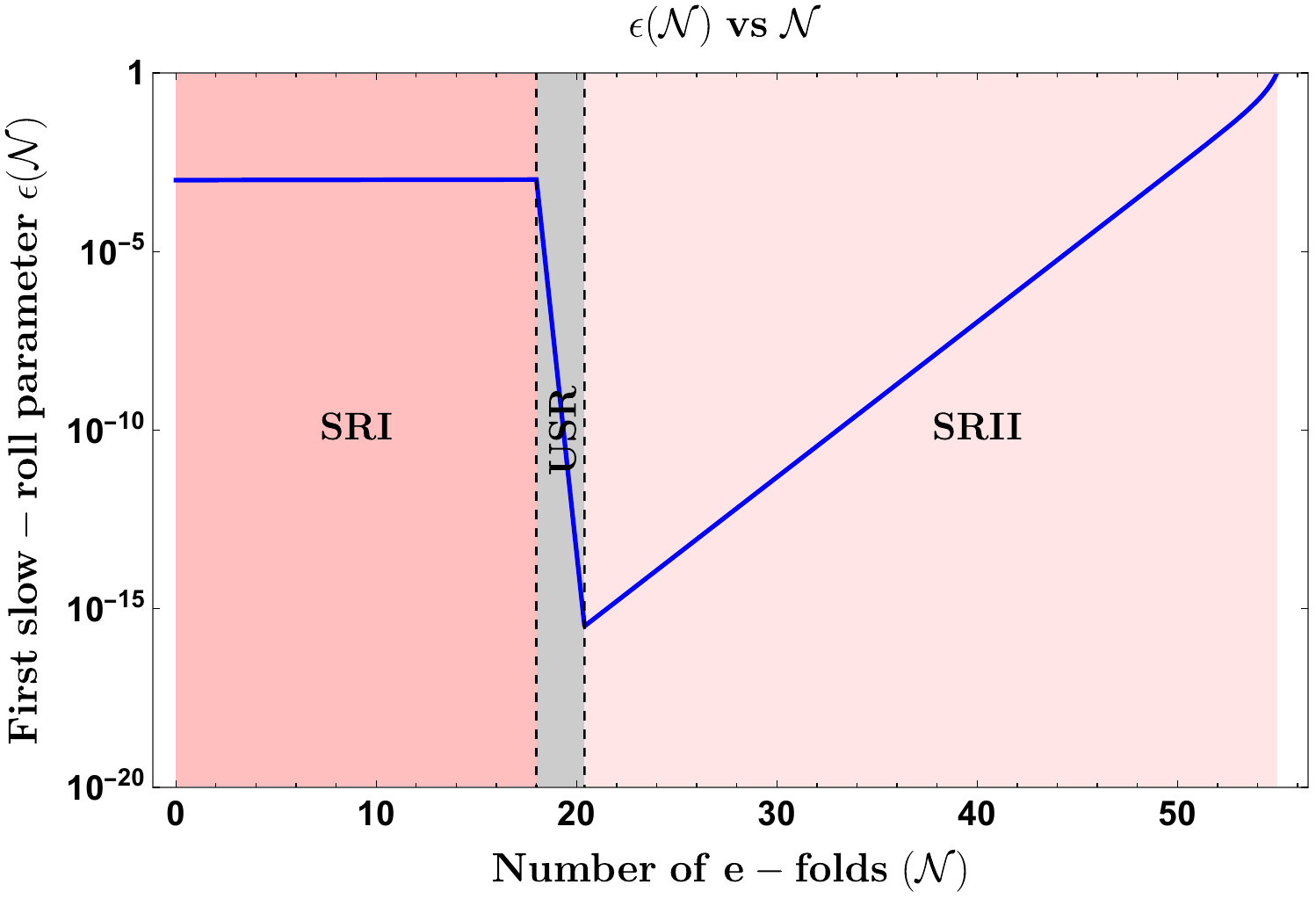}
        \label{epsilon}
    }
    \subfigure[]{
        \includegraphics[width=8.5cm,height=7.5cm] {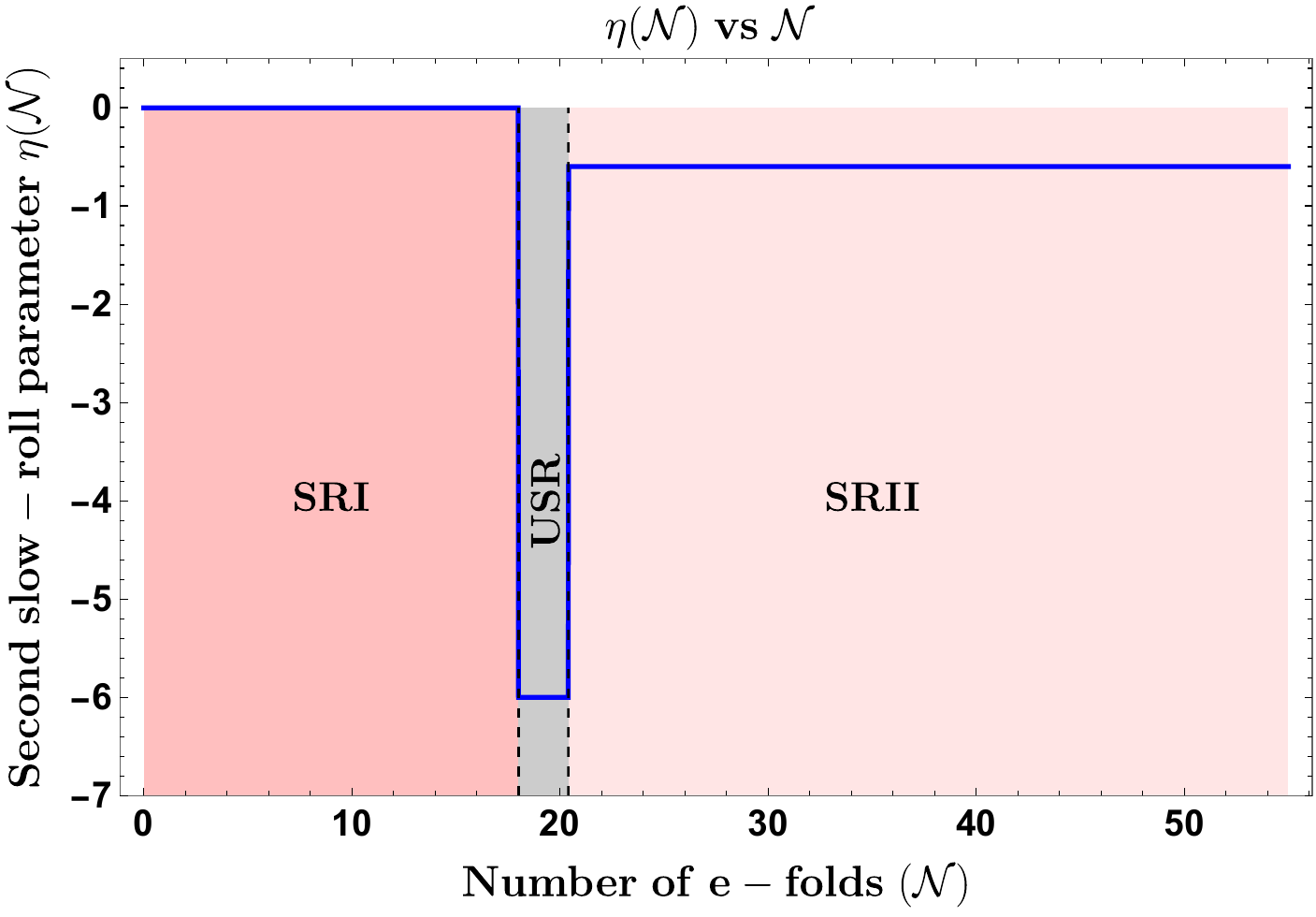}
        \label{eta}
       }
    \subfigure[]{
      	\includegraphics[width=8.5cm,height=7.5cm] {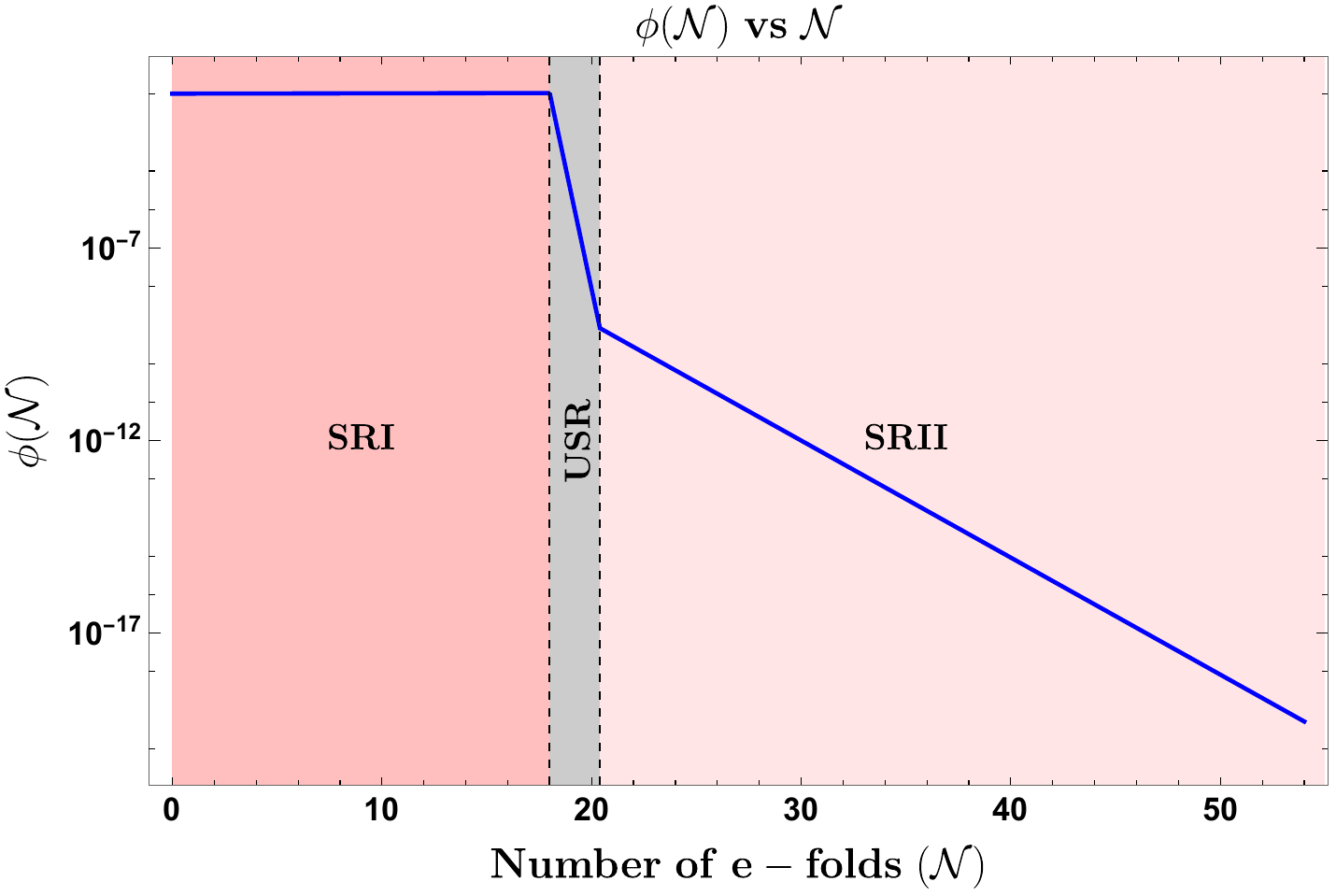}
        \label{phi}
    }
    \subfigure[]{
        \includegraphics[width=8.5cm,height=7.5cm] {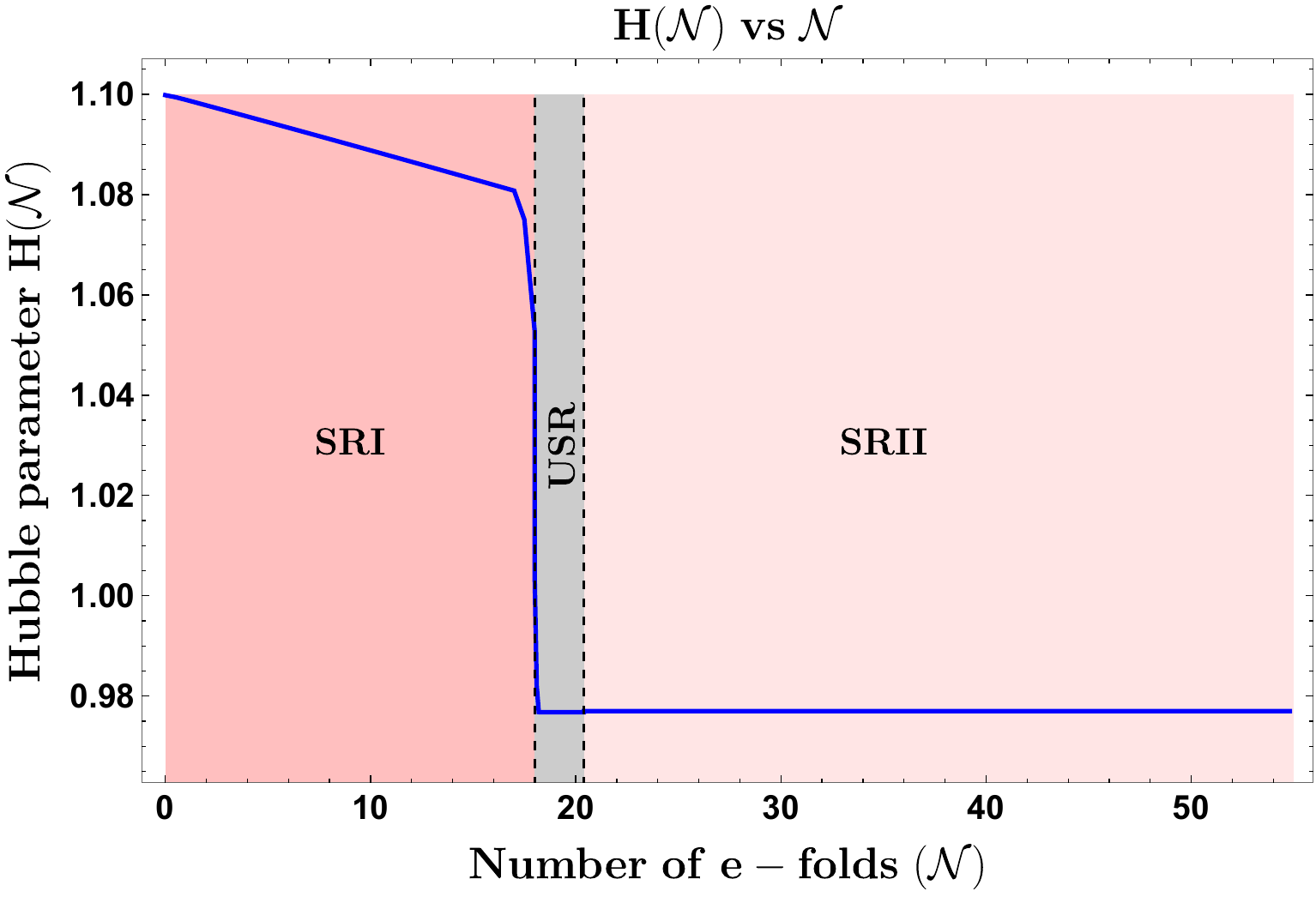}
        \label{hubble}
       }
    	\caption[Optional caption for list of figures]{Figure represents evolution of the various parameters, $\epsilon({\cal N})$ in the top-left panel, $\eta({\cal N})$ in the top-right panel, $\phi({\cal N})$ in the bottom-left panel, $H({\cal N})$ in the bottom-right panel, with respect to the e-folds ${\cal N}$ and throughout the three phases of SRI-USR-SRII. } 
    	\label{paramsevolution}
    \end{figure*}

\textcolor{black}{In fig.(\ref{paramsevolution}) we present the evolution of the respective parameters as stated before. }
\textcolor{black}{From fig.(\ref{paramsevolution}), we visualize how each of the parameters as well the background scalar field behaves across the three phases in our setup. Another crucial parameter involved in our study is the effective sound speed, $c_{s}$ and it is defined in terms of few time-dependent coefficients which are also functions of the $c_{i}\;\forall\;i=2,3,4,5$. The definition of the respective $c_{i}-$dependent functions are as follows:}
\bea \label{coeffA}
{\cal A}&\equiv& \frac{\dot{\bar{\phi}}^2}{2}\Bigg(c_2+12c_3Z+54c_4Z^2+120c_5Z^3\Bigg),\\
\label{coeffB}  {\cal B}&\equiv& 
   \frac{\dot{\bar{\phi}}^2}{2}\Bigg\{c_2+4c_{3}Z\Big(2-\eta\Big)+2c_{4}Z^{2}\Big(13-6\big(\epsilon+2\eta\big)\Big)-24c_5Z^{3}\big(2\epsilon+1\big)\Bigg\}.
\eea
\textcolor{black}{and using this the parameter $c_{s}$ is defined as, $c_{s}^{2} = {\cal B}/{\cal A}$. The effective sound speed is a crucial parameter in that it contains almost all of the coefficients present within the original CGT action, see eqn.(\ref{gaction}).
}

\subsection{Solutions from the perturbed second order action}
\label{s2b}

\textcolor{black}{Following the general discussion about the CGT action and for the background time-dependent Galileon field, we now present the mode solutions which comes from analysing the second-order action for the comoving curvature perturbations. } 
To this effect, we quote the expression of 
the second-order action for the curvature perturbation:
\begin{eqnarray} \label{s2s2} S^{(2)}_{\zeta}=\displaystyle \int d\tau\;d^3x\;a(\tau)^2\frac{{\cal A}}{H^2}\left(\zeta^{'2}-c^2_s\left(\partial_i\zeta\right)^2\right) = \int d\tau\;\frac{d^{3}\mbf{k}}{(2\pi)^{3}}a(\tau)^2\frac{\cal A}{H^{2}}\left(|\zeta_{\mbf{k}}^{'}(\tau)|^{2} - c_{s}^{2}k^{2}|\zeta_{\mbf{k}}(\tau)|^{2}\right)
\end{eqnarray}
where the second equation is written after performing the Fourier transform. 

Variation of this action gives us the following second-order differential equation, commonly known as the Mukhanov-Sasaki equation for the scalar modes in Fourier space which is written as:
\bea 
\zeta^{''}_{\bf k}(\tau)+2\frac{z^{'}(\tau)}{z(\tau)}\zeta^{'}_{\bf k}(\tau) +c^2_sk^2\zeta_{\mbf{k}}(\tau)=0.
\eea
where we introduce a new variable $\displaystyle{z(\tau)=a\sqrt{2{\cal A}}/H^2}$ for the simplification purpose. We then solve the above differential equation in the three regions of interest during inflation, which includes the first slow-roll (SRI), the ultra-slow roll (USR), and the second slow-roll (SRII) phases. This method gives us the respective mode solutions in the three phases of interest written as follows:
\begin{eqnarray} \label{s2modes}
\zeta_{\mbf{k}}(\tau)
&=& \displaystyle
\displaystyle \left(\frac{iH^{2}}{2{\cal A}}\right)\frac{1}{(c_{s}k)^{3/2}}\times\left\{
	\begin{array}{ll}
		\displaystyle \left[\alpha^{(1)}_{\mbf{k}}\left(1+ikc_{s}\tau\right)\exp{\left(-ikc_{s}\tau\right)} - \beta^{(1)}_{\mbf{k}}\left(1-ikc_{s}\tau\right)\exp{\left(ikc_{s}\tau\right)}\right] & \mbox{when}\quad  k < k_s \\ 
			\displaystyle 
			\displaystyle \left[\alpha^{(2)}_{\mbf{k}}\left(1+ikc_{s}\tau\right)\exp{\left(-ikc_{s}\tau\right)} - \beta^{(2)}_{\mbf{k}}\left(1-ikc_{s}\tau\right)\exp{\left(ikc_{s}\tau\right)}\right] & \mbox{when }  k_s\leq k < k_e  \\
   \displaystyle 
			\displaystyle \left[\alpha^{(3)}_{\mbf{k}}\left(1+ikc_{s}\tau\right)\exp{\left(-ikc_{s}\tau\right)} - \beta^{(3)}_{\mbf{k}}\left(1-ikc_{s}\tau\right)\exp{\left(ikc_{s}\tau\right)}\right] & \mbox{when }  k_e\leq k\leq k_{\rm end}  
	\end{array} \right. \end{eqnarray}
The general approach is to start with a quantum initial boundary condition which is taken to be the standard Bunch-Davies initial condition, which is actually a Euclidean vacuum state in this context. This gives us $\alpha_{\mbf{k}}^{(1)} = 1$ and $\beta_{\mbf{k}}^{(1)} = 0$. Then, due to having a sharp transition from one phase to another, new sets of Bogoliubov coefficients for the mode functions in the new phase can be obtained by making use of the Israel matching conditions at the sharp transition scales $k_{s}$ (SRI to USR) and $k_{e}$ (USR to SRII). This change in the behaviour of the Bogoliubov coefficients towards a Non-bunch Davies type vacuum is an important reason for the significant enhancement observed in the scalar power spectrum amplitude. Explicit expressions for the Bogoliubov coefficients are given in Appendix \ref{App:A}. 

\subsection{Impact of the non-renormalization theorem}
\label{s2c}

In this subsection, we briefly describe the implication of the non-renormalization theorem for Galileon inflation.
The theorem states that the couplings of the theory remain protected against any radiative corrections even when the Galilean symmetry gets mildly broken. As a result, we can bypass the need to perform the renormalization and resummation procedures to obtain the one-loop corrected scalar power spectrum in Galileon theory. Other significant consequences of this theorem include the validity of successful inflation with having a prolonged SRII phase and the ability to control large fluctuations and their associated non-Gaussianities produced during the sharp transitions in the USR region.

For a successful inflation in this case, we require a mild breaking of the Galilean shift symmetry, which happens by going from a de-Sitter background to a quasi-de-Sitter one. The symmetry transformation of the terms built from curvature perturbation $\zeta$ is written by following the definition in eqn.(\ref{s1a}):
\bea \label{s1b} \zeta \rightarrow \zeta - \frac{H}{\dot{\bar{\phi}}}a.\delta x, \quad\quad \partial_{i}\zeta \rightarrow \partial_{i}\zeta - \frac{H}{\dot{\bar{\phi}}}a_{i}, \quad\quad \zeta^{'} \rightarrow \zeta^{'} - \frac{H}{\dot{\bar{\phi}}}a_{0}, \quad\quad \partial^{2}\zeta \rightarrow \partial^{2}\zeta. \eea
where $\bar{\phi} \equiv \bar{\phi}(t)$ is the time-dependent background Galileon field. From the above equation, we see that only the term $\partial^{2}\zeta$ remains invariant under Galilean symmetry, and the terms $\zeta, \zeta^{'},$ and $\partial_{i}\zeta$ show mild breaking of the said symmetry. Based on this, let us briefly look into how radiative corrections become unimportant in the scalar power spectrum in Galileon theory. In single-field inflation, the dominant one-loop corrections to the scalar power spectrum come from the operator $\zeta^{'}\zeta^{2}$, due to its coefficient containing the factor $\partial_{\tau}(\eta/c_{s}^{2})$ which is large during a sharp transition. However, this term is absent from the third-order action of Galileon theory even when it breaks the Galilean shift symmetry. This absence becomes evident by the use of eqn.(\ref{s1b}), which converts the operator into a quantity evaluated at the boundary. There exist other terms in the Galileon theory which show mild symmetry breaking but are absent from the final third-order action because of the possibility of performing field redefinition or formation of boundary terms, reducing the allowed number of terms. Detailed discussion on this topic can be found by the authors in \cite{Choudhury:2023hvf}.

Keeping the above discussion in mind, only a few terms are allowed in the third-order action, which are called the bulk self-interaction terms and includes: $\zeta^{'3}, \zeta^{'2}\partial^{2}\zeta, \zeta^{'}(\partial_{i}\zeta)^{2}, \partial^{2}\zeta(\partial_{i}\zeta)^{2}$. These will be used in the next section to describe the third-order action necessary for calculating the one-loop effects.

\subsection{One-loop corrected power spectrum from third order action}
\label{s2d}

In this section, we consider the third-order action in the curvature perturbations which is constructed using the analysis done in the previous section. This action is formed using the Galilean symmetry-breaking terms introduced previously which collectively form the bulk self-interaction terms. The resulting action has the form \cite{Burrage:2010cu,Choudhury:2023hvf, Choudhury:2023kdb}:
\begin{eqnarray}
S^{3}_{\zeta} = \int d\tau\; d^{3}x\frac{a(\tau)^{2}}{H^{3}}\bigg[\frac{{\cal G}_1}{a}\zeta^{'3}+\frac{{\cal G}_2}{a^2}\zeta^{'2}\left(\partial^2\zeta\right)+\frac{{\cal G}_3}{a}\zeta^{'}\left(\partial_i\zeta\right)^2+\frac{{\cal G}_4}{a^2}\left(\partial_i\zeta\right)^2\left(\partial^2\zeta\right)\bigg]
\end{eqnarray}
where the couplings for each operator in the action have the following expressions:
\begin{eqnarray}
    {\cal G}_1:&\equiv& \frac{2H                    \dot{\bar{\phi}}^3}{\Lambda^3}            \Bigg(c_3+9c_4Z+30c_5Z^2\Bigg),\\
    {\cal G}_2:&\equiv& -\frac{2                    \dot{\bar{\phi}}^3}{\Lambda^3}\Bigg(c_3+6c_4Z+18c_5Z^2\Bigg),\\ 
    {\cal G}_3:&\equiv& 
       -\frac{2H\dot{\bar{\phi}}^3}{\Lambda^3}\Bigg(c_3+7c_4Z+18c_5Z^2\Bigg)-\frac{2\dot{\bar{\phi}}^3H\eta}{\Lambda^3}\Bigg(c_3+6c_4Z+18c_5Z^2\Bigg),\\
    {\cal G}_4:&\equiv& 
        \frac{\dot{\bar{\phi}}^3}{\Lambda^3}\bigg\{c_3+3c_4Z+6c_5\bigg[Z^2+\frac{\dot{H}\dot{\bar{\phi}}^2}{\Lambda^6}\bigg]\bigg\}-\frac{3\dot{\bar{\phi}}^4H\eta}{\Lambda^6}\bigg\{c_4+4c_5Z\bigg\},
\end{eqnarray}
and the parameter $Z$ is defined as $Z = H\dot{\bar{\phi}}/\Lambda^{2}$.

The one-loop effects are calculated using the aforementioned third-order action by working with the Schwinger-Keldysh (in-in) formalism. After taking care of all the possible Wick contractions when calculating the one-loop contributions from each interaction operator, we perform the necessary temporal and momentum integrals to arrive at the result for the three phases. We now quote the total scalar power spectrum, which includes the one-loop corrections by combining the individual contributions in the following manner \cite{Choudhury:2023hvf}:   
\bea \label{s2tot}  
\Bigg[\Delta^{2}_{\zeta}(k)\Bigg]_{\bf Total}
&\approx& \displaystyle \Bigg[\Delta^{2}_{\zeta,\bf {Tree}}(k)\Bigg]_{\rm \textbf{SRI}} \bigg\{1 + \left(\frac{k_{e}}{k_{s}}\right)^{6}\left[|\alpha_{\mbf{k}}^{(2)} - \beta_{\mbf{k}}^{(2)}|^{2}\Theta(k-k_{e}) + |\alpha_{\mbf{k}}^{(3)} - \beta_{\mbf{k}}^{(3)}|^{2}\Theta(k-k_{s})\right]\nonumber\\
&+& \Bigg[\Delta^{2}_{\zeta,\bf {Tree}}(k)\Bigg]_{\rm \textbf{SRI}}\times \frac{1}{8{\cal A}^{2}_{*}\pi^{4}}\bigg\{-\sum^{4}_{i=1}{\cal G}_{i,\mbf{SRI}}\mbf{F}_{i,\mbf{SRI}}(k_{s},k_{*}) + \sum^{4}_{i=1}{\cal G}_{i,\mbf{USR}}\mbf{F}_{i,\mbf{USR}}(k_{e},k_{s})\;\Theta(k-k_{s}) \nonumber\\
&+& \sum^{4}_{i=1}{\cal G}_{i,\mbf{SRII}}\mbf{F}_{i,\mbf{SRII}}(k_{\rm end},k_{e})\;\Theta(k-k_{e})\bigg\}\bigg\}. \eea
which involves the tree-level SRI power spectrum: \be \displaystyle{\Bigg[\Delta^{2}_{\zeta,\bf {Tree}}(k)\Bigg]_{\rm \textbf{SRI}} = \left(\frac{H^{4}}{8\pi^{2}{\cal A} c^3_s}\right)_{*}\times\left(1  + \left(\frac{k}{k_{s}}\right)^{2}\right)}.\ee

    \begin{figure*}[htb!]
    	\centering
   {
      	\includegraphics[width=18cm,height=12cm] {
      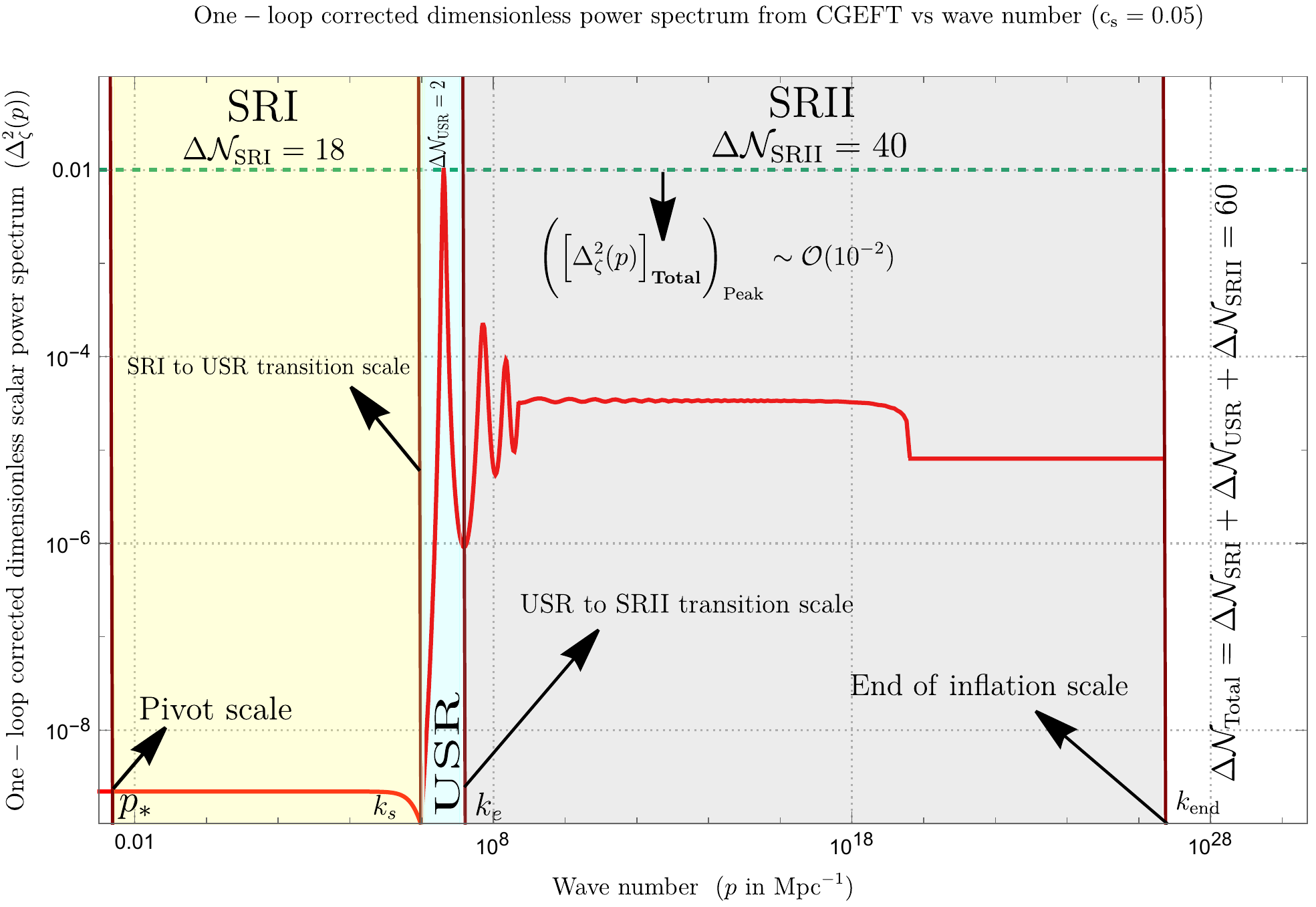}
        \label{fig2}
    }
    	\caption[Optional caption for list of figures]{Figure represents behaviour of the total scalar power spectrum, which includes the one-loop corrections, plotted against the wavenumber. This is obtained for the value of the sound speed parameter $c_{s} = 0.05$. Other important parameters are taken to be as follows: pivot scale at $p_{*} = 0.02{\rm Mpc}^{-1}$, the transition scale $k_{s}$ (SRI to USR) at = $10^{6}{\rm Mpc}^{-1}$ and the transition scale $k_{e}$ (USR to SRII) at = $10^{7}{\rm Mpc}^{-1}$ such that the condition $k_{e}/k_{s} \sim {\cal O}(10)$ is followed, and the end of SRII at $k_{\rm end} = 10^{27}{\rm Mpc}^{-1}$ such that $\Delta{\cal N} = 60$ is achieved.} 
        \label{powerGal}
    \end{figure*}

The explicit equations for the terms describing the one-loop effects above, which include ${\cal G}_{i,\mbf{SRI}}, {\cal G}_{i,\mbf{USR}}, {\cal G}_{i,\mbf{SRII}}$ and the leading order terms in $\mbf{F}_{i,\mbf{SRI}}, \mbf{F}_{i,\mbf{USR}}, \mbf{F}_{i,\mbf{SRII}}$, are mentioned in the appendix \ref{App:B}. Detailed discussion on such terms can be found in a previous work by the authors in \cite{Choudhury:2023hvf}. The total scalar power spectrum hence formed after combining the tree and one-loop contributions, which are also mentioned in eqs.(\ref{Apptree}, \ref{Apploop}) along with discussions, will be used in a later section to evaluate the scalar-induced gravitational wave spectrum, which is the main focus of the next section.

\subsection{PBH production and its comparison with recent studies}
\label{s2e}

Here we address the question of allowed PBH masses, another crucial component of our analysis resting on the properties of Galileon theory before we study the theory behind the scalar-induced GWs.

The mass of PBH is calculated using the formula given as follows:
\bea \label{s4mpbh}
\frac{M_{\rm PBH}}{M_{\odot}} &=& 1.13 \times 10^{15}\times \left(\frac{\gamma}{0.2}\right)\times \left(\frac{g_{*}}{106.75}\right)^{-1/6}\left(\frac{\tilde{c}_{s}k_{s}}{c_{s}k_{*}}\right)^{-2} \nonumber\\
&=& 1.13 \times 10^{15}\times \left(\frac{\gamma}{0.2}\right)\times \left(\frac{g_{*}}{106.75}\right)^{-1/6}\left(\frac{k_{s}}{k_{*}}\right)^{-2}\times c_{s}^{2}(1 \mp 2\delta) \nonumber\\
&=& 1.13 \times 10^{15}\times \left(\frac{\gamma}{0.2}\right)\times \left(\frac{g_{*}}{106.75}\right)^{-1/6}\left(\frac{k_{s}}{k_{*}}\right)^{-2} \times c_{s}^{2},  \eea 
where the leading order term is considered in the last line. To get an estimate of the mass of PBHs produced, we use the fact that $\gamma \sim 0.2$ which is the critical collapse factor, $g_{*} \sim 106.75$ is the number of relativistic d.o.f, the pivot scale is at $k_{*} \sim 0.02 {\rm Mpc}^{-1}$, the transition scale is set at $k_{s} \sim 10^{6}{\rm Mpc}^{-1}$, and using the effective sound speed values within the window of, $0.024 \leq c_{s} < 1$ \cite{Planck:2015sxf}. and the of solar mass value $M_{\odot} \sim 2 \times 10^{30}{\rm Kg}$, we get the resulting range of $M_{\rm PBH} \approx (10^{29} - 10^{30}){\rm Kg}$.
Based on the above formula, one can further compute the evaporation time for the PBHs from galileon inflation as:
\bea
{\rm t}_{\rm PBH}^{\text{evap.}} = 10^{64}\left(\frac{M_{\rm PBH}}{M_{\odot}}\right)^{3}\text{years}=1.4429\times 10^{109}\times\left(\frac{\gamma}{0.2}\right)^3\times \left(\frac{g_{*}}{106.75}\right)^{-1/2}\left(\frac{k_{s}}{k_{*}}\right)^{-6} \times c_{s}^{6} \;\;\text{years}.
\eea

This equation can compute the evaporation time for PBHs within the mass range of solar mass and sub-solar mass. This particular time scale depends on the transition wavenumber and the effective sound speed values. For the case of PBH with $M_{\rm PBH} \sim {\cal O}(M_{\odot})$ and $M_{\rm PBH} \sim {\cal O}(10^{-31}M_{\odot})$, the time scale of evaporation is computed as:
\bea
{\rm t}_{\rm M_{\odot}}^{\text{evap.}} &\approx& 10^{64}\text{years} \sim 10^{71}\text{seconds}\\
{\rm t}_{10^{-31}{\rm M_{\odot}} }^{\text{evap.}} &\approx& 10^{-29}\text{years} \sim 10^{-22}\text{seconds}
\eea
The above calculated time scales represent the extremes of the interval where we can obtain the evaporation time for all PBHs ranging from large mass $M_{\rm PBH} \sim {\cal O}(M_{\odot})$ to the extremely small mass $M_{\rm PBH} \sim {\cal O}(10^{-31}{\rm M_{\odot}} \sim 10^2{\rm gm})$. This fact tells us that solar mass PBH can outlive the universe's current age, making them more helpful to study in cosmology. While, for the sub-solar mass PBH, the shortest time scale is negligibly small; as a result, it evaporated not long after it got produced in the early time. The above analysis is carried out by keeping the effective sound speed constraint $0.024 \leq c_{s} <1$ satisfied, mainly we have taken $c_{s}=0.05$. Hence, in Galileon theory, the causality and unitarity constraints are always respected to give meaningful results for a spectrum of PBHs in the above-mentioned mass range. This is in contrast when working with the EFT framework for single-field slow-roll inflation, where a violation of the said constraints is required to achieve better and more meaningful results relative to those from the causal case scenario \cite{Choudhury:2023jlt}. Now we analyze the PBH mass formula in general to get more insights into the factors controlling their production.
In eqn. (\ref{s4mpbh}), we used the fact that the speed of sound parameter is a time-dependent quantity such that its value at the pivot scale is fixed to be $c_{s}$. It moves abruptly to the value $\tilde{c}_{s} \approx 1 \pm \delta$, where $\delta \ll 1$, at the transition scale $k_{s}$, again reaches the value $c_{s}$ during USR phase and then rapidly falls to the same value of $\tilde{c}_{s}$ at the transition scale $k_{e}$, and finally comes to the value $c_{s}$ when in the SRII phase till the end of inflation. See ref.\cite{Choudhury:2023jlt} for a detailed study.  

The important highlights concerning the mass of PBHs are listed as follows:
\begin{itemize}
    \item Since the mass of PBH depends on the transition scale value $k_{s} \equiv k_{\rm PBH}$, we notice that upon shifting the transition values to higher wavenumbers it is possible to generate PBHs with masses ranging from the almost solar to sub-solar masses $M_{\rm PBH} \approx (1M_{\odot} - 10^{-18}M_{\odot})$.
    \item The amplitude of the perturbations are controllable during the sharp transitions in the USR due to the non-renormalization theorem benefits. Until the perturbativity approximation of $k_{e}/k_{s} \sim {\cal O}(10)$ is satisfied, shifting of the transition scale $k_{s}$ does not affect the existing features which include the amplitudes of the one-loop corrected scalar power spectrum in the SRI (${\cal O}(10^{-9})$), USR (${\cal O}(10^{-2})$), and SRII (${\cal O}(10^{-5})$) phases.
    \item Due to the fact that the shifting of transition scale preserves the total scalar power spectrum features, we are able to cover a wide range of wavenumbers from $k \sim (10^{-2}-10^{27}{\rm Mpc}^{-1})$ which is reflected in the allowed mass of PBH and will be used in later sections to draw important conclusions for the GW spectrum induced by the one-loop corrected scalar power spectrum.
\end{itemize}

Finally, we would like to briefly compare our results on the allowed range of  PBH masses with the recent findings in the literature. The authors in refs.\cite{Riotto:2023gpm, Firouzjahi:2023ahg, Firouzjahi:2023aum} have discussed the properties of having a smooth transition such that it can control the enhanced behavior of the perturbations when going from SRI to USR and USR to SRII phases and also allow for the generation of large mass PBHs when $k_{\rm PBH} \sim 10^{5}{\rm Mpc}^{-1}$. Further studies have explored the same issue in \cite{Choudhury:2023hvf, Choudhury:2023kdb, Franciolini:2023lgy, Tasinato:2023ukp, Motohashi:2023syh}. In other studies \cite{Riotto:2023gpm, Firouzjahi:2023ahg, Firouzjahi:2023aum}, the authors have also shown explicitly that smooth transition can able to suppress the one-loop corrections in the final, loop-corrected, scalar power spectrum. Notably, the nature of the transition, smooth or sharp, is crucial to determine the status of the formed PBHs. In the case of a sharp transition, large mass PBHs are not allowed when renormalization and resummation procedures are included for the one-loop effects \cite{Choudhury:2023vuj, Choudhury:2023rks}, while in the studies \cite{Choudhury:2023hvf, Choudhury:2023kdb, Riotto:2023gpm, Firouzjahi:2023ahg, Firouzjahi:2023aum, Franciolini:2023lgy, Tasinato:2023ukp,Motohashi:2023syh, Motohashi:2023syh}.

It may be noted that, while dealing with a smooth transition \cite{Riotto:2023gpm, Firouzjahi:2023ahg, Firouzjahi:2023aum} and a sharp transition \cite{Kristiano:2022maq, Riotto:2023hoz, Kristiano:2023scm, Franciolini:2023lgy, Motohashi:2023syh} have not mentioned the need for a renormalization and resummation procedure to arrive at the conclusion of controlled one-loop effects and the production of large mass PBHs. 
In the case of Galileon inflation, we have demonstrated that  taking into account the one-loop corrections in the scalar power spectrum along with the non-renormalization theorem, it is possible to both control the large perturbations during sharp transitions and allow for solar mass as well as sub-solar mass PBH production from the underlying theory.   

\subsubsection{Calculation of the PBH mass fraction}
\label{s2e1}

\textcolor{black}{Here we briefly outline the necessary computations required to evaluate the PBH mass fraction at formation time and the corresponding fraction of the current dark matter density present in the form of PBH, also known as its abundance.}

\textcolor{black}{Our computations are based on the Press-Schechter formalism to understand the estimates of PBH mass fraction. We begin by assuming that the perturbations in the primordial overdensity, as soon as it re-enters into the Horizon, collapse into forming the PBHs once it meets a critical condition criterion on its value, and this formation occurs primarily in a radiation-dominated Universe. The mass of the resulting PBH, depending on the volume of the Hubble horizon at the time of formation, is already given by the expression in eqn.(\ref{s4mpbh}). Based on this, one can compute the present day PBH abundance $f_{\rm PBH}$ as follows: }
\bea
\label{fPbh}
f_{\rm PBH}&=&1.68\times 10^8 \bigg[\frac{\gamma}{0.2}\bigg]^{\frac{1}{2}}\bigg[\frac{g_{*}}{106.75}\bigg]^{-\frac{1}{4}}\bigg[\frac{M_{\rm PBH}}{M_{\odot}}\bigg]^{-\frac{1}{2}}\times \beta(M_{\rm PBH}),
\eea
\textcolor{black}{which incorporates the PBH mass fraction $\beta$. This quantity gets evaluated within the Press-Schechter formalism after integrating over a Gaussian distribution for the overdensities $\delta(t,{\bf x})$ which informs about the likelihood that a certain value of overdensity is possible: }
\bea
\beta(M_{\rm PBH}) \simeq \gamma \bigg[\frac{\sigma_{M_{\rm PBH}}}{\sqrt{2 \pi}\delta_{\rm th}}\bigg]\rm exp \bigg(-\frac{\delta_{th}^2}{2\sigma_{M_{\rm PBH}}^2}\bigg),
\eea
\textcolor{black}{where the integration gets performed above some limiting or threshold value of the overdensity denoted above by $\delta_{\rm th}$. Here, we adopt the criterion of the critical collapse theory, which allows one to determine the threshold value as equivalent to the equation of state of the RD fluid, $\delta \geq \delta_{\rm th} = 1/3$. Another quantity essential to complete the estimate for $\beta(M_{\rm PBH})$ is the variance of the distribution $\sigma_{\rm PBH}$, and this is calculated with the aid of the scalar power spectrum amplitude, here it will be the one-loop corrected scalar power spectrum from eqn.(\ref{s2tot}), and this gets implemented as follows:  }
\bea
\label{variance}
\sigma_{M_{\rm PBH}}^2
= \frac{16}{81}\int_{0}^{\infty}\frac{dk}{k}\;(kR)^{4} e^{-k^{2}R^{2}/2}\;\Bigg[\Delta^{2}_{\zeta}(k)\Bigg]_{\bf Total},
\eea
\textcolor{black}{where $R=1/(\tilde{c_{s}}k_{s})$ corresponds to the horizon scale over which the primordial perturbations are coarse-grained with the window function chosen to be a Gaussian in the above formula. The above discussion will prove sufficient to later analyze the PBH abundance resulting from the scalar power spectrum as shown in fig.(\ref{powerGal}). }

\section{Scalar Induced Gravitational Waves}
\label{s3}

In this section, we present a concise review of the theory behind the scalar-induced gravitational waves necessary to understand the calculation of the observationally relevant GW density parameter, which we will finally evaluate in the next section with the case of a one-loop corrected scalar power spectrum.
We follow the refs.\cite{Ananda:2006af, Baumann:2007zm, Inomata:2016rbd} for the derivation presented in this section. 

To this end, let us start with a perturbed version of the spatially flat FLRW metric while working with the conformal Newtonian gauge:
\bea
ds^{2} = a^{2}(\tau)\left\{-(1+2\Phi)d\tau^{2} + \left((1-2\Psi)\delta_{ij}+\frac{1}{2}h_{ij}\right)dx^{i}dx^{j}\right\},
\eea
where $\Phi, \Psi$ are the first-order scalar perturbation modes, commonly referred to as the Bardeen potentials, and $h_{ij}$ is the purely second-order tensor perturbation satisfying the additional property of $\delta^{ij}h_{ij} = \delta^{jk}\partial_{k}h_{ij} = 0$. The basic assumption to be followed in the further analysis of this metric is that the tensor perturbation modes only exist at second-order and no second-order scalar or vector modes are going to be considered. Also, in the absence of anisotropic stress, we can ultimately take $\Phi = \Psi$. 

Using this metric, we proceed to solve the Einstein field equations for the second-order perturbations. Here we require an operator which extracts the transverse, traceless part of this equation since we are ultimately concerned with the evolution of the tensor modes. This is achieved through a projection operator $\hat{\cal T}^{ab}_{ij}$ having the properties:
\bea
\hat{\cal T}^{ab}_{ij}\hat{\cal T}^{ij}_{mn} = \hat{\cal T}^{ab}_{mn},\quad\quad\quad\partial_{j}\hat{\cal T}^{ab}_{ij}{\cal P}_{ab} = \hat{\cal T}^{ab}_{jj}{\cal P}_{ab} = 0. 
\eea
for an arbitrary second-rank tensor ${\cal P}_{ab}$. This operator is implemented onto the Einstein equations as follows:
\bea
\hat{\cal T}^{ab}_{ij}G^{(2)}_{ab} = \hat{\cal T}^{ab}_{ij}T^{(2)}_{ab}
\eea
where we have used the convention $M_{p} = 1/\sqrt{8\pi G} = 1$. This gives us the following equation for the tensor modes:
\bea
\label{s3h}
h^{''}_{ab} + 2{\cal H}h^{'}_{ab} - \nabla^{2}h_{ab} = -4\hat{\cal T}^{ij}_{ab}{\cal S}_{ij},
\eea
where the notation $'$ denotes conformal time derivative throughout this section and ${\cal S}_{ij}$ is a source term that contains quadratic contributions of the first-order scalar perturbations and whose explicit form will be given below shortly. 

The analysis of the aforementioned equation is more illuminating in the Fourier space which includes the polarization information of the tensor modes and would ultimately help in obtaining the tensor power spectrum. For this, we start with the Fourier transform of $h_{ab}$:
\bea
h_{ab}(\tau, \mbf{x}) = \int\frac{d^{3}\mbf{k}}{(2\pi)^{3}}e^{i\mbf{k.x}}\left(e_{ab}h_{\mbf{k}}(\tau) + \bar{e}_{ab}\bar{h}_{\mbf{k}}(\tau)\right),
\eea
which includes the two polarization tensors defined as:
\begin{align}
    e_{ab}(\mbf{k})=\frac{1}{\sqrt{2}}\left(e_{a}(\mbf{k})e_{b}(\mbf{k}) - \bar{e}_{a}(\mbf{k})\bar{e}_{b}(\mbf{k})\right),\quad\quad \bar{e}_{ab}(\mbf{k})=\frac{1}{\sqrt{2}}\left(e_{a}(\mbf{k})\bar{e}_{b}(\mbf{k}) + \bar{e}_{a}(\mbf{k})e_{b}(\mbf{k})\right),
\end{align}
where $e_{a}(\mbf{k}), \bar{e}_{b}(\mbf{k})$ behaves as basis vectors, which are orthonormal and orthogonal to the momentum vector $\mbf{k}$ and hence satisfy $e_{a}(\mbf{k}).\bar{e}_{b}(\mbf{k}) = 0$. 
As for the RHS in eqn.(\ref{s3h}), we can write it using the polarization tensors and Fourier transform of the source as:
\bea
\hat{\cal T}^{ij}_{ab}{\cal S}_{ij} = \int\frac{d^{3}\mbf{k}}{(2\pi)^{3}}e^{i\mbf{k.x}}[e_{ab}(\mbf{k})e^{ij}(\mbf{k}) + \bar{e}_{ab}(\mbf{k})\bar{e}^{ij}(\mbf{k})]{\cal S}_{ij}(\mbf{k})
\eea
The above equations are combined together to give us the following Fourier space version of eqn.(\ref{s3h}):
\bea
\label{s3hk}
h^{''}_{\mbf{k}} + 2{\cal H}h^{'}_{\mbf{k}} + k^{2}h_{\mbf{k}} = {\cal S}(\mbf{k},\tau),
\eea
with the source in RHS defined to be: 
\bea
{\cal S}(\mbf{k},\tau) &=& \int\frac{d^{3}\mbf{q}}{(2\pi)^{3}}e^{ij}(\mbf{k})q_{i}q_{j}F(\mbf{k},\mbf{q},\tau). \eea

Solving eqn.(\ref{s3hk}) requires knowledge of the time-dependent behaviour of the scalar perturbations present inside the function $F(\mbf{k,q}, \tau)$ written explicitly as:
\bea
F(\mbf{k, q},\tau) &=& \frac{4}{3(1+w)}\left(2(5+3w)\Phi_{q}(\tau)\Phi_{k-q}(\tau) + \frac{4}{{\cal H}}(\Phi_{q}(\tau)\Phi^{'}_{k-q}(\tau) + \Phi_{k-q}(\tau)\Phi^{'}_{q}(\tau)) + \frac{4}{{\cal H}^{2}}\Phi^{'}_q(\tau)\Phi^{'}_{k-q}(\tau)\right) \eea
where the above contains the explicit scalar mode couplings to second order and is obtained using the relations from the first-order Einstein equations and the parameter, $w = P/\rho$, denotes the equation of state. The Hubble parameter in this section is defined as, ${\cal H} = \partial_{\tau}\ln{(a(\tau))} = 2/(1+3w)\tau$. In our subsequent analysis, we will work with only one polarization mode.

In general, we solve the inhomogeneous differential equation, eqn.(\ref{s3hk}), using Green's function method. To see this, consider a change in variable $h_{\mbf{k}}(\tau) \longrightarrow a(\tau)h_{\mbf{k}}(\tau)$ which gives us the following solution for the same differential equation:
\bea
h_{\mbf{k}}(\tau) = \frac{1}{a(\tau)}\int^{\tau}_{\tau_{0}}d\tilde{\tau}G_{\mbf{k}}(\tau, \tilde{\tau})a(\bar{\tau}){\cal S}(\mbf{k},\tilde{\tau}). \eea
where the Green's function $G_{\mbf{k}}(\tau, \tilde{\tau})$ satisfies:
\bea
G^{''}_{\mbf{k}}(\tau, \tilde{\tau}) + \left(k^{2} - \frac{a^{''}(\tau)}{a(\tau)} \right)G_{\mbf{k}}(\tau, \tilde{\tau}) = \delta(\tau - \tilde{\tau}).
\eea
Now, from the definition of the tensor power spectrum, we write the following expressions based on our current analysis:
\bea
\langle h_{\mbf{k}}(\tau)h_{\tilde{\mbf{k}}}(\tau)\rangle &=& \frac{2\pi^{2}}{k^{3}}\delta(\mbf{k + \tilde{k}})\Delta^{2}_{h}(k,\tau)\\
&=& \frac{1}{a(\tau)^{2}}\int^{\tau}_{\tau_{0}}d\tilde{\tau}_{1}\int^{\tau}_{\tau_{0}}d\tilde{\tau}_{2}a(\tau_{1})G_{\mbf{k}}(\tau,\tau_{1})a(\tau_{2})G_{\tilde{\mbf{k}}}(\tau,\tau_{2})\langle {\cal S}(\mbf{k},\tilde{\tau}_{1}){\cal S}(\tilde{\mbf{k}},\tilde{\tau}_{2})\rangle. \eea
To simplify this further we use the fact that if the perturbations originated during the inflationary epoch then the behaviour of the gravitational potential, $\Phi_{\mbf{k}}(\tau)$, can be understood using a decomposition into $(a):$ the transfer function $\Phi(k\tau)$ which helps to describe the evolution of the potential, and $(b):$ a component describing the primordial scalar perturbations amplitude, $\phi_{\mbf{k}}$, which follows Gaussian statistics. Hence, we can write the function $F(\mbf{k, q},\tau)$ using the transfer function and its derivatives and pull the primordial perturbation part outside the function. 

We are left to evaluate the correlation function of the source functions in terms of the correlations between the scalar perturbations. After using Wick's theorem to perform the contractions, we get only $2$ connected components which are proportional to  $\delta(\mbf{k + \tilde{k}})\delta(\mbf{q + \tilde{q}})\Delta^{2}_{\phi}(q)\Delta^{2}_{\phi}(|\mbf{k-q}|)$ and $\delta(\mbf{k + \tilde{k}})\delta(\mbf{k-q} + \tilde{\mbf{q}})\Delta^{2}_{\phi}(q)\Delta^{2}_{\phi}(|\mbf{k-q}|)$. An important point to further note is the invariance under the change of variables and $\mbf{k} \longrightarrow -\mbf{k}$ in the function $F(\mbf{k,q},\tau)$ and the invariance of the projection $e^{ij}(\mbf{k})q_{i}q_{j} = q^{2}\sin^{2}\theta\cos 2\beta/\sqrt{2}$ under $\mbf{k} \longrightarrow -\mbf{k}$, where $\theta$ represents angle between $\mbf{k, q}$ with  $\cos\theta = \alpha = \displaystyle{\mbf{k.q}/{kq}}$ and $\beta$ is the azimuth angle. 

Using these facts we write down the complete expression for the tensor power spectrum:
\bea \label{s3ph}
\Delta^{2}_{h}(k,\tau) = \int^{\infty}_{0}dq\int^{1}_{-1}d\alpha \frac{k^{3}q^{3}}{|\mbf{k-q}|^{3}}(1-\alpha^{2})^{2}\Delta^{2}_{\phi}(|\mbf{k-q}|)\Delta^{2}_{\phi}(q){\cal G}(k,q, \tau),
\eea
where the function ${\cal G}$ is of the form:
\bea
{\cal G}(k,q, \tau) = \frac{2}{a(\tau)^{2}}\int^{\tau}_{\tau_{0}}d\tau_{1}\int^{\tau}_{\tau_{0}}d\tau_{2}\; a(\tau_{1})G_{\mbf{k}}(\tau,\tau_{1})a(\tau_{2})G_{\mbf{k}}(\tau,\tau_{2})F(\mbf{k,q},\tau_{1})F(\mbf{k,q},\tau_{2}).
\eea

Let us note that the function $F$ contains the necessary information regarding the evolution of the gravitational potentials through the transfer function. Hence, we require the solutions of the following equation for the time-dependent potentials in the absence of any entropy perturbations: 
\bea \label{s3pot}
\Phi^{''}_{\mbf{k}}(\tau) + 3{\cal H}(1+w)\Phi^{'}_{\mbf{k}}(\tau) + (2{\cal H}^{'}+(1+3w){\cal H}^{2} + wk^{2})\Phi_{\mbf{k}}(\tau) = 0.
\eea
In the super-Hubble limit, it is possible to relate the two-point correlation function of primordial perturbations with the correlation function of the comoving curvature perturbations. To achieve this, we need to consider a particular gauge condition more suitable for outside the horizon. In this gauge, the perturbations in the scalar field satisfy $\delta\phi(\mbf{x},\tau) = 0$, which then enables us to write the comoving curvature perturbation as follows: 
\bea \zeta(\mbf{x},\tau) = -\Phi(\mbf{x},\tau)\left(\frac{5+3w}{3+3w}\right) - \frac{2\Phi^{'}}{3{\cal H}(1+w)}. \eea
From eqn.(\ref{s3pot}), in the super-Hubble limit, the potential $\Phi$ comes out as a constant and this gives us the required relation for the Fourier modes: 
\bea \zeta_{\mbf{k}} = -\frac{5+3w}{3+3w}\phi_{\mbf{k}} \quad\quad\implies\quad\quad
\Delta^{2}_{\zeta}(k) = \left(\frac{5+3w}{3+3w}\right)^{2}\Delta^{2}_{\phi}(k). \eea

The integral in eqn.(\ref{s3ph}) can be simplified even further when considering a variable change from $q, \alpha \longrightarrow u, v$ where $u,v$ are the new dimensionless variables with the form $u = |\mbf{k - q}|/k$ and $v = q/k$. The advantage of this substitution is to make the symmetries of the integrand under variable exchange more explicit which is discussed in the paragraph before eqn.(\ref{s3ph}). The final version of the tensor power spectrum in terms of the newly defined dimensionless variables follows the relation \cite{Kohri:2018awv}:
\bea \label{s3phInt}
\Delta^{2}_{h}(\tau, k) = 4\int^{\infty}_{0}dv\int^{1+v}_{|1-v|}du\left(\frac{4v^{2}-(1+v^{2}-u^{2})^{2}}{4vu}\right)^{2}\left(\frac{3+3w}{5+3w}\right)^{4}{\cal I}^{2}(u,v,x)\Delta^{2}_{\zeta}(kv)\Delta^{2}_{\zeta}(ku),
\eea
with the function ${\cal I}(u,v,x)$ is defined as follows: 
\bea
{\cal I}(u,v,x) &=& \int^{x}_{0}d\tilde{x}\frac{a(\tilde{\tau})}{a(\tau)}kG_{\mbf{k}}(\tau,\tilde{\tau})F(u,v,\tilde{x}) \eea
where $x \equiv k\tau$ substitution is used.

Now, the fraction of the total energy density in GWs is related to the tensor power spectrum through the relation \cite{Kohri:2018awv}:
\bea \label{s3omega}
\Omega_{\rm GW}(\tau, k) = \frac{\rho_{\rm GW}(\tau, k)}{\rho_{\rm total}(\tau)} = \frac{1}{24}\left(\frac{k}{a(\tau)H(\tau)}\right)^{2}\overline{\Delta^{2}_{h}(\tau, k)},
\eea
where the overline represents the time-averaging at a given point within the horizon. We are ultimately interested in the above quantity detected through observations. To determine this quantity at the current time, we first take the late-time limit, $k\tau \longrightarrow \infty$, of the integral ${\cal I}^{2}$ in eqn.(\ref{s3phInt}) and then take the time average of the remaining quantity as given in eqn.(\ref{s3omega}). The final formula for the GW spectrum assumed primarily to be produced during the radiation-dominated (RD) era requires the knowledge of the associated Green's function and the approximation $a(\tilde{\tau})/a(\tau) \approx \tilde{\tau}/\tau$ which, along with the necessary steps mentioned before to observe the spectrum today, later reduce the eqn.(\ref{s3omega}) into the following form \cite{NANOGrav:2023hvm}:
\bea
\Omega_{\rm GW}(f) = 0.39\left(\frac{g_{*}(T_{c})}{106.75}\right)^{-1/3}\Omega_{r,0}\int^{\infty}_{0}dv\int_{|1-v|}^{1+v}du\;{\cal K}(u,v)\Delta^{2}_{\zeta}(ku)\Delta^{2}_{\zeta}(kv)
\eea
where the kernel function from the RD era is \cite{Kohri:2018awv}:
\bea
{\cal K}(u,v) = \frac{3(4v^{2}-(1+v^{2}-u^{2})^{2})^{2}(u^{2}+v^{2}-3)^{4}}{1024u^{8}v^{8}}\Bigg(\bigg(\ln{\frac{|3-(u+v)^{2}|}{|3-(u-v)^{2}|}}-\frac{4uv}{u^{2}+v^{2}-3}\bigg)^{2}+\pi^{2}\Theta(u+v-\sqrt{3})\Bigg)
\eea
and pre-factors outside the integral include the energy density fraction of radiation at present time $\Omega_{r,0}$, the total number of relativistic d.o.f in SM $106.75$, and $g_{*}(T_{c})$ is the total relativistic d.o.f at the temperature $T_{c}$ which is evaluated at the time when the perturbations re-enter the horizon in the RD era. The following conversion between frequency and wavenumber is adopted:
\bea
f = 1.6 \times 10^{-9} {\rm Hz}\left(\frac{k}{10^{6}{\rm Mpc}^{-1}}\right)
\eea


\section{Results for the SIGW spectrum} \label{s4}

\begin{figure*}[htb!]
    	\centering
   {
      	\includegraphics[width=18cm,height=12cm] {
      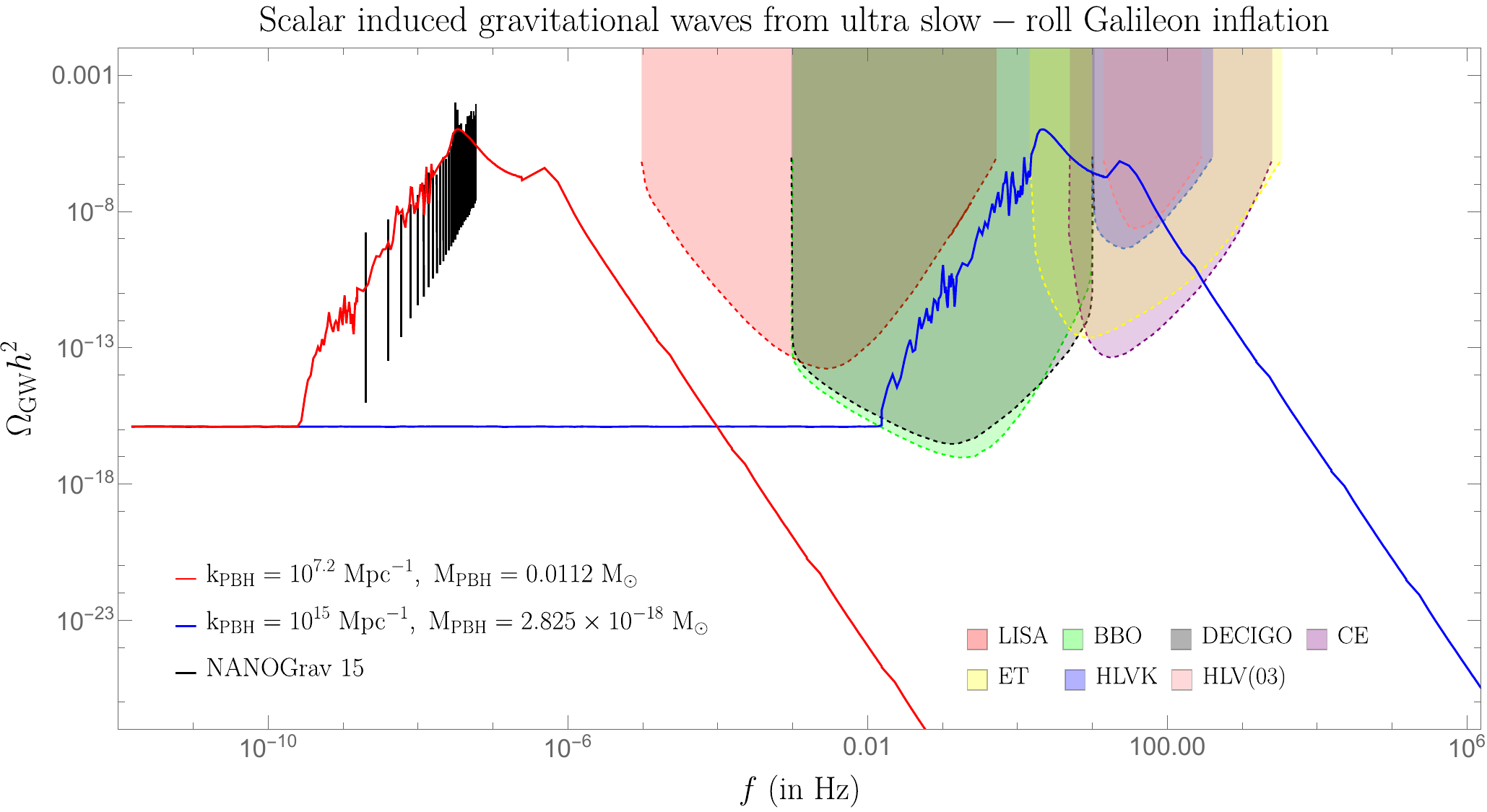}
        \label{fig3}
    }
    	\caption[Optional caption for list of figures]{The plot represents the scalar-induced GW spectrum generated using the one-loop corrected version of the scalar power spectrum obtained in Galileon inflation. This figure contains $2$ separate features each distinguished by colors red and blue which indicate different scenarios. The red spectrum is where the amplitude of the generated signal coincides with the observational result from  NANOGrav 15-year signal and it also indicates the production of $M_{\rm PBH} \sim 0.01M_{\odot}(2 \times 10^{28}{\rm Kg})$ for a transition scale of $f_{\rm PBH} \sim 10^{-8}{\rm Hz}$. The blue spectrum is where the amplitude of the signal lies within the proposed sensitivity curves of the existing and future experiments to detect high-frequency GWs, which include the following: LISA, BBO, DECIGO, Cosmic Explorer(CE), Einstein Telescope(ET), HLVK network (consisting of detectors: aLIGO in Hanford and Livingstone, aVirgo, and KAGRA), and the HLV network during the third observation run (O3). This blue curve also shows a similar spectrum of GWs generated at a high-frequency value, near the transition scale $f_{\rm PBH} \sim 1{\rm Hz}$ where sub-solar PBH, $M_{\rm PBH} \sim 10^{-18}M_{\odot}(2 \times 10^{2}{\rm Kg})$, are generated.} 
        \label{NANOGal}
    \end{figure*}

In this section, we discuss our results for the induced GW spectrum using the one-loop corrected scalar power spectrum obtained in eqn.(\ref{s2tot}) during inflation in the Galileon theory. The behavior of the obtained GW spectrum is analyzed, and the possible range of masses for the PBHs is discussed while covering the whole frequency range allowed by the observations. In fig.(\ref{NANOGal}), we present the scalar-induced GW spectrum results plotted against the allowed frequency range. This range consists of the low-frequency regime where the NANOGrav results are observed and the high-frequency regime where various other observational experiments, both existing and planned, can probe.

    \begin{figure*}[htb!]
    	\centering
   { 
   \includegraphics[width=18cm,height=6cm] {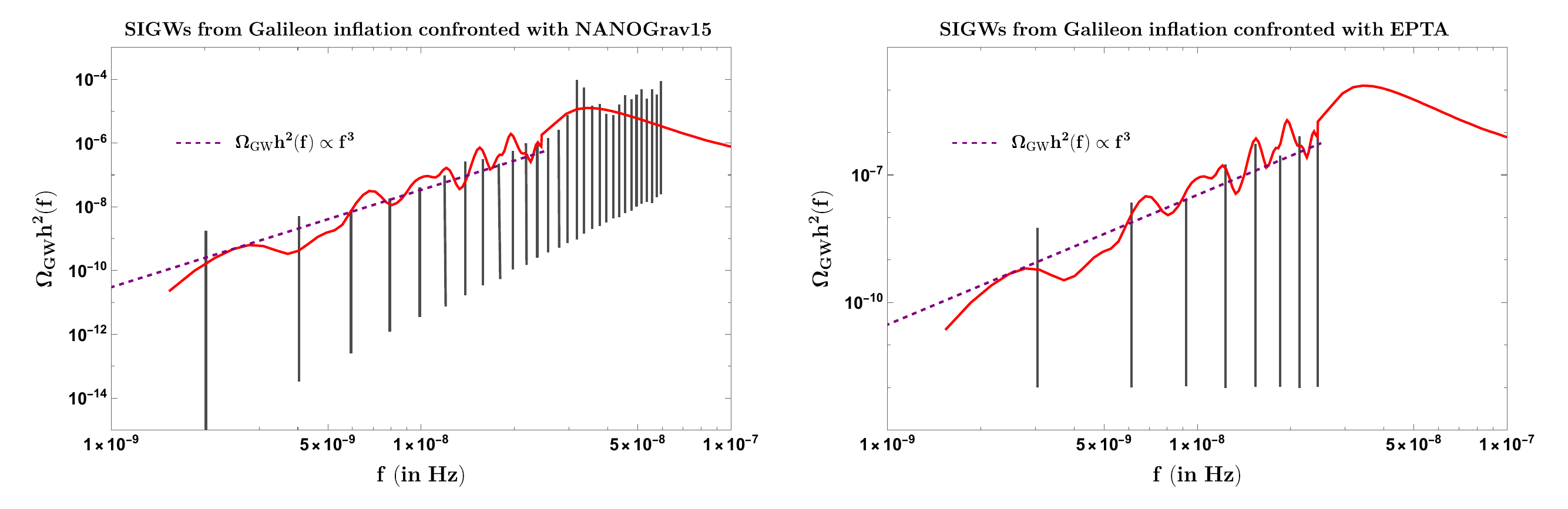}
    }
 	\caption[Optional caption for list of figures]{SIGW spectrum (in red) generated in Galileon inflation where we fix the transition wavenumber, $k_{\rm PBH}=10^{7.2}{\rm Mpc^{-1}}$ and use the effective sound parameter, $c_{s}=0.05$. The background contains the NANOGrav15 (left) and EPTA (right) signals in black and an approximate $k^{3}$ scaling shown via a dashed purple line. } 
    	\label{nanoepta}
    \end{figure*}

The values in range $(10^{-17}{\rm Hz}, 10^{-8}{\rm Hz})$ include part of the spectrum which contains information from the pivot scale $k_{*} \sim (10^{-2}{\rm Mpc}^{-1} \approx 10^{-17}{\rm Hz})$ up to the transition wavenumber $k_{\rm PBH} \sim 10^{7}{\rm Mpc}^{-1}(f_{\rm PBH} \sim 10^{-8}{\rm Hz})$. As we move higher in the frequency range from the frequency at the pivot scale, the spectrum shows a rise in amplitude, including small oscillations, which result from the highly oscillating integrand involved in the spectrum calculation. Up to a specific frequency, the behavior of the power spectrum changes, and we see that for a small region, the spectrum achieves a peak amplitude of $\Omega_{\rm GW}h^{2} \sim {\cal O}(10^{-5})$ which occurs as a result of the sharp transition encountered at the scale $k_{\rm PBH}$.  After the peak, the amplitude of the spectrum starts to fall before going through a bump-like feature visible in the plot right after the frequency $f \sim 10^{-7}{\rm Hz}$. This feature is reminiscent of the similar effect observed in the one-loop corrected scalar power spectrum,  fig.(\ref{powerGal}), for the comoving curvature perturbation, which occurs right after the transition at the wavenumber $k_{e}$ during the end of USR. Then, as we proceed towards completing the necessary e-foldings ${\cal N}= 60$~\footnote{\textcolor{black}{Let us note that the minimum number of e-foldings is dictated by requirement to address causality:}
\bea \label {ef} \textcolor{black} {{\cal N}=\ln\left(\frac{T_0}{H_0}\right)+\ln\left(\frac{H_{inf}}{T_{reh}}\right)=\ln\left(\frac{T_0}{H_0}\right)+\ln\left(\frac{T_{reh}}{M_{\rm pl}}\right)=67+\ln\left(\frac{T_{reh}}{M_{\rm pl}}\right)},\eea \textcolor{black}{where $ H_{inf}\simeq T^2_{reh}/M_{\rm pl}$ and  instantaneous (p)reheating is assumed. $T_0$ and $H_0$ are accurately known from observation, however,  scale of inflation is unknown and in principle and reheating temperature could lie in a window, $1{\rm TeV}<T_{reh}<M_{\rm pl}$, where $M_{\rm pl}\sim 2.43\times 10^{18}{\rm GeV}$. To have the total number of e-foldings, ${\cal N}=60$ from the equation (\ref{ef}), one needs to fix the rehetaing temperature at $T_{reh}\sim 2.215\times 10^{15}{\rm GeV}$. On the other hand, if we fix the number of e-foldings at ${\cal N}=55$ then in that case, $T_{reh}\sim 1.493\times 10^{13}{\rm GeV}$ . However, it is difficult to realize inflation at  low energy scales due to extreme fine-tuning  of equation of state of the inflation \cite{Steinhardt:2004rf}:}
\bea \textcolor{black}{1+\omega_{\phi}\simeq 10^{10}\times\left(\frac{H_{inf}}{T_{reh}}\right)^4=10^{10}\times\left(\frac{T_{reh}}{M_{\rm pl}}\right)^4}.\eea 
\textcolor{black}{Indeed, for $T_{reh}\sim 2.215\times 10^{15}{\rm GeV}$ (${\cal N}=60$) then it gives rise to admissible level of fine-tuning, namely, $1+\omega_{\phi}\simeq 10^{-2}$ compared to the case when $T_{reh}\sim 1.493\times 10^{13}{\rm GeV}$ (${\cal N}=55$) for which, $1+\omega_{\phi}\simeq 10^{-10}$. Thus the total number of e-foldings can be $55$, but in that case one  encounters large fine-tuning, which is not advisable. Hence ${\cal N}=60$ is a feasible choice.}}, the amplitude of the spectrum falls linearly and at a rate faster compared to its rise in the region before the peak, which results from the behavior of the scalar power spectrum in the frequency region corresponding to the SRII phase.

As we can see, the peak value of the GW spectrum shown in the plot coincides with the results observed in the NANOGrav 15-year Data Set. These observations show that inflation in Galileon theory has the required feature to produce the primordial induced GWs and nearly solar mass PBHs, $M_{\rm PBH} \sim 0.01M_{\odot}$. \textcolor{black}{In fig.(\ref{nanoepta}), we mainly focus on the wavenumber region that corresponds to those probed by the recent PTA (NANOGrav15 and EPTA) observations. We fix the transition wavenumber, $k_{\rm PBH}=10^{7.2}{\rm Mpc^{-1}}$, the same as seen in the low wavenumber regime signal of fig.(\ref{NANOGal}). The fig.(\ref{nanoepta}) provides a closer look at the near peak region of the previous SIGW spectrum, where we can observe the growth of the spectrum relative to the PTA signals and its eventual fall off after achieving a peak value close to that from the NANOGrav15 signal in black. The EPTA signal on the right shows the peak value before the maximum of the generated SIGW spectrum. The value for effective sound speed used here is $c_{s}=0.05$ and matches with our previous choice for the same in the scalar power spectrum in fig.(\ref{powerGal}) and fig.(\ref{NANOGal}). Since this choice of the $c_{s}$ falls inside the corresponding causality and unitarity preserving bound which we introduce in (\ref{s2e}), we can infer that in presence of such features in the underlying theory we can obtain the maximum amount of the GW spectrum signal consistent with the PTA observations. We also highlight another important feature present in our result using fig.(\ref{nanoepta}) related to the infrared (IR) scaling of the GW spectrum. It is visible from the figure that the frequency regime probed by the PTA coincides with the IR tail of the generated SIGW before the peak frequency occurs. Now, since we are primarily working under the choice of the induced GWs re-entering in the RD era, the scaling relation for such a spectrum in the IR exhibits a universal $\Omega_{\rm GW}h^{2}\propto k^{3}$ law \cite{Cai:2019cdl,Domenech:2021ztg,Choudhury:2024one}. We notice here that our spectrum in red sufficiently approximates this scaling law, which we illustrate here via a dashed purple line till before the peak frequency, thus remaining consistent with previous studies. }

The interesting feature to note in fig.(\ref{NANOGal}) comes further when we shift the transition scale to higher wavenumbers where $k_{\rm PBH} \sim {\cal O}(10^{15}{\rm Mpc}^{-1} \approx 1{\rm Hz})$. There we observe a similar kind of behavior for the GW spectrum as was with the case when considering almost solar mass PBH, but this time, masses of the produced PBH are tiny, in the extreme sub-solar regime $M_{\rm PBH} \sim 10^{-18}M_{\odot}$. For this case, the peak value of the GW spectrum falls into the sensitivity curves of the proposed and operational space and earth-based probes of GWs, which are in the background of the same  figure for comparison. Maintaining the perturbativity approximation and benefits of the non-renormalization theorem has the GW spectrum preserve its behavior near the sharp transitions and have successful inflation after the end of the USR.

Hence, Galileon theory has the power to accommodate the possibility of producing PBHs within the sub-solar and solar mass range depending on the transition wavenumber $k_{\rm PBH}$ position and can produce enough amount of GW signal which can fall into the experimentally observed low-frequency regime of NANOGrav 15, as well as in the allowed sensitivity of the high-frequency regime probes. 

\begin{figure*}[htb!]
    	\centering
    \subfigure[]{
      	\includegraphics[width=8.5cm,height=7.5cm]{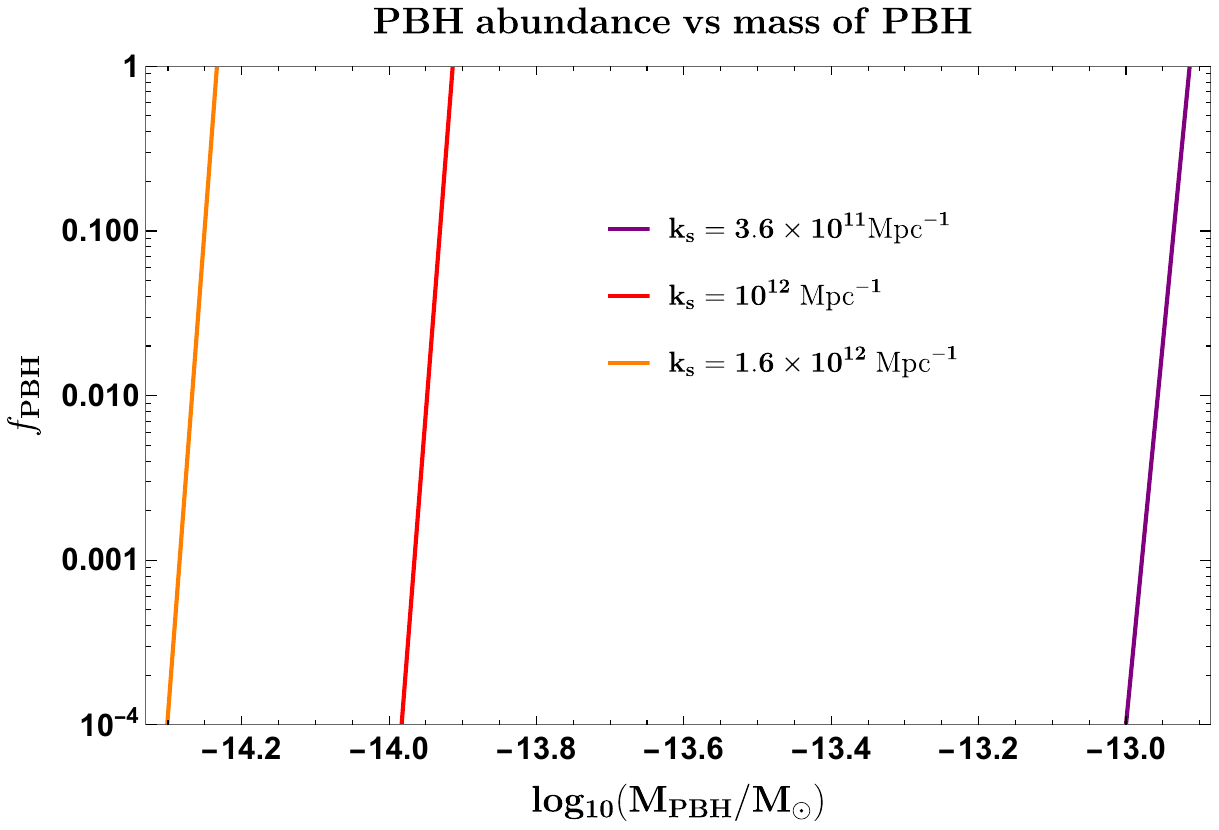}
        \label{smallMFPBH}
    }
    \subfigure[]{
       \includegraphics[width=8.5cm,height=7.5cm]{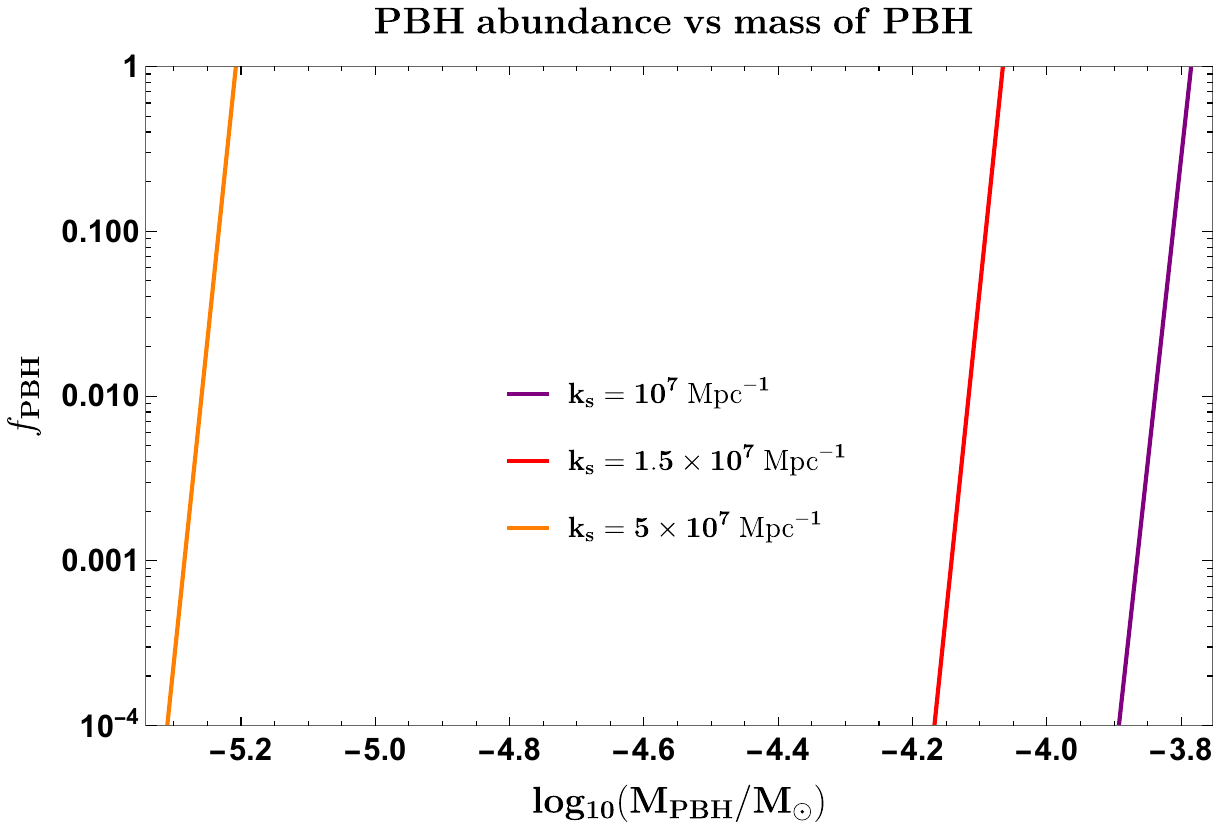}
        \label{NANOFPBH}
    }
    	\caption[Optional caption for list of figures]{PBH abundance as function of their masses. The \textit{left panel} shows PBH abundance for masses in the range ${\cal O}(10^{-15}-10^{-13})M_{\odot}$ corresponding to the small scales. The \textit{right panel} shows PBH abundance for masses in the range ${\cal O}(10^{-6}-10^{-3})M_{\odot}$ corresponding to frequency interval of the NANOGrav15 signal.} 
    	\label{MPBHabundance}
    \end{figure*}

\textcolor{black}{The fig.(\ref{MPBHabundance}) shows behaviour for the PBH abundance as a function of the PBH masses. We analyze the abundance for two regions, one concerning the frequency range corresponding to the NANOGrav 15 signal window (\textit{right panel}) and the other concerning the PBHs formed at large wavenumbers or small scales (\textit{left panel}) where, from the one-loop corrected scalar power spectrum in fig.(\ref{powerGal}), one can notice an amplitude of ${\cal O}(10^{-5}).$ Here, we use a crucial point highlighted through our plot in fig.(\ref{NANOGal}), which shows the invariance of the obtained GW spectrum under translation in wavenumber. This property is also highlighted under discussions in the section \ref{s2e} for working with the Galileon theory. To reiterate, until we can maintain the necessary perturbativity approximation, especially having $k_{e}/k_{s} \sim {\cal O}(10)$, we control the scalar power spectrum amplitude from achieving values greater than ${\cal O}(1)$, and thus, we are free to shift the sharp transition wavenumber to the value of our choosing. By doing so, we will observe successful inflation in our setup, owning to the benefits of the non-renormalization theorem. Since the amplitude from the fig.(\ref{powerGal}) is quite small on the small scales, when PBH formation is of main concern as it requires at least ${\cal O}(10^{-2})$ for the power spectrum amplitude, we utilize here the translation property mentioned above to our advantage. By keeping the wavenumber around $k_{s}\sim {\cal O}(10^{11}-10^{12}){\rm Mpc}^{-1}$, it allows to produce PBHs with masses in the range ${\cal O}(10^{-15}-10^{-13})M_{\odot}$ and there the corresponding abundance is not restricted due to strong constraints coming from PBH evaporation. Such restriction are large as we consider PBHs of even smaller masses such as $M_{\rm PBH} < {\cal O}(10^{-16})$ where PBH evaporation constraints are the most stringent and there, with an amplitude of the scalar power spectrum of the order ${\cal O}(10^{-5})$, produces highly suppressed value of the variance which ultimately produces almost vanishing $f_{\rm PBH}$. In the right panel, we show the PBH abundance for the wavenumbers associated with the frequency interval of the NANOGrav15 signal. The observed mass range corresponding to the sharp transitions at the frequencies significant to the NANOGrav15 is seen to lie within $M_{\rm PBH}\sim {\cal O}(10^{-3}-10^{-5})$ with sufficient abundance for such PBH masses.    }

\section{Conclusion} \label{s5}

In this paper, we have considered the scalar-induced gravitational waves using the one-loop corrected scalar power spectrum from Galileon inflation. We first briefly reviewed the Galileon theory by introducing the action for the Covariantized Galileon Theory, \textcolor{black}{with detailing the implementation of various phases in our setup during inflation} and then outlining the procedure to develop solutions for the comoving curvature perturbations modes in the three regions, SRI, USR, and SRII, by using the second-order action in this theory. The defining feature of this theory, other than the ability to evade the emergence of unwanted ghosts, is its non-renormalization theorem which allows us not to consider any renormalization and resummation procedure for the loop effects in its correlations. This feature helped us to manage a sufficient number of e-foldings of expansion and control the one-loop effects in theory, which later form part of the total scalar power spectrum necessary for further analysis regarding the GWs abundance calculation. We then presented a concise overview of the theory behind the scalar-induced gravitational waves (SIGWs) which tells us that the second-order tensor mode solutions involved quadratic contributions of the first-order scalar perturbations. After deriving the expression for the tensor power spectrum in terms of the scalar power spectrum, the next step was to numerically solve the integral for the GW abundance and plot the resulting behavior against the allowed frequency range. The features of this plot represent the key findings of our overall analysis done in this work.

We observed that the amount of GW abundance, $\Omega_{\rm GW}h^{2}$, produced through the use of one-loop corrected scalar power spectrum in the low-frequency region, $10^{-8}{\rm Hz} \leq f \leq 10^{-6}{\rm Hz}$, is such that it gives a sufficient overlap and thus indicates an agreement with the observational result from the NANOGrav 15-year signal. In the high-frequency region, after the peak value is achieved, the obtained spectrum falls rapidly compared to its ascent near the low-frequency values. The peak enhancement in the signal is due to a sharp transition feature when going from the SRI to USR phase. This behavior is known to exist for the total dimensionless scalar power spectrum in plot  fig.(\ref{powerGal}) near the scales labeled as $k_{s} \sim 10^{6}{\rm Mpc}^{-1}$ and $k_{e} \sim 10^{7}{\rm Mpc}^{-1}$. There the  property of accommodating a long enough SRII region required for successful inflation, $\Delta{\cal N}_{\rm Total} = 60$, is also reflected in the tensor power spectrum, and equivalently the GW spectrum, when it is extended to even higher frequencies and their corresponding lower amplitudes relative to the current values.

The scale of the transition wave number  determines the mass of the PBH being produced. Using the eqn.(\ref{s4mpbh}), we were able to find the allowed range of mass values, $(1M_{\odot} - 10^{-18}M_{\odot})$, of the produced PBH when the transition scale is set between $k_{\rm PBH} \sim (10^{7}{\rm Mpc}^{-1}-10^{15}{\rm Mpc}^{-1})$. Hence, by keeping the transition scale near to the required value in the frequency range of the NANOGrav signal, $k_{\rm PBH} \sim (10^{7}{\rm Mpc}^{-1} \approx 10^{-8}{\rm Hz})$, we have shown that Galileon theory can produce almost solar mass PBH, $M_{\rm PBH} \sim 0.01M_{\odot}$, and along with that generate enough amount of observable signal in the GW spectrum. Next, we examined the case where the transition scale is pushed to the domain of higher frequencies where the existing and proposed GW experiments are able to probe the effect. We found that Galileon theory is also able to produce a sufficient amount of signal in the spectrum such that it falls under the parameter space of these experiments. The observed spectrum exactly mimics the behavior for the case of a transition scale in the lower frequencies. This is important from the perspective of future investigations regarding the observation of SIGWs in higher frequencies as a signal for new physics models. Due to the setting of the transition at such high wave numbers, $k_{\rm PBH} \sim (10^{15}{\rm Mpc}^{-1} \approx 1{\rm Hz})$, we are also able to predict the production of small mass PBHs in the sub-solar category, $M_{\rm PBH} \sim 10^{-18}M_{\odot}$, and also producing an observable amount of signal in the GW spectrum. This is again the result of the intrinsic properties of the Galileon theory,  which does not depend on the scale of transition until we are able to satisfy the perturbativity criteria, which is maintained throughout our analysis by keeping $k_{e}/k_{s} \sim {\cal O}(10)$, where $k_{s}, k_{e}$ are the respective scales at the SRI to USR and USR to SRII transitions. 
Thus, have demonstrated the possibility of generating a measurable spectrum of the GW abundance either in the low-frequency (NANOGrav) region or the high-frequency region of the ground- and space-based GW detectors and the production of corresponding masses of PBHs ranging from $(1M_{\odot}-10^{-18}M_{\odot})$.

\section*{ \large Acknowledgements}

 SC would like to thank the work-friendly environment of The Thanu Padmanabhan Centre For Cosmology and Science Popularization (CCSP), SGT University, Gurugram, for providing tremendous support in research. SC would also like to thank all the members of Quantum Aspects of the Space-Time \& Matter (QASTM) for elaborative discussions. MS is supported by Science and Engineering Research Board (SERB), DST, Government of India under the Grant Agreement number CRG/2022/004120 (Core Research Grant). MS is also partially supported by the Ministry of Education and Science of the Republic of Kazakhstan, Grant
No. 0118RK00935 and CAS President's International Fellowship Initiative (PIFI). Last but not least, we would like to acknowledge our debt to the people belonging to the various parts of the
world for their generous and steady support for research in natural sciences.

\section{Appendix}
\subsection{General Mode Solutions}\label{App:A}

\textcolor{black}{In this appendix, we focus on the calculation of the comoving curvature perturbations modes for our inflationary setup with the phases SRI, USR, and SRII 
involved within the CGT. The contents of this appendix is based on the detailed study provided in the refs.\cite{Choudhury:2023hvf, Choudhury:2023kdb}}.



\textcolor{black}{To begin, we must work with the perturbation theory in second order in the curvature perturbations, and for this, we would need the second-order perturbed action. This action is written in the Fourier space as follows:}
\bea
S^{(2)}_{\zeta} = \int d\tau\;\frac{d^{3}\mbf{k}}{(2\pi)^{3}}a(\tau)^2\frac{\cal A}{H^{2}}\left(|\zeta_{\mbf{k}}^{'}(\tau)|^{2} - c_{s}^{2}k^{2}|\zeta_{\mbf{k}}(\tau)|^{2}\right) = \int d\tau\;\frac{d^{3}\mbf{k}}{(2\pi)^{3}}a(\tau)^2\frac{\cal B}{c_{s}^{2}H^{2}}\left(|\zeta_{\mbf{k}}^{'}(\tau)|^{2} - c_{s}^{2}k^{2}|\zeta_{\mbf{k}}(\tau)|^{2}\right)
\eea
\textcolor{black}{where ${\cal A}$ and ${\cal B}$ are the same time-dependent coefficients as defined in eqns.(\ref{coeffA},\ref{coeffB}), and which form the relation, $c_{s}^{2} = {\cal B}/{\cal A}$, for the effective speed of sound parameter in the CGT as discussed before. Here we can observe the significance of these new coefficients as being directly involved in the action $S_{\zeta}^{(2)}$ and from here they will participate in the mode solutions for the curvature perturbations which further incorporates them into the scalar power spectrum. }
\textcolor{black}{Now, solving the above second order action would give us the following equation of motion:}
\bea
\zeta^{''}_{\bf k}(\tau)+2\frac{z^{'}(\tau)}{z(\tau)}\zeta^{'}_{\bf k}(\tau) +c^2_sk^2\zeta_{\mbf{k}}(\tau)=0.
\eea
with $z(\tau) = a\sqrt{2{\cal A}}/H^{2}$. It is this equation whose solutions in the three phases are to be analyzed. This is performed by setting of the initial conditions for the SRI phase as the Bunch-Davies quantum vacuum state and in the subsequent phases we are required to solve the boundary conditions during the transitions to eventually determine their complete solutions. We now describe the mode solutions for each phase.
 
\begin{itemize}
        \item \underline{\textbf{For the SRI phase:}}\\
        The general mode solution in this phase for the second-order Fourier space equation of motion turns out to be:
        \begin{eqnarray} 
    \zeta_{\mbf{k}}(\tau)&=&\left(\frac{iH^{2}}{2\sqrt{\cal A}}\right)\frac{1}{(c_{s}k)^{3/2}}\times\left[\alpha^{(1)}_{\mbf{k}}\left(1+ikc_{s}\tau\right)\exp{\left(-ikc_{s}\tau\right)} - \beta^{(1)}_{\mbf{k}}\left(1-ikc_{s}\tau\right)\exp{\left(ikc_{s}\tau\right)}\right]
        \end{eqnarray} 
    where the conformal time window is the interval with $\tau < \tau_{s}$. The Bogoliubov coefficients for this phase are obtained as the result of fixing the Bunch-Davies initial condition:
    \bea \label{s21a}\alpha^{(1)}_{\bf k}=1,\\
    \label{s21b}\beta^{(1)}_{\bf k}=0. \eea
    Throughout this phase, the slow-roll parameter $\epsilon(\tau)$ is almost a constant while for the other slow-roll parameter $\eta(\tau) \sim 0$.
    \item \underline{\textbf{For the USR phase:}}\\
    The general mode solution in this phase, for the conformal time window $\tau \in \left[\tau_{s}, \tau_{e}\right)$, is obtained to be as follows:
        \begin{eqnarray}
    \zeta_{\mbf{k}}(\tau)&=&\left(\frac{iH^{2}}{2\sqrt{\cal A}}\right)\left(\frac{\tau_{s}}{\tau}\right)^{3}\frac{1}{(c_{s}k)^{3/2}}\times \left[\alpha^{(2)}_{\bf k}\left(1+ikc_{s}\tau\right)\exp{\left(-ikc_{s}\tau\right)} - \beta^{(2)}_{\bf k}\left(1-ikc_{s}\tau\right)\exp{\left(ikc_{s}\tau\right)} \right]. \end{eqnarray}
    The parameter $\epsilon$ behaves very differently for this phase with having a time-dependent form: $\epsilon(\tau) = \epsilon(\tau/\tau_{s})^{6}$. This effect is reflected in the above equation. To have the complete solution requires the solving of the continuity and differentiability conditions at the transition scale $\tau_{s}$. As a result, we obtain the following Bogoliubov coefficients:
    \begin{eqnarray}
    \label{s22a}\alpha^{(2)}_{\bf k}&=&1+\frac{3k_{s}^{3}}{2 i k^{3}}\left(1+\left(\frac{k}{k_{s}}\right)^{2}\right),\\
    \label{s22b}\beta^{(2)}_{\bf k}&=&\frac{3k_{s}^{3}}{2 i k^{3}}\left(1-i\left(\frac{k}{k_{s}}\right)^{2}\right)^{2}\; \exp{\left(2i\frac{k}{k_{s}}\right)}.
    \end{eqnarray}
    However, the parameter $\eta(\tau)$ here satisfies $\eta \sim -6$ which indicates an extreme jump from the previous value and is the cause for the enhancement in the fluctuations.
        \item \underline{\textbf{For the SRII phase:}}\\
        The general mode solution in this phase, for the conformal time window $\tau \in \left[\tau_{e}, \tau_{{\rm end}}\right]$, is obtained to be as follows:
        \begin{eqnarray}
    \label{s23a}
    \zeta_{\mbf{k}}(\tau)&=&\left(\frac{iH^{2}}{2\sqrt{\cal A}}\right)\left(\frac{\tau_{s}}{\tau_{e}}\right)^{3}\frac{1}{(c_{s}k)^{3/2}}\times \left[\alpha^{(3)}_{\bf k}\left(1+ikc_{s}\tau\right)\exp{\left(-ikc_{s}\tau\right)} - \beta^{(3)}_{\bf k}\left(1-ikc_{s}\tau\right)\exp{\left(ikc_{s}\tau\right)} \right]. \end{eqnarray}
    The parameter $\epsilon$ in this phase has the behaviour where it varies slowly as a result of its form:$\epsilon(\tau) = \epsilon(\tau_{e}/\tau_{s})$. This fact is crucial for the above solution. Again, through the use of the boundary conditions we get the following form of the Bogoliubov coefficients in this phase: 
    \begin{eqnarray}
    \alpha^{(3)}_{\bf k}&=&-\frac{k_{s}^{3}k_{e}^{3}}{4k^6}\Bigg[9 \left(-\frac{k}{k_e}+i\right){}^2 \left(\frac{k}{k_s}+i\right){}^2 \exp{\left(2 i k
   \left(\frac{1}{k_s}-\frac{1}{k_e}\right)\right)}\nonumber\\
    &&\quad\quad\quad\quad\quad\quad\quad\quad\quad\quad\quad\quad\quad\quad-
    \left\{\left(\frac{k}{k_{e}}\right)^2\left(-2\frac{k}{k_{e}}-3i\right)-3i\right\}\left\{\left(\frac{k}{k_{s}}\right)^2\left(-2\frac{k}{k_{s}}+3i\right)+3i\right\}\Bigg],\\
    \label{s23b}\beta^{(3)}_{\bf k}&=&\frac{3k_{s}^{3}k_{e}^{3}}{4k^6}\Bigg[\left(\frac{k}{k_{s}}+i\right)^2\left\{\left(\frac{k}{k_{e}}\right)^{2}\left(2i\frac{k}{k_{e}}\right)+3\right\}\exp{\left(2i\frac{k}{k_{s}}\right)}\nonumber\\
    &&\quad\quad\quad\quad\quad\quad\quad\quad\quad\quad\quad\quad\quad\quad\quad\quad+i\left(\frac{k}{k_{e}}+i\right)^2\left\{3i+\left(\frac{k}{k_{s}}\right)^{2}\left(-2\frac{k}{k_{s}}+3i\right)\right\}\exp{\left(2i\frac{k}{k_{e}}\right)}\Bigg].
    \end{eqnarray}
    For this phase, the parameter $\eta(\tau)$ has an almost constant value throughout. This is quite similar to the behavior of the same parameter in the SRI phase. This nature is also reflected in the final behaviour of the power spectrum during this phase.
\end{itemize}
In all the equations above we have used the horizon crossing condition $-kc_{s}\tau = 1$ which is valid in all three phases during the sharp transitions at $k_{s}$ from the SRI to USR phase and at $k_{e}$ for the USR to SRII phase. These equations are sufficient to write the tree-level scalar power spectrum in the super-horizon limit, which we now mention as:
\begin{eqnarray} \label{Apptree}
\Delta^{2}_{\zeta,{\bf Tree}}(k)
&=& \displaystyle
\displaystyle \left(\frac{H^{4}}{8\pi^{2}{\cal A} c^3_s}\right)_*\times\left\{
	\begin{array}{ll}
		\displaystyle 1+\Bigg(\frac{k}{k_s}\Bigg)^2 & \mbox{when}\quad  k < k_s  \;(\rm SRI)  \\  
			\displaystyle 
			\displaystyle \left(\frac{k_e}{k_s }\right)^{6}\left|\alpha^{(2)}_{\bf k}-\beta^{(2)}_{\bf k}\right|^2 & \mbox{when }  k_s\leq k < k_e  \;(\rm USR)\\ 
   \displaystyle 
			\displaystyle \left(\frac{k_e}{k_s }\right)^{6}\left|\alpha^{(3)}_{\bf k}-\beta^{(3)}_{\bf k}\right|^2 & \mbox{when }  k_e\leq k\leq k_{\rm end}  \;(\rm SRII) 
	\end{array} \right. \end{eqnarray}

\subsection{Couplings and Coefficients in three phases including one-loop effects}\label{App:B}

In this appendix, we present the expressions for the couplings and the momentum-dependent functions which are necessary to describe the one-loop contributions to the total scalar power spectrum.

\begin{itemize}
\item \underline{\textbf{For the SRI phase}}: The one-loop effect from the SRI phase which contributes into the total scalar power spectrum is written as follows:
\bea
\Bigg[\Delta^{2}_{\zeta,\bf {One-Loop}}(k)\Bigg]_{\rm \textbf{SRI}} = \Bigg[\Delta^{2}_{\zeta,\bf {Tree}}(k)\Bigg]_{\rm \textbf{SRI}}\times \left( -\sum^{4}_{i=1}{\cal G}_{i,\mbf{SRI}}\mbf{F}_{i,\mbf{SRI}}(k_{s},k_{*})\right)
\eea
the values for ${\cal G}_{i,\textbf{SRI}}\;\forall i=1,2,3,4$ are defined using the CGT couplings:
\begin{eqnarray} {\cal G}_{1,{\bf SRI}}&=&{\cal G}^2_1(\tau_*)\nonumber\\ 
    {\cal G}_{2,{\bf SRI}}&=&-{\cal G}^2_2(\tau_*)H^2(\tau_*)c^2_s\nonumber\\
    {\cal G}_{3,{\bf SRI}}&=&-\frac{{\cal G}^2_3(\tau_*)}{c^2_s}\nonumber\\
    {\cal G}_{4,{\bf SRI}}&=&\frac{{\cal G}^2_4(\tau_*)H^2(\tau_*)}{c^6_s}\nonumber\\
    \end{eqnarray}
and the momentum-dependent functions $\mbf{F}_{i,\textbf{SRI}}$ are defined using the notation ${\cal K}_{*,s} \equiv k_{*}/k_{s}$ as follows:
\begin{eqnarray} {\bf F}_{1,{\bf SRI}}(k_s,k_*)&=&\frac{1}{2}\Bigg[3+{\cal K}_{*,s}^2\Bigg],\\ 
         {\bf F}_{2,{\bf SRI}}(k_s,k_*)&=&\Bigg[\frac{17}{42}-\frac{2}{3}{\cal K}_{*,s}^6+\frac{24}{7}{\cal K}_{*,s}^7-\frac{9}{2}{\cal K}_{*,s}^8\Bigg],\\ 
         {\bf F}_{3,{\bf SRI}}(k_s,k_*)&=&-\frac{2}{3}\Bigg[1-{\cal K}_{*,s}^6\Bigg],\\ 
         {\bf F}_{4,{\bf SRI}}(k_s,k_*)&=&-\frac{1}{2}\Bigg[1-{\cal K}_{*,s}^2\Bigg].\end{eqnarray}

\item \underline{\textbf{For the USR phase}}: The one-loop effect from the USR phase which contributes into the total scalar power spectrum is written as follows:
\bea
\Bigg[\Delta^{2}_{\zeta,\bf {One-Loop}}(k)\Bigg]_{\rm \textbf{USR}} = \Bigg[\Delta^{2}_{\zeta,\bf {Tree}}(k)\Bigg]_{\rm \textbf{SRI}}\times \left( \sum^{4}_{i=1}{\cal G}_{i,\mbf{USR}}\mbf{F}_{i,\mbf{SRI}}(k_{e},k_{s})\right)
\eea
the values for ${\cal G}_{i,\textbf{USR}}\;\forall i=1,2,3,4$ are defined using the CGT couplings:
\begin{eqnarray} {\cal G}_{1,{\bf USR}}&=&\Bigg(\frac{{\cal G}^2_1(\tau_e)}{c^3_s}{\cal K}_{e,s}^6-\frac{{\cal G}^2_1(\tau_s)}{c^3_s}\Bigg)\nonumber\\
    {\cal G}_{2,{\bf USR}}&=&\Bigg(\frac{{\cal G}^2_2(\tau_e)H^2(\tau_e)}{c^5_s}{\cal K}_{e,s}^4-\frac{{\cal G}^2_2(\tau_s)H^2(\tau_s)}{c^5_s}\Bigg) \nonumber\\
    {\cal G}_{3,{\bf USR}}&=&\Bigg(\frac{{\cal G}^2_3(\tau_e)}{c^6_s}{\cal K}_{e,s}^3-\frac{{\cal G}^2_3(\tau_s)}{c^6_s}\Bigg)\nonumber\\
   {\cal G}_{4,{\bf USR}}&=&\Bigg(\frac{{\cal G}^2_2(\tau_e)H^2(\tau_e)}{c^7_s}{\cal K}_{e,s}^2-\frac{{\cal G}^2_2(\tau_s)H^2(\tau_s)}{c^7_s}\Bigg).\nonumber\\
\end{eqnarray}
where we have used the notation ${\cal K}_{e,s} \equiv k_{e}/k_{s},\;\text{and}\;{\cal K}_{s,e} \equiv k_{s}/k_{e}$. Using these same notations further we write the momentum-dependent functions $\mbf{F}_{i,\textbf{USR}}$ as follows:
 \bea {\bf F}_{1,{\bf USR}}(k_e,k_s)&=&\Bigg[\frac{1}{4}+\frac{9}{4}{\cal K}_{s,e}^2-\frac{1}{4}{\cal K}_{s,e}^4-9{\cal K}_{s,e}^4\ln{\cal K}_{s,e}-6\sin\left(1+{\cal K}_{e,s}\right)\sin\left(1-{\cal K}_{e,s}\right)\nonumber\\
    &&\quad-\frac{9}{8}{\cal K}_{s,e}^2\Bigg\{2\left(\frac{k_e}{k_s}\right)\sin\left(2{\cal K}_{e,s}\right)-\cos\left(2{\cal K}_{e,s}\right)\Bigg\}-6{\cal K}_{s,e}\sin\left(2{\cal K}_{e,s}\right)\nonumber\\
    &&\quad-\frac{69}{16}\cos\left(2{\cal K}_{e,s}\right)-\frac{3}{4}{\cal K}_{s,e}^2\Bigg\{2{\cal K}_{e,s}\cos\left(2{\cal K}_{e,s}\right)-\sin\left(2{\cal K}_{e,s}\right)\Bigg\}\Bigg],\\
         {\bf F}_{2,{\bf USR}}(k_e,k_s)&=&\Bigg[\frac{1}{8}+\frac{9}{12}{\cal K}_{s,e}^2+\frac{9}{4}{\cal K}_{s,e}^4+\frac{9}{4}{\cal K}_{s,e}^6-\frac{43}{8}{\cal K}_{s,e}^8-\frac{9}{8}\cos\left(2{\cal K}_{e,s}\right)\Bigg],\\
    {\bf F}_{3,{\bf USR}}(k_e,k_s)&=&\Bigg[\frac{1}{8}+\frac{9}{12}{\cal K}_{s,e}^2+\frac{9}{4}{\cal K}_{s,e}^4+\frac{9}{4}{\cal K}_{s,e}^6-\frac{43}{8}{\cal K}_{s,e}^8-\frac{9}{8}\cos\left(2{\cal K}_{e,s}\right)\Bigg]\nonumber\\
         &=&{\bf F}_{2,{\bf USR}}(k_e,k_s),\\ 
         {\bf F}_{4,{\bf USR}}(k_e,k_s)&=&\Bigg[\frac{1}{12}+\frac{9}{20}{\cal K}_{s,e}^2+\frac{9}{8}{\cal K}_{s,e}^4+\frac{9}{12}{\cal K}_{s,e}^6-\frac{299}{230}{\cal K}_{s,e}^{12}\Bigg].\eea

\item \underline{\textbf{For the SRII phase}}: The one-loop effect from the SRII phase which contributes into the total scalar power spectrum is written as follows:
\bea
\Bigg[\Delta^{2}_{\zeta,\bf {One-Loop}}(k)\Bigg]_{\rm \textbf{SRII}} = \Bigg[\Delta^{2}_{\zeta,\bf {Tree}}(k)\Bigg]_{\rm \textbf{SRI}}\times \left( \sum^{4}_{i=1}{\cal G}_{i,\mbf{SRII}}\mbf{F}_{i,\mbf{SRII}}(k_{\rm end},k_{e})\right)
\eea
the values for ${\cal G}_{i,\textbf{SRII}}\;\forall i=1,2,3,4$ are defined using the CGT couplings:
\begin{eqnarray} {\cal G}_{1,{\bf SRII}}&=&\Bigg(\frac{{\cal G}^2_1(\tau_{\rm end})}{c^3_s}{\cal K}_{e,s}^6-\frac{{\cal G}^2_1(\tau_e)}{c^3_s}\Bigg)\nonumber\\
    {\cal G}_{2,{\bf SRII}}&=&\Bigg(\frac{{\cal G}^2_2(\tau_{\rm end})H^2(\tau_{\rm end})}{c^5_s}{\cal K}_{e,s}^4-\frac{{\cal G}^2_2(\tau_e)H^2(\tau_e)}{c^5_s}\Bigg) \nonumber\\
   {\cal G}_{3,{\bf SRII}}&=&\Bigg(\frac{{\cal G}^2_3(\tau_{\rm end})}{c^6_s}{\cal K}_{e,s}^3-\frac{{\cal G}^2_3(\tau_e)}{c^6_s}\Bigg) \nonumber\\
   {\cal G}_{4,{\bf SRII}}&=&\Bigg(\frac{{\cal G}^2_2(\tau_{\rm end})H^2(\tau_{\rm end})}{c^7_s}{\cal K}_{e,s}^2-\frac{{\cal G}^2_2(\tau_e)H^2(\tau_e)}{c^7_s}\Bigg)\nonumber\\ \end{eqnarray}
where ${\cal K}_{e,s}\;\text{and}\;{\cal K}_{s,e}$ are previously defined in the USR phase. The momentum-dependent functions $\mbf{F}_{i,\textbf{SRII}}$ are defined here after taking out only the leading order contributions and $"\cdots"$ are used to represent the suppressed quantities. To write the expressions more concisely, we choose the following notation:
\bea
{\cal K}_{{\rm end},e} \equiv \frac{k_{\rm end}}{k_{e}},\quad\quad{\cal K}_{{\rm end},s} \equiv \frac{k_{\rm end}}{k_{s}},\quad\quad{\cal K}_{e,{\rm end}} \equiv \frac{k_{e}}{k_{\rm end}},\quad\quad{\cal K}_{s,{\rm end}} \equiv \frac{k_{s}}{k_{\rm end}}
\eea
The final result is written as follows:
\end{itemize}
 {\allowdisplaybreaks \bea {\bf F}_{1,{\bf SRII}}(k_{\rm end},k_e)&=&\Bigg[\frac{81}{64}+\frac{81}{40}\left(1+{\cal K}_{s,e}\right)-\frac{9}{8}\left(1+{\cal K}_{s,e}^2\right)+\frac{1}{6}\left(\left(1+{\cal K}_{s,e}\right)^2+2{\cal K}_{s,e}\right)+\frac{1}{8}{\cal K}_{s,e}^2+6
 +\frac{2}{7}{\cal K}_{s,e}\left(1+{\cal K}_{s,e}\right)\nonumber\\
 &&+\frac{27}{8}\left(1+{\cal K}_{s,e}^6\right)-\frac{3}{4}{\cal K}_{e,{\rm end}}\cos\left(2{\cal K}_{{\rm end},e}\right)+4{\cal K}_{e,{\rm end}}^2\ln\left({\cal K}_{e,{\rm end}}\right)-\frac{3}{4}\sin\left(1+{\cal K}_{{\rm end},s}\right)\sin\left(1-{\cal K}_{{\rm end},s}\right)\nonumber\\
 &&-\frac{9}{4}\frac{\displaystyle{\cal K}_{s,e}}{\displaystyle\left(1-{\cal K}_{s,e}\right)^2}\Bigg\{\cos\left(2\left({\cal K}_{{\rm end},e}-{\cal K}_{{\rm end},s}\right)\right)\Bigg\}-\frac{9}{4}\frac{\displaystyle 1}{\displaystyle\left(1-{\cal K}_{s,e}\right)}\Bigg\{{\cal K}_{s,{\rm end}}\sin\left(2\left({\cal K}_{{\rm end},e}-{\cal K}_{{\rm end},s}\right)\right)\Bigg\}\nonumber\\
    &&-9\frac{\displaystyle\left(1+{\cal K}_{s,e}\right)^2}{\displaystyle \left(1-{\cal K}_{s,e}\right)}\Bigg\{\sin\left(2\left({\cal K}_{{\rm end},e}-{\cal K}_{{\rm end},s}\right)\right)\Bigg\}+\frac{9}{8}\Bigg\{8-18{\cal K}_{s,e}^4\left(1+{\cal K}_{s,e}\right)\Bigg\}\nonumber\\
    &&\Bigg\{\cos\left(2\left({\cal K}_{{\rm end},e}-{\cal K}_{{\rm end},s}\right)\right)-{\cal K}_{e,{\rm end}}\cos\left(2\left(1-{\cal K}_{e,s}\right)\right)\Bigg\}-\frac{9}{16}{\cal K}_{s,e}^2\left(9+44{\cal K}^{2}_{s,e}\right)\ln{\cal K}_{e,{\rm end}}\nonumber\\
    &&+9{\cal K}_{s,e}^4\left(1+{\cal K}_{s,e}^2\right)+\frac{9}{7}{\cal K}_{s,e}^6+22{\cal K}_{s,e}^2\left(1-{\cal K}_{e,{\rm end}}\right)\nonumber\\
    &&-\frac{27}{16}{\cal K}_{s,e}^4{\cal K}_{e,{\rm end}}\Bigg\{\sin\left(2{\cal K}_{{\rm end},e}\right)-{\cal K}_{e,{\rm end}}\cos\left(2{\cal K}_{{\rm end},e}\right)\Bigg\}\cdots\Bigg],\\
    {\bf F}_{2,{\bf SRII}}(k_{\rm end},k_e)&=&\Bigg[\frac{81}{16}\ln{\cal K}_{e,{\rm end}}+\frac{81}{8}\left(1-{\cal K}_{e,{\rm end}}\right)\left(1+{\cal K}_{s,e}\right)+5{\cal K}_{s,e}^6\ln{\cal K}_{e,{\rm end}}\nonumber\\
    &&+\frac{1}{2}\left(\left(1+{\cal K}_{s,e}\right)^2+2{\cal K}_{s,e}\right)+\frac{2}{3}\left(1-{\cal K}_{e,{\rm end}}^3\right){\cal K}_{s,e}\left(1+{\cal K}_{s,e}\right)+\frac{1}{4}{\cal K}_{s,e}^2+\frac{1}{4}+\frac{27}{64}-\frac{27}{8}\left(1+{\cal K}_{s,e}^6\right)\nonumber\\
    &&+\frac{3}{8}\left(1+{\cal K}_{s,e}^2\right)-\frac{9}{16}\frac{\displaystyle{\cal K}_{s,e}^3}{\displaystyle\left(1-{\cal K}_{s,e}\right)^2}\Bigg\{\cos\left(2\left({\cal K}_{{\rm end},e}-{\cal K}_{{\rm end},s}\right)\right)\Bigg\}\nonumber\\
    &&-\frac{27}{4}\frac{\displaystyle 1}{\displaystyle\left(1-{\cal K}_{s,e}\right)}\Bigg\{{\cal K}_{s,{\rm end}}\sin\left(2\left({\cal K}_{{\rm end},e}-{\cal K}_{{\rm end},s}\right)\right)\Bigg\}+\frac{81}{16}\frac{\displaystyle\left(1+{\cal K}_{s,e}^2\right)}{\displaystyle \left(1-{\cal K}_{s,e}\right)}\Bigg\{\cos\left(2\left({\cal K}_{{\rm end},e}-{\cal K}_{{\rm end},s}\right)\right)\Bigg\}\nonumber\\
    &&-\frac{9}{16}\frac{\displaystyle{\cal K}_{s,e}^3}{\displaystyle\left(1-{\cal K}_{s,e}\right)}\Bigg\{\sin\left(2\left({\cal K}_{{\rm end},e}-{\cal K}_{{\rm end},s}\right)\right)\Bigg\}-\frac{9}{16}\left(1+\left(\frac{k_e}{k_s}\right)\right)\Bigg\{{\cal K}_{e,{\rm end}}\cos\left(2\left({\cal K}_{{\rm end},e}-{\cal K}_{{\rm end},s}\right)\right)\Bigg\}\nonumber\\
    &&+\frac{27}{16}{\cal K}_{s,e}^6+\frac{81}{64}{\cal K}_{s,e}^6-\frac{9}{32}{\cal K}_{s,e}^2\left(9+8{\cal K}_{s,e}^2+36{\cal K}_{s,e}^4\right)\left(1-{\cal K}_{e,{\rm end}}\right)-\frac{81}{16}{\cal K}_{s,e}^4\left(1+{\cal K}_{s,e}^2\right)\nonumber\\
    &&+\left(1+4{\cal K}_{s,e}^2+{\cal K}_{s,e}^4\right)\ln{\cal K}_{e,{\rm end}}-\frac{3}{4}{\cal K}_{e,{\rm end}}\cos\left(2{\cal K}_{{\rm end},e}\right)\cdots\Bigg]={\bf F}_{3,{\bf SRII}}(k_{\rm end},k_e),\\
         {\bf F}_{4,{\bf SRII}}(k_{\rm end},k_e)&=&\Bigg[\frac{81}{16}+\frac{27}{8}\left(1+{\cal K}_{s,e}\right)+\frac{81}{16}{\cal K}_{s,e}^6\ln{\cal K}_{e,{\rm end}}+\frac{81}{32}\left(\left(1+{\cal K}_{s,e}\right)^2+2{\cal K}_{s,e}\right)\nonumber\\
         &&+10{\cal K}_{s,e}\left(1+{\cal K}_{s,e}\right)+\frac{1}{12}+\frac{9}{40}\left(1+{\cal K}_{s,e}^2\right)+\frac{27}{64}\left(1+{\cal K}_{s,e}^6\right)+\frac{81}{128}+6{\cal K}_{s,e}^2\ln{\cal K}_{e,{\rm end}}\nonumber\\
         &&-\frac{9}{16}\frac{\displaystyle\left(1+{\cal K}_{s,e}\right)^2}{\displaystyle\left(1-{\cal K}_{s,e}\right)^2}\Bigg\{\cos\left(2\left({\cal K}_{{\rm end},e}-{\cal K}_{{\rm end},s}\right)\right)\Bigg\}-\frac{27}{8}\frac{\displaystyle \displaystyle\left(1+{\cal K}_{s,e}\right)^2}{\displaystyle\left(1-{\cal K}_{s,e}\right)^3}\Bigg\{\sin\left(2\left({\cal K}_{{\rm end},e}-{\cal K}_{{\rm end},s}\right)\right)\Bigg\}\nonumber\\
    &&-\frac{27}{4}\frac{\displaystyle \displaystyle\left(1+{\cal K}_{s,e}\right)^2}{\displaystyle\left(1-{\cal K}_{s,e}\right)}\Bigg\{{\cal K}_{e,{\rm end}}^2\sin\left(2\left({\cal K}_{{\rm end},e}-{\cal K}_{{\rm end},s}\right)\right)\Bigg\}+\frac{9}{80}{\cal K}_{s,e}^2\left(9+44{\cal K}_{s,e}^2\right)+\frac{27}{8}{\cal K}_{s,e}^4\left(1+{\cal K}_{s,e}^2\right)\nonumber\\
    &&+\frac{81}{16}{\cal K}_{s,e}^6\left(1-{\cal K}_{e,{\rm end}}\right)+\frac{99}{56}+\frac{1}{4}{\cal K}_{e,{\rm end}}^2
   +\frac{81}{32}{\cal K}_{s,e}^6{\cal K}_{e,{\rm end}}\Bigg\{\sin\left(2{\cal K}_{{\rm end},e}\right)-{\cal K}_{e,{\rm end}}\cos\left(2{\cal K}_{{\rm end},e}\right)\Bigg\}\cdots\Bigg].\quad\quad
   \quad\eea

In terms of the above equations defined for all the three phases, the one-loop corrections to the scalar power spectrum from each of these phases can be written using the following expression:
\begin{eqnarray} \label{Apploop}
\Delta^{2}_{\zeta,{\bf One-Loop}}(k)
&=& \displaystyle
\displaystyle \Bigg[\Delta^{2}_{\zeta,\bf {Tree}}(k)\Bigg]_{\rm \textbf{SRI}} \times\left\{
	\begin{array}{ll}
		\displaystyle -\sum^{4}_{i=1}{\cal G}_{i,\mbf{SRI}}\mbf{F}_{i,\mbf{SRI}}(k_{s},k_{*})  & \mbox{when}\quad  k < k_s  \;(\rm SRI)  \\  
			\displaystyle 
			\displaystyle \sum^{4}_{i=1}{\cal G}_{i,\mbf{USR}}\mbf{F}_{i,\mbf{USR}}(k_{e},k_{s})\;\Theta(k-k_{s}) & \mbox{when }  k_s\leq k < k_e  \;(\rm USR)\\ 
   \displaystyle 
			\displaystyle \sum^{4}_{i=1}{\cal G}_{i,\mbf{SRII}}\mbf{F}_{i,\mbf{SRII}}(k_{\rm end},k_{e})\;\Theta(k-k_{e}) & \mbox{when }  k_e\leq k\leq k_{\rm end}  \;(\rm SRII) 
	\end{array} \right. \end{eqnarray}

Combining Eqs.(\ref{Apptree}, \ref{Apploop}) gives us the total dimensionless one-loop corrected version of the primordial scalar power spectrum which we ultimately utilize to generate the scalar-induced GW spectrum in Galileon inflation.

\bibliography{references4}

\providecommand{\href}[2]{#2}\begingroup\raggedright\begin{thebibliography}{100}

\bibitem{LIGOScientific:2016aoc}
{\bfseries LIGO Scientific, Virgo} Collaboration, B.~P. Abbott {\em et~al.},
  ``{Observation of Gravitational Waves from a Binary Black Hole Merger},''
  \href{http://dx.doi.org/10.1103/PhysRevLett.116.061102}{{\em Phys. Rev.
  Lett.} {\bfseries 116} no.~6, (2016) 061102},
  \href{http://arxiv.org/abs/1602.03837}{{\ttfamily arXiv:1602.03837 [gr-qc]}}.

\bibitem{Zu:2023olm}
L.~Zu, C.~Zhang, Y.-Y. Li, Y.-C. Gu, Y.-L.~S. Tsai, and Y.-Z. Fan, ``{Mirror
  QCD phase transition as the origin of the nanohertz Stochastic
  Gravitational-Wave Background},''
  \href{http://arxiv.org/abs/2306.16769}{{\ttfamily arXiv:2306.16769
  [astro-ph.HE]}}.

\bibitem{Abe:2023yrw}
K.~T. Abe and Y.~Tada, ``{Translating nano-Hertz gravitational wave background
  into primordial perturbations taking account of the cosmological QCD phase
  transition},'' \href{http://arxiv.org/abs/2307.01653}{{\ttfamily
  arXiv:2307.01653 [astro-ph.CO]}}.

\bibitem{Gouttenoire:2023bqy}
Y.~Gouttenoire, ``{First-order Phase Transition interpretation of PTA signal
  produces solar-mass Black Holes},''
  \href{http://arxiv.org/abs/2307.04239}{{\ttfamily arXiv:2307.04239
  [hep-ph]}}.

\bibitem{NANOGrav:2021flc}
{\bfseries NANOGrav} Collaboration, Z.~Arzoumanian {\em et~al.}, ``{Searching
  for Gravitational Waves from Cosmological Phase Transitions with the NANOGrav
  12.5-Year Dataset},''
  \href{http://dx.doi.org/10.1103/PhysRevLett.127.251302}{{\em Phys. Rev.
  Lett.} {\bfseries 127} no.~25, (2021) 251302},
  \href{http://arxiv.org/abs/2104.13930}{{\ttfamily arXiv:2104.13930
  [astro-ph.CO]}}.

\bibitem{Xue:2021gyq}
X.~Xue {\em et~al.}, ``{Constraining Cosmological Phase Transitions with the
  Parkes Pulsar Timing Array},''
  \href{http://dx.doi.org/10.1103/PhysRevLett.127.251303}{{\em Phys. Rev.
  Lett.} {\bfseries 127} no.~25, (2021) 251303},
  \href{http://arxiv.org/abs/2110.03096}{{\ttfamily arXiv:2110.03096
  [astro-ph.CO]}}.

\bibitem{Nakai:2020oit}
Y.~Nakai, M.~Suzuki, F.~Takahashi, and M.~Yamada, ``{Gravitational Waves and
  Dark Radiation from Dark Phase Transition: Connecting NANOGrav Pulsar Timing
  Data and Hubble Tension},''
  \href{http://dx.doi.org/10.1016/j.physletb.2021.136238}{{\em Phys. Lett. B}
  {\bfseries 816} (2021) 136238},
  \href{http://arxiv.org/abs/2009.09754}{{\ttfamily arXiv:2009.09754
  [astro-ph.CO]}}.

\bibitem{Athron:2023mer}
P.~Athron, A.~Fowlie, C.-T. Lu, L.~Morris, L.~Wu, Y.~Wu, and Z.~Xu, ``{Can
  supercooled phase transitions explain the gravitational wave background
  observed by pulsar timing arrays?},''
  \href{http://arxiv.org/abs/2306.17239}{{\ttfamily arXiv:2306.17239
  [hep-ph]}}.

\bibitem{Madge:2023cak}
E.~Madge, E.~Morgante, C.~Puchades-Ib\'a\~nez, N.~Ramberg, W.~Ratzinger,
  S.~Schenk, and P.~Schwaller, ``{Primordial gravitational waves in the
  nano-Hertz regime and PTA data -- towards solving the GW inverse problem},''
  \href{http://arxiv.org/abs/2306.14856}{{\ttfamily arXiv:2306.14856
  [hep-ph]}}.

\bibitem{Kitajima:2023cek}
N.~Kitajima, J.~Lee, K.~Murai, F.~Takahashi, and W.~Yin, ``{Nanohertz
  Gravitational Waves from Axion Domain Walls Coupled to QCD},''
  \href{http://arxiv.org/abs/2306.17146}{{\ttfamily arXiv:2306.17146
  [hep-ph]}}.

\bibitem{Babichev:2023pbf}
E.~Babichev, D.~Gorbunov, S.~Ramazanov, R.~Samanta, and A.~Vikman, ``{NANOGrav
  spectral index $\gamma=3$ from melting domain walls},''
  \href{http://arxiv.org/abs/2307.04582}{{\ttfamily arXiv:2307.04582
  [hep-ph]}}.

\bibitem{Zhang:2023nrs}
Z.~Zhang, C.~Cai, Y.-H. Su, S.~Wang, Z.-H. Yu, and H.-H. Zhang, ``{Nano-Hertz
  gravitational waves from collapsing domain walls associated with freeze-in
  dark matter in light of pulsar timing array observations},''
  \href{http://arxiv.org/abs/2307.11495}{{\ttfamily arXiv:2307.11495
  [hep-ph]}}.

\bibitem{Zeng:2023jut}
Z.-M. Zeng, J.~Liu, and Z.-K. Guo, ``{Enhanced curvature perturbations from
  spherical domain walls nucleated during inflation},''
  \href{http://arxiv.org/abs/2301.07230}{{\ttfamily arXiv:2301.07230
  [astro-ph.CO]}}.

\bibitem{Ferreira:2022zzo}
R.~Z. Ferreira, A.~Notari, O.~Pujolas, and F.~Rompineve, ``{Gravitational waves
  from domain walls in Pulsar Timing Array datasets},''
  \href{http://dx.doi.org/10.1088/1475-7516/2023/02/001}{{\em JCAP} {\bfseries
  02} (2023) 001}, \href{http://arxiv.org/abs/2204.04228}{{\ttfamily
  arXiv:2204.04228 [astro-ph.CO]}}.

\bibitem{An:2023idh}
H.~An and C.~Yang, ``{Gravitational Waves Produced by Domain Walls During
  Inflation},'' \href{http://arxiv.org/abs/2304.02361}{{\ttfamily
  arXiv:2304.02361 [hep-ph]}}.

\bibitem{Li:2023tdx}
X.-F. Li, ``{Probing the high temperature symmetry breaking with gravitational
  waves from domain walls},'' \href{http://arxiv.org/abs/2307.03163}{{\ttfamily
  arXiv:2307.03163 [hep-ph]}}.

\bibitem{Ellis:2020ena}
J.~Ellis and M.~Lewicki, ``{Cosmic String Interpretation of NANOGrav Pulsar
  Timing Data},'' \href{http://dx.doi.org/10.1103/PhysRevLett.126.041304}{{\em
  Phys. Rev. Lett.} {\bfseries 126} no.~4, (2021) 041304},
  \href{http://arxiv.org/abs/2009.06555}{{\ttfamily arXiv:2009.06555
  [astro-ph.CO]}}.

\bibitem{Blasi:2020mfx}
S.~Blasi, V.~Brdar, and K.~Schmitz, ``{Has NANOGrav found first evidence for
  cosmic strings?},''
  \href{http://dx.doi.org/10.1103/PhysRevLett.126.041305}{{\em Phys. Rev.
  Lett.} {\bfseries 126} no.~4, (2021) 041305},
  \href{http://arxiv.org/abs/2009.06607}{{\ttfamily arXiv:2009.06607
  [astro-ph.CO]}}.

\bibitem{Buchmuller:2020lbh}
W.~Buchmuller, V.~Domcke, and K.~Schmitz, ``{From NANOGrav to LIGO with
  metastable cosmic strings},''
  \href{http://dx.doi.org/10.1016/j.physletb.2020.135914}{{\em Phys. Lett. B}
  {\bfseries 811} (2020) 135914},
  \href{http://arxiv.org/abs/2009.10649}{{\ttfamily arXiv:2009.10649
  [astro-ph.CO]}}.

\bibitem{Blanco-Pillado:2021ygr}
J.~J. Blanco-Pillado, K.~D. Olum, and J.~M. Wachter, ``{Comparison of cosmic
  string and superstring models to NANOGrav 12.5-year results},''
  \href{http://dx.doi.org/10.1103/PhysRevD.103.103512}{{\em Phys. Rev. D}
  {\bfseries 103} no.~10, (2021) 103512},
  \href{http://arxiv.org/abs/2102.08194}{{\ttfamily arXiv:2102.08194
  [astro-ph.CO]}}.

\bibitem{Buchmuller:2021mbb}
W.~Buchmuller, V.~Domcke, and K.~Schmitz, ``{Stochastic gravitational-wave
  background from metastable cosmic strings},''
  \href{http://dx.doi.org/10.1088/1475-7516/2021/12/006}{{\em JCAP} {\bfseries
  12} no.~12, (2021) 006}, \href{http://arxiv.org/abs/2107.04578}{{\ttfamily
  arXiv:2107.04578 [hep-ph]}}.

\bibitem{Inomata:2023zup}
K.~Inomata, K.~Kohri, and T.~Terada, ``{The Detected Stochastic Gravitational
  Waves and Subsolar-Mass Primordial Black Holes},''
  \href{http://arxiv.org/abs/2306.17834}{{\ttfamily arXiv:2306.17834
  [astro-ph.CO]}}.

\bibitem{Zhu:2023faa}
Q.-H. Zhu, Z.-C. Zhao, and S.~Wang, ``{Joint implications of BBN, CMB, and PTA
  Datasets for Scalar-Induced Gravitational Waves of Second and Third
  orders},'' \href{http://arxiv.org/abs/2307.03095}{{\ttfamily arXiv:2307.03095
  [astro-ph.CO]}}.

\bibitem{HosseiniMansoori:2023mqh}
S.~A. Hosseini~Mansoori, F.~Felegray, A.~Talebian, and M.~Sami, ``{PBHs and GWs
  from $\mathbb{T}^2$-inflation and NANOGrav 15-year data},''
  \href{http://arxiv.org/abs/2307.06757}{{\ttfamily arXiv:2307.06757
  [astro-ph.CO]}}.

\bibitem{Das:2023nmm}
B.~Das, N.~Jaman, and M.~Sami, ``{Gravitational Waves Background (NANOGrav)
  from Quintessential Inflation},''
  \href{http://arxiv.org/abs/2307.12913}{{\ttfamily arXiv:2307.12913 [gr-qc]}}.

\bibitem{Balaji:2023ehk}
S.~Balaji, G.~Dom\`enech, and G.~Franciolini, ``{Scalar-induced gravitational
  wave interpretation of PTA data: the role of scalar fluctuation propagation
  speed},'' \href{http://arxiv.org/abs/2307.08552}{{\ttfamily arXiv:2307.08552
  [gr-qc]}}.

\bibitem{Cai:2023dls}
Y.-F. Cai, X.-C. He, X.~Ma, S.-F. Yan, and G.-W. Yuan, ``{Limits on
  scalar-induced gravitational waves from the stochastic background by pulsar
  timing array observations},''
  \href{http://arxiv.org/abs/2306.17822}{{\ttfamily arXiv:2306.17822 [gr-qc]}}.

\bibitem{Wang:2023ost}
S.~Wang, Z.-C. Zhao, J.-P. Li, and Q.-H. Zhu, ``{Exploring the Implications of
  2023 Pulsar Timing Array Datasets for Scalar-Induced Gravitational Waves and
  Primordial Black Holes},'' \href{http://arxiv.org/abs/2307.00572}{{\ttfamily
  arXiv:2307.00572 [astro-ph.CO]}}.

\bibitem{Yi:2023mbm}
Z.~Yi, Q.~Gao, Y.~Gong, Y.~Wang, and F.~Zhang, ``{The waveform of the scalar
  induced gravitational waves in light of Pulsar Timing Array data},''
  \href{http://arxiv.org/abs/2307.02467}{{\ttfamily arXiv:2307.02467 [gr-qc]}}.

\bibitem{Choudhury:2013woa}
S.~Choudhury and A.~Mazumdar, ``{Primordial blackholes and gravitational waves
  for an inflection-point model of inflation},''
  \href{http://dx.doi.org/10.1016/j.physletb.2014.04.050}{{\em Phys. Lett. B}
  {\bfseries 733} (2014) 270--275},
  \href{http://arxiv.org/abs/1307.5119}{{\ttfamily arXiv:1307.5119
  [astro-ph.CO]}}.

\bibitem{Choudhury:2023kam}
S.~Choudhury, ``{Single field inflation in the light of NANOGrav 15-year Data:
  Quintessential interpretation of blue tilted tensor spectrum through
  Non-Bunch Davies initial condition},''
  \href{http://arxiv.org/abs/2307.03249}{{\ttfamily arXiv:2307.03249
  [astro-ph.CO]}}.

\bibitem{Choudhury:2023tcn}
S.~Choudhury, A.~Karde, K.~Dey, S.~Panda, and M.~Sami, ``{Primordial
  non-Gaussianity as a saviour for PBH overproduction in SIGWs generated by
  Pulsar Timing Arrays for Galileon inflation},''
  \href{http://arxiv.org/abs/2310.11034}{{\ttfamily arXiv:2310.11034
  [astro-ph.CO]}}.

\bibitem{Bhattacharya:2023ysp}
G.~Bhattacharya, S.~Choudhury, K.~Dey, S.~Ghosh, A.~Karde, and N.~S. Mishra,
  ``{Evading no-go for PBH formation and production of SIGWs using Multiple
  Sharp Transitions in EFT of single field inflation},''
  \href{http://arxiv.org/abs/2309.00973}{{\ttfamily arXiv:2309.00973
  [astro-ph.CO]}}.

\bibitem{Vagnozzi:2023lwo}
S.~Vagnozzi, ``{Inflationary interpretation of the stochastic gravitational
  wave background signal detected by pulsar timing array experiments},''
  \href{http://dx.doi.org/10.1016/j.jheap.2023.07.001}{{\em JHEAp} {\bfseries
  39} (2023) 81--98}, \href{http://arxiv.org/abs/2306.16912}{{\ttfamily
  arXiv:2306.16912 [astro-ph.CO]}}.

\bibitem{Franciolini:2023pbf}
G.~Franciolini, A.~Iovino, Junior., V.~Vaskonen, and H.~Veermae, ``{The recent
  gravitational wave observation by pulsar timing arrays and primordial black
  holes: the importance of non-gaussianities},''
  \href{http://arxiv.org/abs/2306.17149}{{\ttfamily arXiv:2306.17149
  [astro-ph.CO]}}.

\bibitem{Gorji:2023sil}
M.~A. Gorji, M.~Sasaki, and T.~Suyama, ``{Extra-tensor-induced origin for the
  PTA signal: No primordial black hole production},''
  \href{http://arxiv.org/abs/2307.13109}{{\ttfamily arXiv:2307.13109
  [astro-ph.CO]}}.

\bibitem{DeLuca:2023tun}
V.~De~Luca, A.~Kehagias, and A.~Riotto, ``{How Well Do We Know the Primordial
  Black Hole Abundance? The Crucial Role of Non-Linearities when Approaching
  the Horizon},'' \href{http://arxiv.org/abs/2307.13633}{{\ttfamily
  arXiv:2307.13633 [astro-ph.CO]}}.

\bibitem{Frosina:2023nxu}
L.~Frosina and A.~Urbano, ``{On the inflationary interpretation of the nHz
  gravitational-wave background},''
  \href{http://arxiv.org/abs/2308.06915}{{\ttfamily arXiv:2308.06915
  [astro-ph.CO]}}.

\bibitem{Chen:2019xse}
Z.-C. Chen, C.~Yuan, and Q.-G. Huang, ``{Pulsar Timing Array Constraints on
  Primordial Black Holes with NANOGrav 11-Year Dataset},''
  \href{http://dx.doi.org/10.1103/PhysRevLett.124.251101}{{\em Phys. Rev.
  Lett.} {\bfseries 124} no.~25, (2020) 251101},
  \href{http://arxiv.org/abs/1910.12239}{{\ttfamily arXiv:1910.12239
  [astro-ph.CO]}}.

\bibitem{Cai:2023uhc}
Y.~Cai, M.~Zhu, and Y.-S. Piao, ``{Primordial black holes from null energy
  condition violation during inflation},''
  \href{http://arxiv.org/abs/2305.10933}{{\ttfamily arXiv:2305.10933 [gr-qc]}}.

\bibitem{Huang:2023chx}
H.-L. Huang, Y.~Cai, J.-Q. Jiang, J.~Zhang, and Y.-S. Piao, ``{Supermassive
  primordial black holes in multiverse: for nano-Hertz gravitational wave and
  high-redshift JWST galaxies},''
  \href{http://arxiv.org/abs/2306.17577}{{\ttfamily arXiv:2306.17577 [gr-qc]}}.

\bibitem{Cang:2023ysz}
J.~Cang, Y.~Gao, Y.~Liu, and S.~Sun, ``{High Frequency Gravitational Waves from
  Pulsar Timing Arrays},'' \href{http://arxiv.org/abs/2309.15069}{{\ttfamily
  arXiv:2309.15069 [astro-ph.CO]}}.

\bibitem{Domenech:2021ztg}
G.~Dom\`enech, ``{Scalar Induced Gravitational Waves Review},''
  \href{http://dx.doi.org/10.3390/universe7110398}{{\em Universe} {\bfseries 7}
  no.~11, (2021) 398}, \href{http://arxiv.org/abs/2109.01398}{{\ttfamily
  arXiv:2109.01398 [gr-qc]}}.

\bibitem{Choudhury:2023fjs}
S.~Choudhury, K.~Dey, and A.~Karde, ``{Untangling PBH overproduction in
  $w$-SIGWs generated by Pulsar Timing Arrays for MST-EFT of single field
  inflation},'' \href{http://arxiv.org/abs/2311.15065}{{\ttfamily
  arXiv:2311.15065 [astro-ph.CO]}}.

\bibitem{Choudhury:2023fwk}
S.~Choudhury, K.~Dey, A.~Karde, S.~Panda, and M.~Sami, ``{Primordial
  non-Gaussianity as a saviour for PBH overproduction in SIGWs generated by
  Pulsar Timing Arrays for Galileon inflation},''
  \href{http://arxiv.org/abs/2310.11034}{{\ttfamily arXiv:2310.11034
  [astro-ph.CO]}}.

\bibitem{Baumann:2009ds}
D.~Baumann,
  \href{http://dx.doi.org/10.1142/9789814327183_0010}{``{Inflation},''} in {\em
  {Theoretical Advanced Study Institute in Elementary Particle Physics}:
  {Physics of the Large and the Small}}, pp.~523--686.
\newblock 2011.
\newblock \href{http://arxiv.org/abs/0907.5424}{{\ttfamily arXiv:0907.5424
  [hep-th]}}.

\bibitem{Baumann:2022mni}
D.~Baumann, \href{http://dx.doi.org/10.1017/9781108937092}{{\em {Cosmology}}}.
\newblock Cambridge University Press, 7, 2022.

\bibitem{Baumann:2018muz}
D.~Baumann, ``{Primordial Cosmology},''
  \href{http://dx.doi.org/10.22323/1.305.0009}{{\em PoS} {\bfseries TASI2017}
  (2018) 009}, \href{http://arxiv.org/abs/1807.03098}{{\ttfamily
  arXiv:1807.03098 [hep-th]}}.

\bibitem{Senatore:2016aui}
L.~Senatore, \href{http://dx.doi.org/10.1142/9789813149441_0008}{``{Lectures on
  Inflation},''} in {\em {Theoretical Advanced Study Institute in Elementary
  Particle Physics}: {New Frontiers in Fields and Strings}}, pp.~447--543.
\newblock 2017.
\newblock \href{http://arxiv.org/abs/1609.00716}{{\ttfamily arXiv:1609.00716
  [hep-th]}}.

\bibitem{Martin:2013tda}
J.~Martin, C.~Ringeval, and V.~Vennin, ``{Encyclop\ae{}dia Inflationaris},''
  \href{http://dx.doi.org/10.1016/j.dark.2014.01.003}{{\em Phys. Dark Univ.}
  {\bfseries 5-6} (2014) 75--235},
  \href{http://arxiv.org/abs/1303.3787}{{\ttfamily arXiv:1303.3787
  [astro-ph.CO]}}.

\bibitem{Martin:2013nzq}
J.~Martin, C.~Ringeval, R.~Trotta, and V.~Vennin, ``{The Best Inflationary
  Models After Planck},''
  \href{http://dx.doi.org/10.1088/1475-7516/2014/03/039}{{\em JCAP} {\bfseries
  03} (2014) 039}, \href{http://arxiv.org/abs/1312.3529}{{\ttfamily
  arXiv:1312.3529 [astro-ph.CO]}}.

\bibitem{Mazumdar:2010sa}
A.~Mazumdar and J.~Rocher, ``{Particle physics models of inflation and curvaton
  scenarios},'' \href{http://dx.doi.org/10.1016/j.physrep.2010.08.001}{{\em
  Phys. Rept.} {\bfseries 497} (2011) 85--215},
  \href{http://arxiv.org/abs/1001.0993}{{\ttfamily arXiv:1001.0993 [hep-ph]}}.

\bibitem{Lyth:1998xn}
D.~H. Lyth and A.~Riotto, ``{Particle physics models of inflation and the
  cosmological density perturbation},''
  \href{http://dx.doi.org/10.1016/S0370-1573(98)00128-8}{{\em Phys. Rept.}
  {\bfseries 314} (1999) 1--146},
  \href{http://arxiv.org/abs/hep-ph/9807278}{{\ttfamily arXiv:hep-ph/9807278}}.

\bibitem{Planck:2018jri}
{\bfseries Planck} Collaboration, Y.~Akrami {\em et~al.}, ``{Planck 2018
  results. X. Constraints on inflation},''
  \href{http://dx.doi.org/10.1051/0004-6361/201833887}{{\em Astron. Astrophys.}
  {\bfseries 641} (2020) A10},
  \href{http://arxiv.org/abs/1807.06211}{{\ttfamily arXiv:1807.06211
  [astro-ph.CO]}}.

\bibitem{Planck:2018vyg}
{\bfseries Planck} Collaboration, N.~Aghanim {\em et~al.}, ``{Planck 2018
  results. VI. Cosmological parameters},''
  \href{http://dx.doi.org/10.1051/0004-6361/201833910}{{\em Astron. Astrophys.}
  {\bfseries 641} (2020) A6}, \href{http://arxiv.org/abs/1807.06209}{{\ttfamily
  arXiv:1807.06209 [astro-ph.CO]}}. [Erratum: Astron.Astrophys. 652, C4
  (2021)].

\bibitem{CMB-S4:2016ple}
{\bfseries CMB-S4} Collaboration, K.~N. Abazajian {\em et~al.}, ``{CMB-S4
  Science Book, First Edition},''
  \href{http://arxiv.org/abs/1610.02743}{{\ttfamily arXiv:1610.02743
  [astro-ph.CO]}}.

\bibitem{NANOGrav:2023gor}
{\bfseries NANOGrav} Collaboration, G.~Agazie {\em et~al.}, ``{The NANOGrav 15
  yr Data Set: Evidence for a Gravitational-wave Background},''
  \href{http://dx.doi.org/10.3847/2041-8213/acdac6}{{\em Astrophys. J. Lett.}
  {\bfseries 951} no.~1, (2023) L8},
  \href{http://arxiv.org/abs/2306.16213}{{\ttfamily arXiv:2306.16213
  [astro-ph.HE]}}.

\bibitem{NANOGrav:2023hde}
{\bfseries NANOGrav} Collaboration, G.~Agazie {\em et~al.}, ``{The NANOGrav 15
  yr Data Set: Observations and Timing of 68 Millisecond Pulsars},''
  \href{http://dx.doi.org/10.3847/2041-8213/acda9a}{{\em Astrophys. J. Lett.}
  {\bfseries 951} no.~1, (2023) L9},
  \href{http://arxiv.org/abs/2306.16217}{{\ttfamily arXiv:2306.16217
  [astro-ph.HE]}}.

\bibitem{NANOGrav:2023ctt}
{\bfseries NANOGrav} Collaboration, G.~Agazie {\em et~al.}, ``{The NANOGrav 15
  yr Data Set: Detector Characterization and Noise Budget},''
  \href{http://dx.doi.org/10.3847/2041-8213/acda88}{{\em Astrophys. J. Lett.}
  {\bfseries 951} no.~1, (2023) L10},
  \href{http://arxiv.org/abs/2306.16218}{{\ttfamily arXiv:2306.16218
  [astro-ph.HE]}}.

\bibitem{NANOGrav:2023hvm}
{\bfseries NANOGrav} Collaboration, A.~Afzal {\em et~al.}, ``{The NANOGrav 15
  yr Data Set: Search for Signals from New Physics},''
  \href{http://dx.doi.org/10.3847/2041-8213/acdc91}{{\em Astrophys. J. Lett.}
  {\bfseries 951} no.~1, (2023) L11},
  \href{http://arxiv.org/abs/2306.16219}{{\ttfamily arXiv:2306.16219
  [astro-ph.HE]}}.

\bibitem{NANOGrav:2023hfp}
{\bfseries NANOGrav} Collaboration, G.~Agazie {\em et~al.}, ``{The NANOGrav 15
  yr Data Set: Constraints on Supermassive Black Hole Binaries from the
  Gravitational-wave Background},''
  \href{http://dx.doi.org/10.3847/2041-8213/ace18b}{{\em Astrophys. J. Lett.}
  {\bfseries 952} no.~2, (2023) L37},
  \href{http://arxiv.org/abs/2306.16220}{{\ttfamily arXiv:2306.16220
  [astro-ph.HE]}}.

\bibitem{NANOGrav:2023tcn}
{\bfseries NANOGrav} Collaboration, G.~Agazie {\em et~al.}, ``{The NANOGrav
  15-year Data Set: Search for Anisotropy in the Gravitational-Wave
  Background},'' \href{http://arxiv.org/abs/2306.16221}{{\ttfamily
  arXiv:2306.16221 [astro-ph.HE]}}.

\bibitem{NANOGrav:2023pdq}
{\bfseries NANOGrav} Collaboration, G.~Agazie {\em et~al.}, ``{The NANOGrav 15
  yr Data Set: Bayesian Limits on Gravitational Waves from Individual
  Supermassive Black Hole Binaries},''
  \href{http://dx.doi.org/10.3847/2041-8213/ace18a}{{\em Astrophys. J. Lett.}
  {\bfseries 951} no.~2, (2023) L50},
  \href{http://arxiv.org/abs/2306.16222}{{\ttfamily arXiv:2306.16222
  [astro-ph.HE]}}.

\bibitem{NANOGrav:2023icp}
{\bfseries NANOGrav} Collaboration, A.~D. Johnson {\em et~al.}, ``{The NANOGrav
  15-year Gravitational-Wave Background Analysis Pipeline},''
  \href{http://arxiv.org/abs/2306.16223}{{\ttfamily arXiv:2306.16223
  [astro-ph.HE]}}.

\bibitem{EPTA:2023fyk}
{\bfseries EPTA} Collaboration, J.~Antoniadis {\em et~al.}, ``{The second data
  release from the European Pulsar Timing Array III. Search for gravitational
  wave signals},'' \href{http://arxiv.org/abs/2306.16214}{{\ttfamily
  arXiv:2306.16214 [astro-ph.HE]}}.

\bibitem{EPTA:2023sfo}
{\bfseries EPTA} Collaboration, J.~Antoniadis {\em et~al.}, ``{The second data
  release from the European Pulsar Timing Array I. The dataset and timing
  analysis},'' \href{http://arxiv.org/abs/2306.16224}{{\ttfamily
  arXiv:2306.16224 [astro-ph.HE]}}.

\bibitem{EPTA:2023akd}
{\bfseries EPTA} Collaboration, J.~Antoniadis {\em et~al.}, ``{The second data
  release from the European Pulsar Timing Array II. Customised pulsar noise
  models for spatially correlated gravitational waves},''
  \href{http://arxiv.org/abs/2306.16225}{{\ttfamily arXiv:2306.16225
  [astro-ph.HE]}}.

\bibitem{EPTA:2023gyr}
{\bfseries EPTA} Collaboration, J.~Antoniadis {\em et~al.}, ``{The second data
  release from the European Pulsar Timing Array IV. Search for continuous
  gravitational wave signals},''
  \href{http://arxiv.org/abs/2306.16226}{{\ttfamily arXiv:2306.16226
  [astro-ph.HE]}}.

\bibitem{EPTA:2023xxk}
{\bfseries EPTA} Collaboration, J.~Antoniadis {\em et~al.}, ``{The second data
  release from the European Pulsar Timing Array: V. Implications for massive
  black holes, dark matter and the early Universe},''
  \href{http://arxiv.org/abs/2306.16227}{{\ttfamily arXiv:2306.16227
  [astro-ph.CO]}}.

\bibitem{EPTA:2023xiy}
{\bfseries EPTA} Collaboration, C.~Smarra {\em et~al.}, ``{The second data
  release from the European Pulsar Timing Array: VI. Challenging the ultralight
  dark matter paradigm},'' \href{http://arxiv.org/abs/2306.16228}{{\ttfamily
  arXiv:2306.16228 [astro-ph.HE]}}.

\bibitem{Reardon:2023gzh}
D.~J. Reardon {\em et~al.}, ``{Search for an Isotropic Gravitational-wave
  Background with the Parkes Pulsar Timing Array},''
  \href{http://dx.doi.org/10.3847/2041-8213/acdd02}{{\em Astrophys. J. Lett.}
  {\bfseries 951} no.~1, (2023) L6},
  \href{http://arxiv.org/abs/2306.16215}{{\ttfamily arXiv:2306.16215
  [astro-ph.HE]}}.

\bibitem{Reardon:2023zen}
D.~J. Reardon {\em et~al.}, ``{The Gravitational-wave Background Null
  Hypothesis: Characterizing Noise in Millisecond Pulsar Arrival Times with the
  Parkes Pulsar Timing Array},''
  \href{http://dx.doi.org/10.3847/2041-8213/acdd03}{{\em Astrophys. J. Lett.}
  {\bfseries 951} no.~1, (2023) L7},
  \href{http://arxiv.org/abs/2306.16229}{{\ttfamily arXiv:2306.16229
  [astro-ph.HE]}}.

\bibitem{Zic:2023gta}
A.~Zic {\em et~al.}, ``{The Parkes Pulsar Timing Array Third Data Release},''
  \href{http://arxiv.org/abs/2306.16230}{{\ttfamily arXiv:2306.16230
  [astro-ph.HE]}}.

\bibitem{Xu:2023wog}
H.~Xu {\em et~al.}, ``{Searching for the Nano-Hertz Stochastic Gravitational
  Wave Background with the Chinese Pulsar Timing Array Data Release I},''
  \href{http://dx.doi.org/10.1088/1674-4527/acdfa5}{{\em Res. Astron.
  Astrophys.} {\bfseries 23} no.~7, (2023) 075024},
  \href{http://arxiv.org/abs/2306.16216}{{\ttfamily arXiv:2306.16216
  [astro-ph.HE]}}.

\bibitem{Matarrese:1992rp}
S.~Matarrese, O.~Pantano, and D.~Saez, ``{A General relativistic approach to
  the nonlinear evolution of collisionless matter},''
  \href{http://dx.doi.org/10.1103/PhysRevD.47.1311}{{\em Phys. Rev. D}
  {\bfseries 47} (1993) 1311--1323}.

\bibitem{Matarrese:1993zf}
S.~Matarrese, O.~Pantano, and D.~Saez, ``{General relativistic dynamics of
  irrotational dust: Cosmological implications},''
  \href{http://dx.doi.org/10.1103/PhysRevLett.72.320}{{\em Phys. Rev. Lett.}
  {\bfseries 72} (1994) 320--323},
  \href{http://arxiv.org/abs/astro-ph/9310036}{{\ttfamily
  arXiv:astro-ph/9310036}}.

\bibitem{Matarrese:1997ay}
S.~Matarrese, S.~Mollerach, and M.~Bruni, ``{Second order perturbations of the
  Einstein-de Sitter universe},''
  \href{http://dx.doi.org/10.1103/PhysRevD.58.043504}{{\em Phys. Rev. D}
  {\bfseries 58} (1998) 043504},
  \href{http://arxiv.org/abs/astro-ph/9707278}{{\ttfamily
  arXiv:astro-ph/9707278}}.

\bibitem{Ananda:2006af}
K.~N. Ananda, C.~Clarkson, and D.~Wands, ``{The Cosmological gravitational wave
  background from primordial density perturbations},''
  \href{http://dx.doi.org/10.1103/PhysRevD.75.123518}{{\em Phys. Rev. D}
  {\bfseries 75} (2007) 123518},
  \href{http://arxiv.org/abs/gr-qc/0612013}{{\ttfamily arXiv:gr-qc/0612013}}.

\bibitem{Baumann:2007zm}
D.~Baumann, P.~J. Steinhardt, K.~Takahashi, and K.~Ichiki, ``{Gravitational
  Wave Spectrum Induced by Primordial Scalar Perturbations},''
  \href{http://dx.doi.org/10.1103/PhysRevD.76.084019}{{\em Phys. Rev. D}
  {\bfseries 76} (2007) 084019},
  \href{http://arxiv.org/abs/hep-th/0703290}{{\ttfamily arXiv:hep-th/0703290}}.

\bibitem{Saito:2008jc}
R.~Saito and J.~Yokoyama, ``{Gravitational wave background as a probe of the
  primordial black hole abundance},''
  \href{http://dx.doi.org/10.1103/PhysRevLett.102.161101}{{\em Phys. Rev.
  Lett.} {\bfseries 102} (2009) 161101},
  \href{http://arxiv.org/abs/0812.4339}{{\ttfamily arXiv:0812.4339
  [astro-ph]}}. [Erratum: Phys.Rev.Lett. 107, 069901 (2011)].

\bibitem{saito2010gravitational}
R.~Saito and J.~Yokoyama, ``Gravitational-wave constraints on the abundance of
  primordial black holes,'' {\em Progress of theoretical physics} {\bfseries
  123} no.~5, (2010) 867--886.

\bibitem{Hawking:1974rv}
S.~W. Hawking, ``{Black hole explosions},''
  \href{http://dx.doi.org/10.1038/248030a0}{{\em Nature} {\bfseries 248} (1974)
  30--31}.

\bibitem{Carr:1974nx}
B.~J. Carr and S.~W. Hawking, ``{Black holes in the early Universe},''
  \href{http://dx.doi.org/10.1093/mnras/168.2.399}{{\em Mon. Not. Roy. Astron.
  Soc.} {\bfseries 168} (1974) 399--415}.

\bibitem{Carr:1975qj}
B.~J. Carr, ``{The Primordial black hole mass spectrum},''
  \href{http://dx.doi.org/10.1086/153853}{{\em Astrophys. J.} {\bfseries 201}
  (1975) 1--19}.

\bibitem{Chapline:1975ojl}
G.~F. Chapline, ``{Cosmological effects of primordial black holes},''
  \href{http://dx.doi.org/10.1038/253251a0}{{\em Nature} {\bfseries 253}
  no.~5489, (1975) 251--252}.

\bibitem{Carr:1993aq}
B.~J. Carr and J.~E. Lidsey, ``{Primordial black holes and generalized
  constraints on chaotic inflation},''
  \href{http://dx.doi.org/10.1103/PhysRevD.48.543}{{\em Phys. Rev. D}
  {\bfseries 48} (1993) 543--553}.

\bibitem{Yokoyama:1998pt}
J.~Yokoyama, ``{Chaotic new inflation and formation of primordial black
  holes},'' \href{http://dx.doi.org/10.1103/PhysRevD.58.083510}{{\em Phys. Rev.
  D} {\bfseries 58} (1998) 083510},
  \href{http://arxiv.org/abs/astro-ph/9802357}{{\ttfamily
  arXiv:astro-ph/9802357}}.

\bibitem{Rubin:2001yw}
S.~G. Rubin, A.~S. Sakharov, and M.~Y. Khlopov, ``{The Formation of primary
  galactic nuclei during phase transitions in the early universe},''
  \href{http://dx.doi.org/10.1134/1.1385631}{{\em J. Exp. Theor. Phys.}
  {\bfseries 91} (2001) 921--929},
  \href{http://arxiv.org/abs/hep-ph/0106187}{{\ttfamily arXiv:hep-ph/0106187}}.

\bibitem{Khlopov:2002yi}
M.~Y. Khlopov, S.~G. Rubin, and A.~S. Sakharov, ``{Strong primordial
  inhomogeneities and galaxy formation},''
  \href{http://arxiv.org/abs/astro-ph/0202505}{{\ttfamily
  arXiv:astro-ph/0202505}}.

\bibitem{Khlopov:2004sc}
M.~Y. Khlopov, S.~G. Rubin, and A.~S. Sakharov, ``{Primordial structure of
  massive black hole clusters},''
  \href{http://dx.doi.org/10.1016/j.astropartphys.2004.12.002}{{\em Astropart.
  Phys.} {\bfseries 23} (2005) 265},
  \href{http://arxiv.org/abs/astro-ph/0401532}{{\ttfamily
  arXiv:astro-ph/0401532}}.

\bibitem{Saito:2008em}
R.~Saito, J.~Yokoyama, and R.~Nagata, ``{Single-field inflation, anomalous
  enhancement of superhorizon fluctuations, and non-Gaussianity in primordial
  black hole formation},''
  \href{http://dx.doi.org/10.1088/1475-7516/2008/06/024}{{\em JCAP} {\bfseries
  06} (2008) 024}, \href{http://arxiv.org/abs/0804.3470}{{\ttfamily
  arXiv:0804.3470 [astro-ph]}}.

\bibitem{Khlopov:2008qy}
M.~Y. Khlopov, ``{Primordial Black Holes},''
  \href{http://dx.doi.org/10.1088/1674-4527/10/6/001}{{\em Res. Astron.
  Astrophys.} {\bfseries 10} (2010) 495--528},
  \href{http://arxiv.org/abs/0801.0116}{{\ttfamily arXiv:0801.0116
  [astro-ph]}}.

\bibitem{Carr:2009jm}
B.~J. Carr, K.~Kohri, Y.~Sendouda, and J.~Yokoyama, ``{New cosmological
  constraints on primordial black holes},''
  \href{http://dx.doi.org/10.1103/PhysRevD.81.104019}{{\em Phys. Rev. D}
  {\bfseries 81} (2010) 104019},
  \href{http://arxiv.org/abs/0912.5297}{{\ttfamily arXiv:0912.5297
  [astro-ph.CO]}}.

\bibitem{Choudhury:2011jt}
S.~Choudhury and S.~Pal, ``{Fourth level MSSM inflation from new flat
  directions},'' \href{http://dx.doi.org/10.1088/1475-7516/2012/04/018}{{\em
  JCAP} {\bfseries 04} (2012) 018},
  \href{http://arxiv.org/abs/1111.3441}{{\ttfamily arXiv:1111.3441 [hep-ph]}}.

\bibitem{Lyth:2011kj}
D.~H. Lyth, ``{Primordial black hole formation and hybrid inflation},''
  \href{http://arxiv.org/abs/1107.1681}{{\ttfamily arXiv:1107.1681
  [astro-ph.CO]}}.

\bibitem{Drees:2011yz}
M.~Drees and E.~Erfani, ``{Running Spectral Index and Formation of Primordial
  Black Hole in Single Field Inflation Models},''
  \href{http://dx.doi.org/10.1088/1475-7516/2012/01/035}{{\em JCAP} {\bfseries
  01} (2012) 035}, \href{http://arxiv.org/abs/1110.6052}{{\ttfamily
  arXiv:1110.6052 [astro-ph.CO]}}.

\bibitem{Drees:2011hb}
M.~Drees and E.~Erfani, ``{Running-Mass Inflation Model and Primordial Black
  Holes},'' \href{http://dx.doi.org/10.1088/1475-7516/2011/04/005}{{\em JCAP}
  {\bfseries 04} (2011) 005}, \href{http://arxiv.org/abs/1102.2340}{{\ttfamily
  arXiv:1102.2340 [hep-ph]}}.

\bibitem{Hertzberg:2017dkh}
M.~P. Hertzberg and M.~Yamada, ``{Primordial Black Holes from Polynomial
  Potentials in Single Field Inflation},''
  \href{http://dx.doi.org/10.1103/PhysRevD.97.083509}{{\em Phys. Rev. D}
  {\bfseries 97} no.~8, (2018) 083509},
  \href{http://arxiv.org/abs/1712.09750}{{\ttfamily arXiv:1712.09750
  [astro-ph.CO]}}.

\bibitem{Cicoli:2018asa}
M.~Cicoli, V.~A. Diaz, and F.~G. Pedro, ``{Primordial Black Holes from String
  Inflation},'' \href{http://dx.doi.org/10.1088/1475-7516/2018/06/034}{{\em
  JCAP} {\bfseries 06} (2018) 034},
  \href{http://arxiv.org/abs/1803.02837}{{\ttfamily arXiv:1803.02837
  [hep-th]}}.

\bibitem{Ozsoy:2018flq}
O.~\"Ozsoy, S.~Parameswaran, G.~Tasinato, and I.~Zavala, ``{Mechanisms for
  Primordial Black Hole Production in String Theory},''
  \href{http://dx.doi.org/10.1088/1475-7516/2018/07/005}{{\em JCAP} {\bfseries
  07} (2018) 005}, \href{http://arxiv.org/abs/1803.07626}{{\ttfamily
  arXiv:1803.07626 [hep-th]}}.

\bibitem{Byrnes:2018txb}
C.~T. Byrnes, P.~S. Cole, and S.~P. Patil, ``{Steepest growth of the power
  spectrum and primordial black holes},''
  \href{http://dx.doi.org/10.1088/1475-7516/2019/06/028}{{\em JCAP} {\bfseries
  06} (2019) 028}, \href{http://arxiv.org/abs/1811.11158}{{\ttfamily
  arXiv:1811.11158 [astro-ph.CO]}}.

\bibitem{Martin:2019nuw}
J.~Martin, T.~Papanikolaou, and V.~Vennin, ``{Primordial black holes from the
  preheating instability in single-field inflation},''
  \href{http://dx.doi.org/10.1088/1475-7516/2020/01/024}{{\em JCAP} {\bfseries
  01} (2020) 024}, \href{http://arxiv.org/abs/1907.04236}{{\ttfamily
  arXiv:1907.04236 [astro-ph.CO]}}.

\bibitem{Ezquiaga:2019ftu}
J.~M. Ezquiaga, J.~Garc\'\i{}a-Bellido, and V.~Vennin, ``{The exponential tail
  of inflationary fluctuations: consequences for primordial black holes},''
  \href{http://dx.doi.org/10.1088/1475-7516/2020/03/029}{{\em JCAP} {\bfseries
  03} (2020) 029}, \href{http://arxiv.org/abs/1912.05399}{{\ttfamily
  arXiv:1912.05399 [astro-ph.CO]}}.

\bibitem{Motohashi:2019rhu}
H.~Motohashi, S.~Mukohyama, and M.~Oliosi, ``{Constant Roll and Primordial
  Black Holes},'' \href{http://dx.doi.org/10.1088/1475-7516/2020/03/002}{{\em
  JCAP} {\bfseries 03} (2020) 002},
  \href{http://arxiv.org/abs/1910.13235}{{\ttfamily arXiv:1910.13235 [gr-qc]}}.

\bibitem{Ashoorioon:2019xqc}
A.~Ashoorioon, A.~Rostami, and J.~T. Firouzjaee, ``{EFT compatible PBHs:
  effective spawning of the seeds for primordial black holes during
  inflation},'' \href{http://dx.doi.org/10.1007/JHEP07(2021)087}{{\em JHEP}
  {\bfseries 07} (2021) 087}, \href{http://arxiv.org/abs/1912.13326}{{\ttfamily
  arXiv:1912.13326 [astro-ph.CO]}}.

\bibitem{Auclair:2020csm}
P.~Auclair and V.~Vennin, ``{Primordial black holes from metric preheating:
  mass fraction in the excursion-set approach},''
  \href{http://dx.doi.org/10.1088/1475-7516/2021/02/038}{{\em JCAP} {\bfseries
  02} (2021) 038}, \href{http://arxiv.org/abs/2011.05633}{{\ttfamily
  arXiv:2011.05633 [astro-ph.CO]}}.

\bibitem{Vennin:2020kng}
V.~Vennin, {\em {Stochastic inflation and primordial black holes}}.
\newblock PhD thesis, U. Paris-Saclay, 6, 2020.
\newblock \href{http://arxiv.org/abs/2009.08715}{{\ttfamily arXiv:2009.08715
  [astro-ph.CO]}}.

\bibitem{Inomata:2021uqj}
K.~Inomata, E.~McDonough, and W.~Hu, ``{Primordial black holes arise when the
  inflaton falls},'' \href{http://dx.doi.org/10.1103/PhysRevD.104.123553}{{\em
  Phys. Rev. D} {\bfseries 104} no.~12, (2021) 123553},
  \href{http://arxiv.org/abs/2104.03972}{{\ttfamily arXiv:2104.03972
  [astro-ph.CO]}}.

\bibitem{Ng:2021hll}
K.-W. Ng and Y.-P. Wu, ``{Constant-rate inflation: primordial black holes from
  conformal weight transitions},''
  \href{http://dx.doi.org/10.1007/JHEP11(2021)076}{{\em JHEP} {\bfseries 11}
  (2021) 076}, \href{http://arxiv.org/abs/2102.05620}{{\ttfamily
  arXiv:2102.05620 [astro-ph.CO]}}.

\bibitem{Wang:2021kbh}
Q.~Wang, Y.-C. Liu, B.-Y. Su, and N.~Li, ``{Primordial black holes from the
  perturbations in the inflaton potential in peak theory},''
  \href{http://dx.doi.org/10.1103/PhysRevD.104.083546}{{\em Phys. Rev. D}
  {\bfseries 104} no.~8, (2021) 083546},
  \href{http://arxiv.org/abs/2111.10028}{{\ttfamily arXiv:2111.10028
  [astro-ph.CO]}}.

\bibitem{Kawai:2021edk}
S.~Kawai and J.~Kim, ``{Primordial black holes from Gauss-Bonnet-corrected
  single field inflation},''
  \href{http://dx.doi.org/10.1103/PhysRevD.104.083545}{{\em Phys. Rev. D}
  {\bfseries 104} no.~8, (2021) 083545},
  \href{http://arxiv.org/abs/2108.01340}{{\ttfamily arXiv:2108.01340
  [astro-ph.CO]}}.

\bibitem{Solbi:2021rse}
M.~Solbi and K.~Karami, ``{Primordial black holes formation in the inflationary
  model with field-dependent kinetic term for quartic and natural
  potentials},'' \href{http://dx.doi.org/10.1140/epjc/s10052-021-09690-9}{{\em
  Eur. Phys. J. C} {\bfseries 81} no.~10, (2021) 884},
  \href{http://arxiv.org/abs/2106.02863}{{\ttfamily arXiv:2106.02863
  [astro-ph.CO]}}.

\bibitem{Ballesteros:2021fsp}
G.~Ballesteros, S.~C\'espedes, and L.~Santoni, ``{Large power spectrum and
  primordial black holes in the effective theory of inflation},''
  \href{http://dx.doi.org/10.1007/JHEP01(2022)074}{{\em JHEP} {\bfseries 01}
  (2022) 074}, \href{http://arxiv.org/abs/2109.00567}{{\ttfamily
  arXiv:2109.00567 [hep-th]}}.

\bibitem{Rigopoulos:2021nhv}
G.~Rigopoulos and A.~Wilkins, ``{Inflation is always semi-classical: diffusion
  domination overproduces Primordial Black Holes},''
  \href{http://dx.doi.org/10.1088/1475-7516/2021/12/027}{{\em JCAP} {\bfseries
  12} no.~12, (2021) 027}, \href{http://arxiv.org/abs/2107.05317}{{\ttfamily
  arXiv:2107.05317 [astro-ph.CO]}}.

\bibitem{Animali:2022otk}
C.~Animali and V.~Vennin, ``{Primordial black holes from stochastic
  tunnelling},'' \href{http://arxiv.org/abs/2210.03812}{{\ttfamily
  arXiv:2210.03812 [astro-ph.CO]}}.

\bibitem{Frolovsky:2022ewg}
D.~Frolovsky, S.~V. Ketov, and S.~Saburov, ``{Formation of primordial black
  holes after Starobinsky inflation},''
  \href{http://dx.doi.org/10.1142/S0217732322501358}{{\em Mod. Phys. Lett. A}
  {\bfseries 37} no.~21, (2022) 2250135},
  \href{http://arxiv.org/abs/2205.00603}{{\ttfamily arXiv:2205.00603
  [astro-ph.CO]}}.

\bibitem{Escriva:2022duf}
A.~Escriv\`a, F.~Kuhnel, and Y.~Tada, ``{Primordial Black Holes},''
  \href{http://arxiv.org/abs/2211.05767}{{\ttfamily arXiv:2211.05767
  [astro-ph.CO]}}.

\bibitem{Kristiano:2022maq}
J.~Kristiano and J.~Yokoyama, ``{Ruling Out Primordial Black Hole Formation
  From Single-Field Inflation},''
  \href{http://arxiv.org/abs/2211.03395}{{\ttfamily arXiv:2211.03395
  [hep-th]}}.

\bibitem{Kristiano:2023scm}
J.~Kristiano and J.~Yokoyama, ``{Response to criticism on ''Ruling Out
  Primordial Black Hole Formation From Single-Field Inflation'': A note on
  bispectrum and one-loop correction in single-field inflation with primordial
  black hole formation},'' \href{http://arxiv.org/abs/2303.00341}{{\ttfamily
  arXiv:2303.00341 [hep-th]}}.

\bibitem{Karam:2022nym}
A.~Karam, N.~Koivunen, E.~Tomberg, V.~Vaskonen, and H.~Veerm\"ae, ``{Anatomy of
  single-field inflationary models for primordial black holes},''
  \href{http://arxiv.org/abs/2205.13540}{{\ttfamily arXiv:2205.13540
  [astro-ph.CO]}}.

\bibitem{Riotto:2023hoz}
A.~Riotto, ``{The Primordial Black Hole Formation from Single-Field Inflation
  is Not Ruled Out},'' \href{http://arxiv.org/abs/2301.00599}{{\ttfamily
  arXiv:2301.00599 [astro-ph.CO]}}.

\bibitem{Riotto:2023gpm}
A.~Riotto, ``{The Primordial Black Hole Formation from Single-Field Inflation
  is Still Not Ruled Out},'' \href{http://arxiv.org/abs/2303.01727}{{\ttfamily
  arXiv:2303.01727 [astro-ph.CO]}}.

\bibitem{Ozsoy:2023ryl}
O.~\"Ozsoy and G.~Tasinato, ``{Inflation and Primordial Black Holes},''
  \href{http://arxiv.org/abs/2301.03600}{{\ttfamily arXiv:2301.03600
  [astro-ph.CO]}}.

\bibitem{Choudhury:2023vuj}
S.~Choudhury, M.~R. Gangopadhyay, and M.~Sami, ``{No-go for the formation of
  heavy mass Primordial Black Holes in Single Field Inflation},''
  \href{http://arxiv.org/abs/2301.10000}{{\ttfamily arXiv:2301.10000
  [astro-ph.CO]}}.

\bibitem{Choudhury:2023jlt}
S.~Choudhury, S.~Panda, and M.~Sami, ``{PBH formation in EFT of single field
  inflation with sharp transition},''
  \href{http://dx.doi.org/10.1016/j.physletb.2023.138123}{{\em Phys. Lett. B}
  {\bfseries 845} (2023) 138123},
  \href{http://arxiv.org/abs/2302.05655}{{\ttfamily arXiv:2302.05655
  [astro-ph.CO]}}.

\bibitem{Choudhury:2023rks}
S.~Choudhury, S.~Panda, and M.~Sami, ``{Quantum loop effects on the power
  spectrum and constraints on primordial black holes},''
  \href{http://dx.doi.org/10.1088/1475-7516/2023/11/066}{{\em JCAP} {\bfseries
  11} (2023) 066}, \href{http://arxiv.org/abs/2303.06066}{{\ttfamily
  arXiv:2303.06066 [astro-ph.CO]}}.

\bibitem{Choudhury:2023hvf}
S.~Choudhury, S.~Panda, and M.~Sami, ``{Galileon inflation evades the no-go for
  PBH formation in the single-field framework},''
  \href{http://dx.doi.org/10.1088/1475-7516/2023/08/078}{{\em JCAP} {\bfseries
  08} (2023) 078}, \href{http://arxiv.org/abs/2304.04065}{{\ttfamily
  arXiv:2304.04065 [astro-ph.CO]}}.

\bibitem{Choudhury:2023kdb}
S.~Choudhury, A.~Karde, S.~Panda, and M.~Sami, ``{Primordial non-Gaussianity
  from ultra slow-roll Galileon inflation},''
  \href{http://dx.doi.org/10.1088/1475-7516/2024/01/012}{{\em JCAP} {\bfseries
  01} (2024) 012}, \href{http://arxiv.org/abs/2306.12334}{{\ttfamily
  arXiv:2306.12334 [astro-ph.CO]}}.

\bibitem{Choudhury:2024ybk}
S.~Choudhury, ``{Large fluctuations in the Sky},''
  \href{http://arxiv.org/abs/2403.07343}{{\ttfamily arXiv:2403.07343
  [astro-ph.CO]}}.

\bibitem{Choudhury:2024jlz}
S.~Choudhury, A.~Karde, P.~Padiyar, and M.~Sami, ``{Primordial Black Holes from
  Effective Field Theory of Stochastic Single Field Inflation at NNNLO},''
  \href{http://arxiv.org/abs/2403.13484}{{\ttfamily arXiv:2403.13484
  [astro-ph.CO]}}.

\bibitem{Choudhury:2024dei}
S.~Choudhury, A.~Karde, S.~Panda, and S.~SenGupta,
  ``{Regularized-Renormalized-Resummed loop corrected power spectrum of
  non-singular bounce with Primordial Black Hole formation},''
  \href{http://arxiv.org/abs/2405.06882}{{\ttfamily arXiv:2405.06882
  [astro-ph.CO]}}.

\bibitem{Choudhury:2024aji}
S.~Choudhury and M.~Sami, ``{Large fluctuations and Primordial Black Holes},''
  \href{http://arxiv.org/abs/2407.17006}{{\ttfamily arXiv:2407.17006 [gr-qc]}}.

\bibitem{Choudhury:2024dzw}
S.~Choudhury, S.~Ganguly, S.~Panda, S.~SenGupta, and P.~Tiwari, ``{Obviating
  PBH overproduction for SIGWs generated by Pulsar Timing Arrays in loop
  corrected EFT of bounce},''
  \href{http://dx.doi.org/10.1088/1475-7516/2024/09/013}{{\em JCAP} {\bfseries
  09} (2024) 013}, \href{http://arxiv.org/abs/2407.18976}{{\ttfamily
  arXiv:2407.18976 [astro-ph.CO]}}.

\bibitem{Banerjee:2021lqu}
S.~Banerjee, S.~Choudhury, S.~Chowdhury, J.~Knaute, S.~Panda, and K.~Shirish,
  ``{Thermalization in quenched open quantum cosmology},''
  \href{http://dx.doi.org/10.1016/j.nuclphysb.2023.116368}{{\em Nucl. Phys. B}
  {\bfseries 996} (2023) 116368},
  \href{http://arxiv.org/abs/2104.10692}{{\ttfamily arXiv:2104.10692
  [hep-th]}}.

\bibitem{Firouzjahi:2023ahg}
H.~Firouzjahi and A.~Riotto, ``{Primordial Black Holes and Loops in
  Single-Field Inflation},'' \href{http://arxiv.org/abs/2304.07801}{{\ttfamily
  arXiv:2304.07801 [astro-ph.CO]}}.

\bibitem{Firouzjahi:2023aum}
H.~Firouzjahi, ``{One-loop Corrections in Power Spectrum in Single Field
  Inflation},'' \href{http://arxiv.org/abs/2303.12025}{{\ttfamily
  arXiv:2303.12025 [astro-ph.CO]}}.

\bibitem{Franciolini:2023lgy}
G.~Franciolini, A.~Iovino, Junior., M.~Taoso, and A.~Urbano, ``{One loop to
  rule them all: Perturbativity in the presence of ultra slow-roll dynamics},''
  \href{http://arxiv.org/abs/2305.03491}{{\ttfamily arXiv:2305.03491
  [astro-ph.CO]}}.

\bibitem{Tasinato:2023ukp}
G.~Tasinato, ``{A large $|\eta|$ approach to single field inflation},''
  \href{http://arxiv.org/abs/2305.11568}{{\ttfamily arXiv:2305.11568
  [hep-th]}}.

\bibitem{Motohashi:2023syh}
H.~Motohashi and Y.~Tada, ``{Squeezed bispectrum and one-loop corrections in
  transient constant-roll inflation},''
  \href{http://arxiv.org/abs/2303.16035}{{\ttfamily arXiv:2303.16035
  [astro-ph.CO]}}.

\bibitem{Afshordi:2003zb}
N.~Afshordi, P.~McDonald, and D.~N. Spergel, ``{Primordial black holes as dark
  matter: The Power spectrum and evaporation of early structures},''
  \href{http://dx.doi.org/10.1086/378763}{{\em Astrophys. J. Lett.} {\bfseries
  594} (2003) L71--L74},
  \href{http://arxiv.org/abs/astro-ph/0302035}{{\ttfamily
  arXiv:astro-ph/0302035}}.

\bibitem{Frampton:2010sw}
P.~H. Frampton, M.~Kawasaki, F.~Takahashi, and T.~T. Yanagida, ``{Primordial
  Black Holes as All Dark Matter},''
  \href{http://dx.doi.org/10.1088/1475-7516/2010/04/023}{{\em JCAP} {\bfseries
  04} (2010) 023}, \href{http://arxiv.org/abs/1001.2308}{{\ttfamily
  arXiv:1001.2308 [hep-ph]}}.

\bibitem{Carr:2016drx}
B.~Carr, F.~Kuhnel, and M.~Sandstad, ``{Primordial Black Holes as Dark
  Matter},'' \href{http://dx.doi.org/10.1103/PhysRevD.94.083504}{{\em Phys.
  Rev. D} {\bfseries 94} no.~8, (2016) 083504},
  \href{http://arxiv.org/abs/1607.06077}{{\ttfamily arXiv:1607.06077
  [astro-ph.CO]}}.

\bibitem{Kawasaki:2016pql}
M.~Kawasaki, A.~Kusenko, Y.~Tada, and T.~T. Yanagida, ``{Primordial black holes
  as dark matter in supergravity inflation models},''
  \href{http://dx.doi.org/10.1103/PhysRevD.94.083523}{{\em Phys. Rev. D}
  {\bfseries 94} no.~8, (2016) 083523},
  \href{http://arxiv.org/abs/1606.07631}{{\ttfamily arXiv:1606.07631
  [astro-ph.CO]}}.

\bibitem{Inomata:2017okj}
K.~Inomata, M.~Kawasaki, K.~Mukaida, Y.~Tada, and T.~T. Yanagida,
  ``{Inflationary Primordial Black Holes as All Dark Matter},''
  \href{http://dx.doi.org/10.1103/PhysRevD.96.043504}{{\em Phys. Rev. D}
  {\bfseries 96} no.~4, (2017) 043504},
  \href{http://arxiv.org/abs/1701.02544}{{\ttfamily arXiv:1701.02544
  [astro-ph.CO]}}.

\bibitem{Espinosa:2017sgp}
J.~R. Espinosa, D.~Racco, and A.~Riotto, ``{Cosmological Signature of the
  Standard Model Higgs Vacuum Instability: Primordial Black Holes as Dark
  Matter},'' \href{http://dx.doi.org/10.1103/PhysRevLett.120.121301}{{\em Phys.
  Rev. Lett.} {\bfseries 120} no.~12, (2018) 121301},
  \href{http://arxiv.org/abs/1710.11196}{{\ttfamily arXiv:1710.11196
  [hep-ph]}}.

\bibitem{Ballesteros:2017fsr}
G.~Ballesteros and M.~Taoso, ``{Primordial black hole dark matter from single
  field inflation},'' \href{http://dx.doi.org/10.1103/PhysRevD.97.023501}{{\em
  Phys. Rev. D} {\bfseries 97} no.~2, (2018) 023501},
  \href{http://arxiv.org/abs/1709.05565}{{\ttfamily arXiv:1709.05565
  [hep-ph]}}.

\bibitem{Sasaki:2018dmp}
M.~Sasaki, T.~Suyama, T.~Tanaka, and S.~Yokoyama, ``{Primordial black
  holes\textemdash{}perspectives in gravitational wave astronomy},''
  \href{http://dx.doi.org/10.1088/1361-6382/aaa7b4}{{\em Class. Quant. Grav.}
  {\bfseries 35} no.~6, (2018) 063001},
  \href{http://arxiv.org/abs/1801.05235}{{\ttfamily arXiv:1801.05235
  [astro-ph.CO]}}.

\bibitem{Ballesteros:2019hus}
G.~Ballesteros, J.~Rey, and F.~Rompineve, ``{Detuning primordial black hole
  dark matter with early matter domination and axion monodromy},''
  \href{http://dx.doi.org/10.1088/1475-7516/2020/06/014}{{\em JCAP} {\bfseries
  06} (2020) 014}, \href{http://arxiv.org/abs/1912.01638}{{\ttfamily
  arXiv:1912.01638 [astro-ph.CO]}}.

\bibitem{Dalianis:2019asr}
I.~Dalianis and G.~Tringas, ``{Primordial black hole remnants as dark matter
  produced in thermal, matter, and runaway-quintessence postinflationary
  scenarios},'' \href{http://dx.doi.org/10.1103/PhysRevD.100.083512}{{\em Phys.
  Rev. D} {\bfseries 100} no.~8, (2019) 083512},
  \href{http://arxiv.org/abs/1905.01741}{{\ttfamily arXiv:1905.01741
  [astro-ph.CO]}}.

\bibitem{Cheong:2019vzl}
D.~Y. Cheong, S.~M. Lee, and S.~C. Park, ``{Primordial black holes in
  Higgs-$R^2$ inflation as the whole of dark matter},''
  \href{http://dx.doi.org/10.1088/1475-7516/2021/01/032}{{\em JCAP} {\bfseries
  01} (2021) 032}, \href{http://arxiv.org/abs/1912.12032}{{\ttfamily
  arXiv:1912.12032 [hep-ph]}}.

\bibitem{Green:2020jor}
A.~M. Green and B.~J. Kavanagh, ``{Primordial Black Holes as a dark matter
  candidate},'' \href{http://dx.doi.org/10.1088/1361-6471/abc534}{{\em J. Phys.
  G} {\bfseries 48} no.~4, (2021) 043001},
  \href{http://arxiv.org/abs/2007.10722}{{\ttfamily arXiv:2007.10722
  [astro-ph.CO]}}.

\bibitem{Carr:2020xqk}
B.~Carr and F.~Kuhnel, ``{Primordial Black Holes as Dark Matter: Recent
  Developments},''
  \href{http://dx.doi.org/10.1146/annurev-nucl-050520-125911}{{\em Ann. Rev.
  Nucl. Part. Sci.} {\bfseries 70} (2020) 355--394},
  \href{http://arxiv.org/abs/2006.02838}{{\ttfamily arXiv:2006.02838
  [astro-ph.CO]}}.

\bibitem{Ballesteros:2020qam}
G.~Ballesteros, J.~Rey, M.~Taoso, and A.~Urbano, ``{Primordial black holes as
  dark matter and gravitational waves from single-field polynomial
  inflation},'' \href{http://dx.doi.org/10.1088/1475-7516/2020/07/025}{{\em
  JCAP} {\bfseries 07} (2020) 025},
  \href{http://arxiv.org/abs/2001.08220}{{\ttfamily arXiv:2001.08220
  [astro-ph.CO]}}.

\bibitem{Carr:2020gox}
B.~Carr, K.~Kohri, Y.~Sendouda, and J.~Yokoyama, ``{Constraints on primordial
  black holes},'' \href{http://dx.doi.org/10.1088/1361-6633/ac1e31}{{\em Rept.
  Prog. Phys.} {\bfseries 84} no.~11, (2021) 116902},
  \href{http://arxiv.org/abs/2002.12778}{{\ttfamily arXiv:2002.12778
  [astro-ph.CO]}}.

\bibitem{Ozsoy:2020kat}
O.~\"Ozsoy and Z.~Lalak, ``{Primordial black holes as dark matter and
  gravitational waves from bumpy axion inflation},''
  \href{http://dx.doi.org/10.1088/1475-7516/2021/01/040}{{\em JCAP} {\bfseries
  01} (2021) 040}, \href{http://arxiv.org/abs/2008.07549}{{\ttfamily
  arXiv:2008.07549 [astro-ph.CO]}}.

\bibitem{Saito:2009jt}
R.~Saito and J.~Yokoyama, ``{Gravitational-Wave Constraints on the Abundance of
  Primordial Black Holes},'' \href{http://dx.doi.org/10.1143/PTP.126.351}{{\em
  Prog. Theor. Phys.} {\bfseries 123} (2010) 867--886},
  \href{http://arxiv.org/abs/0912.5317}{{\ttfamily arXiv:0912.5317
  [astro-ph.CO]}}. [Erratum: Prog.Theor.Phys. 126, 351--352 (2011)].

\bibitem{Sasaki:2016jop}
M.~Sasaki, T.~Suyama, T.~Tanaka, and S.~Yokoyama, ``{Primordial Black Hole
  Scenario for the Gravitational-Wave Event GW150914},''
  \href{http://dx.doi.org/10.1103/PhysRevLett.117.061101}{{\em Phys. Rev.
  Lett.} {\bfseries 117} no.~6, (2016) 061101},
  \href{http://arxiv.org/abs/1603.08338}{{\ttfamily arXiv:1603.08338
  [astro-ph.CO]}}. [Erratum: Phys.Rev.Lett. 121, 059901 (2018)].

\bibitem{Raidal:2017mfl}
M.~Raidal, V.~Vaskonen, and H.~Veerm\"ae, ``{Gravitational Waves from
  Primordial Black Hole Mergers},''
  \href{http://dx.doi.org/10.1088/1475-7516/2017/09/037}{{\em JCAP} {\bfseries
  09} (2017) 037}, \href{http://arxiv.org/abs/1707.01480}{{\ttfamily
  arXiv:1707.01480 [astro-ph.CO]}}.

\bibitem{Ali-Haimoud:2017rtz}
Y.~Ali-Ha\"\i{}moud, E.~D. Kovetz, and M.~Kamionkowski, ``{Merger rate of
  primordial black-hole binaries},''
  \href{http://dx.doi.org/10.1103/PhysRevD.96.123523}{{\em Phys. Rev. D}
  {\bfseries 96} no.~12, (2017) 123523},
  \href{http://arxiv.org/abs/1709.06576}{{\ttfamily arXiv:1709.06576
  [astro-ph.CO]}}.

\bibitem{Di:2017ndc}
H.~Di and Y.~Gong, ``{Primordial black holes and second order gravitational
  waves from ultra-slow-roll inflation},''
  \href{http://dx.doi.org/10.1088/1475-7516/2018/07/007}{{\em JCAP} {\bfseries
  07} (2018) 007}, \href{http://arxiv.org/abs/1707.09578}{{\ttfamily
  arXiv:1707.09578 [astro-ph.CO]}}.

\bibitem{Cheng:2018yyr}
S.-L. Cheng, W.~Lee, and K.-W. Ng, ``{Primordial black holes and associated
  gravitational waves in axion monodromy inflation},''
  \href{http://dx.doi.org/10.1088/1475-7516/2018/07/001}{{\em JCAP} {\bfseries
  07} (2018) 001}, \href{http://arxiv.org/abs/1801.09050}{{\ttfamily
  arXiv:1801.09050 [astro-ph.CO]}}.

\bibitem{Vaskonen:2019jpv}
V.~Vaskonen and H.~Veerm\"ae, ``{Lower bound on the primordial black hole
  merger rate},'' \href{http://dx.doi.org/10.1103/PhysRevD.101.043015}{{\em
  Phys. Rev. D} {\bfseries 101} no.~4, (2020) 043015},
  \href{http://arxiv.org/abs/1908.09752}{{\ttfamily arXiv:1908.09752
  [astro-ph.CO]}}.

\bibitem{Drees:2019xpp}
M.~Drees and Y.~Xu, ``{Overshooting, Critical Higgs Inflation and Second Order
  Gravitational Wave Signatures},''
  \href{http://dx.doi.org/10.1140/epjc/s10052-021-08976-2}{{\em Eur. Phys. J.
  C} {\bfseries 81} no.~2, (2021) 182},
  \href{http://arxiv.org/abs/1905.13581}{{\ttfamily arXiv:1905.13581
  [hep-ph]}}.

\bibitem{Hall:2020daa}
A.~Hall, A.~D. Gow, and C.~T. Byrnes, ``{Bayesian analysis of LIGO-Virgo
  mergers: Primordial vs. astrophysical black hole populations},''
  \href{http://dx.doi.org/10.1103/PhysRevD.102.123524}{{\em Phys. Rev. D}
  {\bfseries 102} (2020) 123524},
  \href{http://arxiv.org/abs/2008.13704}{{\ttfamily arXiv:2008.13704
  [astro-ph.CO]}}.

\bibitem{Ragavendra:2020sop}
H.~V. Ragavendra, P.~Saha, L.~Sriramkumar, and J.~Silk, ``{Primordial black
  holes and secondary gravitational waves from ultraslow roll and punctuated
  inflation},'' \href{http://dx.doi.org/10.1103/PhysRevD.103.083510}{{\em Phys.
  Rev. D} {\bfseries 103} no.~8, (2021) 083510},
  \href{http://arxiv.org/abs/2008.12202}{{\ttfamily arXiv:2008.12202
  [astro-ph.CO]}}.

\bibitem{Ashoorioon:2020hln}
A.~Ashoorioon, A.~Rostami, and J.~T. Firouzjaee, ``{Examining the end of
  inflation with primordial black holes mass distribution and gravitational
  waves},'' \href{http://dx.doi.org/10.1103/PhysRevD.103.123512}{{\em Phys.
  Rev. D} {\bfseries 103} (2021) 123512},
  \href{http://arxiv.org/abs/2012.02817}{{\ttfamily arXiv:2012.02817
  [astro-ph.CO]}}.

\bibitem{Ragavendra:2020vud}
H.~V. Ragavendra, L.~Sriramkumar, and J.~Silk, ``{Could PBHs and secondary GWs
  have originated from squeezed initial states?},''
  \href{http://dx.doi.org/10.1088/1475-7516/2021/05/010}{{\em JCAP} {\bfseries
  05} (2021) 010}, \href{http://arxiv.org/abs/2011.09938}{{\ttfamily
  arXiv:2011.09938 [astro-ph.CO]}}.

\bibitem{Papanikolaou:2020qtd}
T.~Papanikolaou, V.~Vennin, and D.~Langlois, ``{Gravitational waves from a
  universe filled with primordial black holes},''
  \href{http://dx.doi.org/10.1088/1475-7516/2021/03/053}{{\em JCAP} {\bfseries
  03} (2021) 053}, \href{http://arxiv.org/abs/2010.11573}{{\ttfamily
  arXiv:2010.11573 [astro-ph.CO]}}.

\bibitem{Teimoori:2021pte}
Z.~Teimoori, K.~Rezazadeh, M.~A. Rasheed, and K.~Karami, ``{Mechanism of
  primordial black holes production and secondary gravitational waves in
  $\alpha$-attractor Galileon inflationary scenario},''
  \href{http://arxiv.org/abs/2107.07620}{{\ttfamily arXiv:2107.07620
  [astro-ph.CO]}}.

\bibitem{Cicoli:2022sih}
M.~Cicoli, F.~G. Pedro, and N.~Pedron, ``{Secondary GWs and PBHs in string
  inflation: formation and detectability},''
  \href{http://dx.doi.org/10.1088/1475-7516/2022/08/030}{{\em JCAP} {\bfseries
  08} no.~08, (2022) 030}, \href{http://arxiv.org/abs/2203.00021}{{\ttfamily
  arXiv:2203.00021 [hep-th]}}.

\bibitem{Ashoorioon:2022raz}
A.~Ashoorioon, K.~Rezazadeh, and A.~Rostami, ``{NANOGrav signal from the end of
  inflation and the LIGO mass and heavier primordial black holes},''
  \href{http://dx.doi.org/10.1016/j.physletb.2022.137542}{{\em Phys. Lett. B}
  {\bfseries 835} (2022) 137542},
  \href{http://arxiv.org/abs/2202.01131}{{\ttfamily arXiv:2202.01131
  [astro-ph.CO]}}.

\bibitem{Papanikolaou:2022chm}
T.~Papanikolaou, ``{Gravitational waves induced from primordial black hole
  fluctuations: the~effect of an extended mass function},''
  \href{http://dx.doi.org/10.1088/1475-7516/2022/10/089}{{\em JCAP} {\bfseries
  10} (2022) 089}, \href{http://arxiv.org/abs/2207.11041}{{\ttfamily
  arXiv:2207.11041 [astro-ph.CO]}}.

\bibitem{Papanikolaou:2023crz}
T.~Papanikolaou, ``{Primordial black holes in loop quantum cosmology: the
  effect on the threshold},''
  \href{http://dx.doi.org/10.1088/1361-6382/acd97d}{{\em Class. Quant. Grav.}
  {\bfseries 40} no.~13, (2023) 134001},
  \href{http://arxiv.org/abs/2301.11439}{{\ttfamily arXiv:2301.11439 [gr-qc]}}.

\bibitem{Papanikolaou:2022did}
T.~Papanikolaou, A.~Lymperis, S.~Lola, and E.~N. Saridakis, ``{Primordial black
  holes and gravitational waves from non-canonical inflation},''
  \href{http://dx.doi.org/10.1088/1475-7516/2023/03/003}{{\em JCAP} {\bfseries
  03} (2023) 003}, \href{http://arxiv.org/abs/2211.14900}{{\ttfamily
  arXiv:2211.14900 [astro-ph.CO]}}.

\bibitem{Wang:2022nml}
X.~Wang, Y.-l. Zhang, R.~Kimura, and M.~Yamaguchi, ``{Reconstruction of Power
  Spectrum of Primordial Curvature Perturbations on small scales from
  Primordial Black Hole Binaries scenario of LIGO/VIRGO detection},''
  \href{http://arxiv.org/abs/2209.12911}{{\ttfamily arXiv:2209.12911
  [astro-ph.CO]}}.

\bibitem{Ahmed:2021ucx}
W.~Ahmed, M.~Junaid, and U.~Zubair, ``{Primordial black holes and gravitational
  waves in hybrid inflation with chaotic potentials},''
  \href{http://dx.doi.org/10.1016/j.nuclphysb.2022.115968}{{\em Nucl. Phys. B}
  {\bfseries 984} (2022) 115968},
  \href{http://arxiv.org/abs/2109.14838}{{\ttfamily arXiv:2109.14838
  [astro-ph.CO]}}.

\bibitem{Yi:2023npi}
Z.~Yi, Z.-Q. You, Y.~Wu, Z.-C. Chen, and L.~Liu, ``{Exploring the NANOGrav
  Signal and Planet-mass Primordial Black Holes through Higgs Inflation},''
  \href{http://arxiv.org/abs/2308.14688}{{\ttfamily arXiv:2308.14688
  [astro-ph.CO]}}.

\bibitem{Yuan:2021qgz}
C.~Yuan and Q.-G. Huang, ``{A topic review on probing primordial black hole
  dark matter with scalar induced gravitational waves},''
  \href{http://arxiv.org/abs/2103.04739}{{\ttfamily arXiv:2103.04739
  [astro-ph.GA]}}.

\bibitem{Aghaie:2023lan}
M.~Aghaie, G.~Armando, A.~Dondarini, and P.~Panci, ``{Bounds on Ultralight Dark
  Matter from NANOGrav},'' \href{http://arxiv.org/abs/2308.04590}{{\ttfamily
  arXiv:2308.04590 [astro-ph.CO]}}.

\bibitem{Burrage:2010cu}
C.~Burrage, C.~de~Rham, D.~Seery, and A.~J. Tolley, ``{Galileon inflation},''
  \href{http://dx.doi.org/10.1088/1475-7516/2011/01/014}{{\em JCAP} {\bfseries
  01} (2011) 014}, \href{http://arxiv.org/abs/1009.2497}{{\ttfamily
  arXiv:1009.2497 [hep-th]}}.

\bibitem{Jain:2010ka}
B.~Jain and J.~Khoury, ``{Cosmological Tests of Gravity},''
  \href{http://dx.doi.org/10.1016/j.aop.2010.04.002}{{\em Annals Phys.}
  {\bfseries 325} (2010) 1479--1516},
  \href{http://arxiv.org/abs/1004.3294}{{\ttfamily arXiv:1004.3294
  [astro-ph.CO]}}.

\bibitem{Gannouji:2010au}
R.~Gannouji and M.~Sami, ``{Galileon gravity and its relevance to late time
  cosmic acceleration},''
  \href{http://dx.doi.org/10.1103/PhysRevD.82.024011}{{\em Phys. Rev. D}
  {\bfseries 82} (2010) 024011},
  \href{http://arxiv.org/abs/1004.2808}{{\ttfamily arXiv:1004.2808 [gr-qc]}}.

\bibitem{Ali:2010gr}
A.~Ali, R.~Gannouji, and M.~Sami, ``{Modified gravity a la Galileon: Late time
  cosmic acceleration and observational constraints},''
  \href{http://dx.doi.org/10.1103/PhysRevD.82.103015}{{\em Phys. Rev. D}
  {\bfseries 82} (2010) 103015},
  \href{http://arxiv.org/abs/1008.1588}{{\ttfamily arXiv:1008.1588
  [astro-ph.CO]}}.

\bibitem{deRham:2011by}
C.~de~Rham and L.~Heisenberg, ``{Cosmology of the Galileon from Massive
  Gravity},'' \href{http://dx.doi.org/10.1103/PhysRevD.84.043503}{{\em Phys.
  Rev. D} {\bfseries 84} (2011) 043503},
  \href{http://arxiv.org/abs/1106.3312}{{\ttfamily arXiv:1106.3312 [hep-th]}}.

\bibitem{Burrage:2010rs}
C.~Burrage and D.~Seery, ``{Revisiting fifth forces in the Galileon model},''
  \href{http://dx.doi.org/10.1088/1475-7516/2010/08/011}{{\em JCAP} {\bfseries
  08} (2010) 011}, \href{http://arxiv.org/abs/1005.1927}{{\ttfamily
  arXiv:1005.1927 [astro-ph.CO]}}.

\bibitem{DeFelice:2010jn}
A.~De~Felice and S.~Tsujikawa, ``{Generalized Brans-Dicke theories},''
  \href{http://dx.doi.org/10.1088/1475-7516/2010/07/024}{{\em JCAP} {\bfseries
  07} (2010) 024}, \href{http://arxiv.org/abs/1005.0868}{{\ttfamily
  arXiv:1005.0868 [astro-ph.CO]}}.

\bibitem{DeFelice:2010gb}
A.~De~Felice, S.~Mukohyama, and S.~Tsujikawa, ``{Density perturbations in
  general modified gravitational theories},''
  \href{http://dx.doi.org/10.1103/PhysRevD.82.023524}{{\em Phys. Rev. D}
  {\bfseries 82} (2010) 023524},
  \href{http://arxiv.org/abs/1006.0281}{{\ttfamily arXiv:1006.0281
  [astro-ph.CO]}}.

\bibitem{Babichev:2010jd}
E.~Babichev, C.~Deffayet, and R.~Ziour, ``{The Recovery of General Relativity
  in massive gravity via the Vainshtein mechanism},''
  \href{http://dx.doi.org/10.1103/PhysRevD.82.104008}{{\em Phys. Rev. D}
  {\bfseries 82} (2010) 104008},
  \href{http://arxiv.org/abs/1007.4506}{{\ttfamily arXiv:1007.4506 [gr-qc]}}.

\bibitem{DeFelice:2010pv}
A.~De~Felice and S.~Tsujikawa, ``{Cosmology of a covariant Galileon field},''
  \href{http://dx.doi.org/10.1103/PhysRevLett.105.111301}{{\em Phys. Rev.
  Lett.} {\bfseries 105} (2010) 111301},
  \href{http://arxiv.org/abs/1007.2700}{{\ttfamily arXiv:1007.2700
  [astro-ph.CO]}}.

\bibitem{DeFelice:2010nf}
A.~De~Felice and S.~Tsujikawa, ``{Generalized Galileon cosmology},''
  \href{http://dx.doi.org/10.1103/PhysRevD.84.124029}{{\em Phys. Rev. D}
  {\bfseries 84} (2011) 124029},
  \href{http://arxiv.org/abs/1008.4236}{{\ttfamily arXiv:1008.4236 [hep-th]}}.

\bibitem{Hinterbichler:2010xn}
K.~Hinterbichler, M.~Trodden, and D.~Wesley, ``{Multi-field galileons and
  higher co-dimension branes},''
  \href{http://dx.doi.org/10.1103/PhysRevD.82.124018}{{\em Phys. Rev. D}
  {\bfseries 82} (2010) 124018},
  \href{http://arxiv.org/abs/1008.1305}{{\ttfamily arXiv:1008.1305 [hep-th]}}.

\bibitem{Kobayashi:2010cm}
T.~Kobayashi, M.~Yamaguchi, and J.~Yokoyama, ``{G-inflation: Inflation driven
  by the Galileon field},''
  \href{http://dx.doi.org/10.1103/PhysRevLett.105.231302}{{\em Phys. Rev.
  Lett.} {\bfseries 105} (2010) 231302},
  \href{http://arxiv.org/abs/1008.0603}{{\ttfamily arXiv:1008.0603 [hep-th]}}.

\bibitem{Deffayet:2010qz}
C.~Deffayet, O.~Pujolas, I.~Sawicki, and A.~Vikman, ``{Imperfect Dark Energy
  from Kinetic Gravity Braiding},''
  \href{http://dx.doi.org/10.1088/1475-7516/2010/10/026}{{\em JCAP} {\bfseries
  10} (2010) 026}, \href{http://arxiv.org/abs/1008.0048}{{\ttfamily
  arXiv:1008.0048 [hep-th]}}.

\bibitem{Mizuno:2010ag}
S.~Mizuno and K.~Koyama, ``{Primordial non-Gaussianity from the DBI
  Galileons},'' \href{http://dx.doi.org/10.1103/PhysRevD.82.103518}{{\em Phys.
  Rev. D} {\bfseries 82} (2010) 103518},
  \href{http://arxiv.org/abs/1009.0677}{{\ttfamily arXiv:1009.0677 [hep-th]}}.

\bibitem{Khoury:2010xi}
J.~Khoury, ``{Theories of Dark Energy with Screening Mechanisms},''
  \href{http://arxiv.org/abs/1011.5909}{{\ttfamily arXiv:1011.5909
  [astro-ph.CO]}}.

\bibitem{DeFelice:2010as}
A.~De~Felice, R.~Kase, and S.~Tsujikawa, ``{Matter perturbations in Galileon
  cosmology},'' \href{http://dx.doi.org/10.1103/PhysRevD.83.043515}{{\em Phys.
  Rev. D} {\bfseries 83} (2011) 043515},
  \href{http://arxiv.org/abs/1011.6132}{{\ttfamily arXiv:1011.6132
  [astro-ph.CO]}}.

\bibitem{Kamada:2010qe}
K.~Kamada, T.~Kobayashi, M.~Yamaguchi, and J.~Yokoyama, ``{Higgs
  G-inflation},'' \href{http://dx.doi.org/10.1103/PhysRevD.83.083515}{{\em
  Phys. Rev. D} {\bfseries 83} (2011) 083515},
  \href{http://arxiv.org/abs/1012.4238}{{\ttfamily arXiv:1012.4238
  [astro-ph.CO]}}.

\bibitem{Kobayashi:2011pc}
T.~Kobayashi, M.~Yamaguchi, and J.~Yokoyama, ``{Primordial non-Gaussianity from
  G-inflation},'' \href{http://dx.doi.org/10.1103/PhysRevD.83.103524}{{\em
  Phys. Rev. D} {\bfseries 83} (2011) 103524},
  \href{http://arxiv.org/abs/1103.1740}{{\ttfamily arXiv:1103.1740 [hep-th]}}.

\bibitem{DeFelice:2011zh}
A.~De~Felice and S.~Tsujikawa, ``{Primordial non-Gaussianities in general
  modified gravitational models of inflation},''
  \href{http://dx.doi.org/10.1088/1475-7516/2011/04/029}{{\em JCAP} {\bfseries
  04} (2011) 029}, \href{http://arxiv.org/abs/1103.1172}{{\ttfamily
  arXiv:1103.1172 [astro-ph.CO]}}.

\bibitem{Khoury:2011da}
J.~Khoury, J.-L. Lehners, and B.~A. Ovrut, ``{Supersymmetric Galileons},''
  \href{http://dx.doi.org/10.1103/PhysRevD.84.043521}{{\em Phys. Rev. D}
  {\bfseries 84} (2011) 043521},
  \href{http://arxiv.org/abs/1103.0003}{{\ttfamily arXiv:1103.0003 [hep-th]}}.

\bibitem{Trodden:2011xh}
M.~Trodden and K.~Hinterbichler, ``{Generalizing Galileons},''
  \href{http://dx.doi.org/10.1088/0264-9381/28/20/204003}{{\em Class. Quant.
  Grav.} {\bfseries 28} (2011) 204003},
  \href{http://arxiv.org/abs/1104.2088}{{\ttfamily arXiv:1104.2088 [hep-th]}}.

\bibitem{Burrage:2011bt}
C.~Burrage, C.~de~Rham, and L.~Heisenberg, ``{de Sitter Galileon},''
  \href{http://dx.doi.org/10.1088/1475-7516/2011/05/025}{{\em JCAP} {\bfseries
  05} (2011) 025}, \href{http://arxiv.org/abs/1104.0155}{{\ttfamily
  arXiv:1104.0155 [hep-th]}}.

\bibitem{Kobayashi:2011nu}
T.~Kobayashi, M.~Yamaguchi, and J.~Yokoyama, ``{Generalized G-inflation:
  Inflation with the most general second-order field equations},''
  \href{http://dx.doi.org/10.1143/PTP.126.511}{{\em Prog. Theor. Phys.}
  {\bfseries 126} (2011) 511--529},
  \href{http://arxiv.org/abs/1105.5723}{{\ttfamily arXiv:1105.5723 [hep-th]}}.

\bibitem{PerreaultLevasseur:2011wto}
L.~Perreault~Levasseur, R.~Brandenberger, and A.-C. Davis, ``{Defrosting in an
  Emergent Galileon Cosmology},''
  \href{http://dx.doi.org/10.1103/PhysRevD.84.103512}{{\em Phys. Rev. D}
  {\bfseries 84} (2011) 103512},
  \href{http://arxiv.org/abs/1105.5649}{{\ttfamily arXiv:1105.5649
  [astro-ph.CO]}}.

\bibitem{Brax:2011sv}
P.~Brax, C.~Burrage, and A.-C. Davis, ``{Laboratory Tests of the Galileon},''
  \href{http://dx.doi.org/10.1088/1475-7516/2011/09/020}{{\em JCAP} {\bfseries
  09} (2011) 020}, \href{http://arxiv.org/abs/1106.1573}{{\ttfamily
  arXiv:1106.1573 [hep-ph]}}.

\bibitem{DeFelice:2011uc}
A.~De~Felice and S.~Tsujikawa, ``{Inflationary non-Gaussianities in the most
  general second-order scalar-tensor theories},''
  \href{http://dx.doi.org/10.1103/PhysRevD.84.083504}{{\em Phys. Rev. D}
  {\bfseries 84} (2011) 083504},
  \href{http://arxiv.org/abs/1107.3917}{{\ttfamily arXiv:1107.3917 [gr-qc]}}.

\bibitem{Gao:2011qe}
X.~Gao and D.~A. Steer, ``{Inflation and primordial non-Gaussianities of
  'generalized Galileons'},''
  \href{http://dx.doi.org/10.1088/1475-7516/2011/12/019}{{\em JCAP} {\bfseries
  12} (2011) 019}, \href{http://arxiv.org/abs/1107.2642}{{\ttfamily
  arXiv:1107.2642 [astro-ph.CO]}}.

\bibitem{Babichev:2011iz}
E.~Babichev, C.~Deffayet, and G.~Esposito-Farese, ``{Constraints on
  Shift-Symmetric Scalar-Tensor Theories with a Vainshtein Mechanism from
  Bounds on the Time Variation of G},''
  \href{http://dx.doi.org/10.1103/PhysRevLett.107.251102}{{\em Phys. Rev.
  Lett.} {\bfseries 107} (2011) 251102},
  \href{http://arxiv.org/abs/1107.1569}{{\ttfamily arXiv:1107.1569 [gr-qc]}}.

\bibitem{DeFelice:2011hq}
A.~De~Felice, T.~Kobayashi, and S.~Tsujikawa, ``{Effective gravitational
  couplings for cosmological perturbations in the most general scalar-tensor
  theories with second-order field equations},''
  \href{http://dx.doi.org/10.1016/j.physletb.2011.11.028}{{\em Phys. Lett. B}
  {\bfseries 706} (2011) 123--133},
  \href{http://arxiv.org/abs/1108.4242}{{\ttfamily arXiv:1108.4242 [gr-qc]}}.

\bibitem{Khoury:2011ay}
J.~Khoury, G.~E.~J. Miller, and A.~J. Tolley, ``{Spatially Covariant Theories
  of a Transverse, Traceless Graviton, Part I: Formalism},''
  \href{http://dx.doi.org/10.1103/PhysRevD.85.084002}{{\em Phys. Rev. D}
  {\bfseries 85} (2012) 084002},
  \href{http://arxiv.org/abs/1108.1397}{{\ttfamily arXiv:1108.1397 [hep-th]}}.

\bibitem{Qiu:2011cy}
T.~Qiu, J.~Evslin, Y.-F. Cai, M.~Li, and X.~Zhang, ``{Bouncing Galileon
  Cosmologies},'' \href{http://dx.doi.org/10.1088/1475-7516/2011/10/036}{{\em
  JCAP} {\bfseries 10} (2011) 036},
  \href{http://arxiv.org/abs/1108.0593}{{\ttfamily arXiv:1108.0593 [hep-th]}}.

\bibitem{Renaux-Petel:2011rmu}
S.~Renaux-Petel, S.~Mizuno, and K.~Koyama, ``{Primordial fluctuations and
  non-Gaussianities from multifield DBI Galileon inflation},''
  \href{http://dx.doi.org/10.1088/1475-7516/2011/11/042}{{\em JCAP} {\bfseries
  11} (2011) 042}, \href{http://arxiv.org/abs/1108.0305}{{\ttfamily
  arXiv:1108.0305 [astro-ph.CO]}}.

\bibitem{DeFelice:2011bh}
A.~De~Felice and S.~Tsujikawa, ``{Conditions for the cosmological viability of
  the most general scalar-tensor theories and their applications to extended
  Galileon dark energy models},''
  \href{http://dx.doi.org/10.1088/1475-7516/2012/02/007}{{\em JCAP} {\bfseries
  02} (2012) 007}, \href{http://arxiv.org/abs/1110.3878}{{\ttfamily
  arXiv:1110.3878 [gr-qc]}}.

\bibitem{DeFelice:2011th}
A.~De~Felice, R.~Kase, and S.~Tsujikawa, ``{Vainshtein mechanism in
  second-order scalar-tensor theories},''
  \href{http://dx.doi.org/10.1103/PhysRevD.85.044059}{{\em Phys. Rev. D}
  {\bfseries 85} (2012) 044059},
  \href{http://arxiv.org/abs/1111.5090}{{\ttfamily arXiv:1111.5090 [gr-qc]}}.

\bibitem{DeFelice:2011aa}
A.~De~Felice and S.~Tsujikawa, ``{Cosmological constraints on extended Galileon
  models},'' \href{http://dx.doi.org/10.1088/1475-7516/2012/03/025}{{\em JCAP}
  {\bfseries 03} (2012) 025}, \href{http://arxiv.org/abs/1112.1774}{{\ttfamily
  arXiv:1112.1774 [astro-ph.CO]}}.

\bibitem{Zhou:2011ix}
S.-Y. Zhou and E.~J. Copeland, ``{Galileons with Gauge Symmetries},''
  \href{http://dx.doi.org/10.1103/PhysRevD.85.065002}{{\em Phys. Rev. D}
  {\bfseries 85} (2012) 065002},
  \href{http://arxiv.org/abs/1112.0968}{{\ttfamily arXiv:1112.0968 [hep-th]}}.

\bibitem{Goon:2012mu}
G.~Goon, K.~Hinterbichler, A.~Joyce, and M.~Trodden, ``{Gauged Galileons From
  Branes},'' \href{http://dx.doi.org/10.1016/j.physletb.2012.06.065}{{\em Phys.
  Lett. B} {\bfseries 714} (2012) 115--119},
  \href{http://arxiv.org/abs/1201.0015}{{\ttfamily arXiv:1201.0015 [hep-th]}}.

\bibitem{Shirai:2012iw}
N.~Shirai, K.~Bamba, S.~Kumekawa, J.~Matsumoto, and S.~Nojiri, ``{Generalized
  Galileon Model: Cosmological reconstruction and the Vainshtein mechanism},''
  \href{http://dx.doi.org/10.1103/PhysRevD.86.043006}{{\em Phys. Rev. D}
  {\bfseries 86} (2012) 043006},
  \href{http://arxiv.org/abs/1203.4962}{{\ttfamily arXiv:1203.4962 [hep-th]}}.

\bibitem{Goon:2012dy}
G.~Goon, K.~Hinterbichler, A.~Joyce, and M.~Trodden, ``{Galileons as
  Wess-Zumino Terms},'' \href{http://dx.doi.org/10.1007/JHEP06(2012)004}{{\em
  JHEP} {\bfseries 06} (2012) 004},
  \href{http://arxiv.org/abs/1203.3191}{{\ttfamily arXiv:1203.3191 [hep-th]}}.

\bibitem{deRham:2012az}
C.~de~Rham, ``{Galileons in the Sky},''
  \href{http://dx.doi.org/10.1016/j.crhy.2012.04.006}{{\em Comptes Rendus
  Physique} {\bfseries 13} (2012) 666--681},
  \href{http://arxiv.org/abs/1204.5492}{{\ttfamily arXiv:1204.5492
  [astro-ph.CO]}}.

\bibitem{Ali:2012cv}
A.~Ali, R.~Gannouji, M.~W. Hossain, and M.~Sami, ``{Light mass galileons:
  Cosmological dynamics, mass screening and observational constraints},''
  \href{http://dx.doi.org/10.1016/j.physletb.2012.10.009}{{\em Phys. Lett. B}
  {\bfseries 718} (2012) 5--14},
  \href{http://arxiv.org/abs/1207.3959}{{\ttfamily arXiv:1207.3959 [gr-qc]}}.

\bibitem{Liu:2012ww}
Z.-G. Liu and Y.-S. Piao, ``{A Galileon Design of Slow Expansion: Emergent
  universe},'' \href{http://dx.doi.org/10.1016/j.physletb.2012.11.068}{{\em
  Phys. Lett. B} {\bfseries 718} (2013) 734--739},
  \href{http://arxiv.org/abs/1207.2568}{{\ttfamily arXiv:1207.2568 [gr-qc]}}.

\bibitem{Choudhury:2012yh}
S.~Choudhury and S.~Pal, ``{DBI Galileon inflation in background SUGRA},''
  \href{http://dx.doi.org/10.1016/j.nuclphysb.2013.05.010}{{\em Nucl. Phys. B}
  {\bfseries 874} (2013) 85--114},
  \href{http://arxiv.org/abs/1208.4433}{{\ttfamily arXiv:1208.4433 [hep-th]}}.

\bibitem{Choudhury:2012whm}
S.~Choudhury and S.~Pal, ``{Primordial non-Gaussian features from DBI Galileon
  inflation},'' \href{http://dx.doi.org/10.1140/epjc/s10052-015-3452-3}{{\em
  Eur. Phys. J. C} {\bfseries 75} no.~6, (2015) 241},
  \href{http://arxiv.org/abs/1210.4478}{{\ttfamily arXiv:1210.4478 [hep-th]}}.

\bibitem{Barreira:2012kk}
A.~Barreira, B.~Li, C.~M. Baugh, and S.~Pascoli, ``{Linear perturbations in
  Galileon gravity models},''
  \href{http://dx.doi.org/10.1103/PhysRevD.86.124016}{{\em Phys. Rev. D}
  {\bfseries 86} (2012) 124016},
  \href{http://arxiv.org/abs/1208.0600}{{\ttfamily arXiv:1208.0600
  [astro-ph.CO]}}.

\bibitem{deFromont:2013iwa}
P.~de~Fromont, C.~de~Rham, L.~Heisenberg, and A.~Matas, ``{Superluminality in
  the Bi- and Multi- Galileon},''
  \href{http://dx.doi.org/10.1007/JHEP07(2013)067}{{\em JHEP} {\bfseries 07}
  (2013) 067}, \href{http://arxiv.org/abs/1303.0274}{{\ttfamily arXiv:1303.0274
  [hep-th]}}.

\bibitem{Arroja:2013dya}
F.~Arroja, N.~Bartolo, E.~Dimastrogiovanni, and M.~Fasiello, ``{On the
  Trispectrum of Galileon Inflation},''
  \href{http://dx.doi.org/10.1088/1475-7516/2013/11/005}{{\em JCAP} {\bfseries
  11} (2013) 005}, \href{http://arxiv.org/abs/1307.5371}{{\ttfamily
  arXiv:1307.5371 [astro-ph.CO]}}.

\bibitem{Sami:2013ssa}
M.~Sami and R.~Myrzakulov, ``{Late time cosmic acceleration: ABCD of dark
  energy and modified theories of gravity},''
  \href{http://dx.doi.org/10.1142/S0218271816300317}{{\em Int. J. Mod. Phys. D}
  {\bfseries 25} no.~12, (2016) 1630031},
  \href{http://arxiv.org/abs/1309.4188}{{\ttfamily arXiv:1309.4188 [hep-th]}}.

\bibitem{Khoury:2013tda}
J.~Khoury, ``{Les Houches Lectures on Physics Beyond the Standard Model of
  Cosmology},'' \href{http://arxiv.org/abs/1312.2006}{{\ttfamily
  arXiv:1312.2006 [astro-ph.CO]}}.

\bibitem{Burrage:2015lla}
C.~Burrage, D.~Parkinson, and D.~Seery, ``{Beyond the growth rate of cosmic
  structure: Testing modified gravity models with an extra degree of
  freedom},'' \href{http://dx.doi.org/10.1103/PhysRevD.96.043509}{{\em Phys.
  Rev. D} {\bfseries 96} no.~4, (2017) 043509},
  \href{http://arxiv.org/abs/1502.03710}{{\ttfamily arXiv:1502.03710
  [astro-ph.CO]}}.

\bibitem{Koyama:2015vza}
K.~Koyama, ``{Cosmological Tests of Modified Gravity},''
  \href{http://dx.doi.org/10.1088/0034-4885/79/4/046902}{{\em Rept. Prog.
  Phys.} {\bfseries 79} no.~4, (2016) 046902},
  \href{http://arxiv.org/abs/1504.04623}{{\ttfamily arXiv:1504.04623
  [astro-ph.CO]}}.

\bibitem{Brax:2015dma}
P.~Brax, C.~Burrage, and A.-C. Davis, ``{The Speed of Galileon Gravity},''
  \href{http://dx.doi.org/10.1088/1475-7516/2016/03/004}{{\em JCAP} {\bfseries
  03} (2016) 004}, \href{http://arxiv.org/abs/1510.03701}{{\ttfamily
  arXiv:1510.03701 [gr-qc]}}.

\bibitem{Saltas:2016nkg}
I.~D. Saltas and V.~Vitagliano, ``{Covariantly Quantum Galileon},''
  \href{http://dx.doi.org/10.1103/PhysRevD.95.105002}{{\em Phys. Rev. D}
  {\bfseries 95} no.~10, (2017) 105002},
  \href{http://arxiv.org/abs/1611.07984}{{\ttfamily arXiv:1611.07984
  [hep-th]}}.

\bibitem{LISA:2017pwj}
{\bfseries LISA} Collaboration, P.~Amaro-Seoane {\em et~al.}, ``{Laser
  Interferometer Space Antenna},''
  \href{http://arxiv.org/abs/1702.00786}{{\ttfamily arXiv:1702.00786
  [astro-ph.IM]}}.

\bibitem{Crowder:2005nr}
J.~Crowder and N.~J. Cornish, ``{Beyond LISA: Exploring future gravitational
  wave missions},'' \href{http://dx.doi.org/10.1103/PhysRevD.72.083005}{{\em
  Phys. Rev. D} {\bfseries 72} (2005) 083005},
  \href{http://arxiv.org/abs/gr-qc/0506015}{{\ttfamily arXiv:gr-qc/0506015}}.

\bibitem{Kawamura:2011zz}
S.~Kawamura {\em et~al.}, ``{The Japanese space gravitational wave antenna:
  DECIGO},'' \href{http://dx.doi.org/10.1088/0264-9381/28/9/094011}{{\em Class.
  Quant. Grav.} {\bfseries 28} (2011) 094011}.

\bibitem{Reitze:2019iox}
D.~Reitze {\em et~al.}, ``{Cosmic Explorer: The U.S. Contribution to
  Gravitational-Wave Astronomy beyond LIGO},'' {\em Bull. Am. Astron. Soc.}
  {\bfseries 51} no.~7, (2019) 035,
  \href{http://arxiv.org/abs/1907.04833}{{\ttfamily arXiv:1907.04833
  [astro-ph.IM]}}.

\bibitem{Punturo:2010zz}
M.~Punturo {\em et~al.}, ``{The Einstein Telescope: A third-generation
  gravitational wave observatory},''
  \href{http://dx.doi.org/10.1088/0264-9381/27/19/194002}{{\em Class. Quant.
  Grav.} {\bfseries 27} (2010) 194002}.

\bibitem{LIGOScientific:2014pky}
{\bfseries LIGO Scientific} Collaboration, J.~Aasi {\em et~al.}, ``{Advanced
  LIGO},'' \href{http://dx.doi.org/10.1088/0264-9381/32/7/074001}{{\em Class.
  Quant. Grav.} {\bfseries 32} (2015) 074001},
  \href{http://arxiv.org/abs/1411.4547}{{\ttfamily arXiv:1411.4547 [gr-qc]}}.

\bibitem{VIRGO:2014yos}
{\bfseries VIRGO} Collaboration, F.~Acernese {\em et~al.}, ``{Advanced Virgo: a
  second-generation interferometric gravitational wave detector},''
  \href{http://dx.doi.org/10.1088/0264-9381/32/2/024001}{{\em Class. Quant.
  Grav.} {\bfseries 32} no.~2, (2015) 024001},
  \href{http://arxiv.org/abs/1408.3978}{{\ttfamily arXiv:1408.3978 [gr-qc]}}.

\bibitem{KAGRA:2018plz}
{\bfseries KAGRA} Collaboration, T.~Akutsu {\em et~al.}, ``{KAGRA: 2.5
  Generation Interferometric Gravitational Wave Detector},''
  \href{http://dx.doi.org/10.1038/s41550-018-0658-y}{{\em Nature Astron.}
  {\bfseries 3} no.~1, (2019) 35--40},
  \href{http://arxiv.org/abs/1811.08079}{{\ttfamily arXiv:1811.08079 [gr-qc]}}.

\bibitem{Nicolis:2008in}
A.~Nicolis, R.~Rattazzi, and E.~Trincherini, ``{The Galileon as a local
  modification of gravity},''
  \href{http://dx.doi.org/10.1103/PhysRevD.79.064036}{{\em Phys. Rev. D}
  {\bfseries 79} (2009) 064036},
  \href{http://arxiv.org/abs/0811.2197}{{\ttfamily arXiv:0811.2197 [hep-th]}}.

\bibitem{Deffayet:2009wt}
C.~Deffayet, G.~Esposito-Farese, and A.~Vikman, ``{Covariant Galileon},''
  \href{http://dx.doi.org/10.1103/PhysRevD.79.084003}{{\em Phys. Rev. D}
  {\bfseries 79} (2009) 084003},
  \href{http://arxiv.org/abs/0901.1314}{{\ttfamily arXiv:0901.1314 [hep-th]}}.

\bibitem{Planck:2015sxf}
{\bfseries Planck} Collaboration, P.~A.~R. Ade {\em et~al.}, ``{Planck 2015
  results. XX. Constraints on inflation},''
  \href{http://dx.doi.org/10.1051/0004-6361/201525898}{{\em Astron. Astrophys.}
  {\bfseries 594} (2016) A20},
  \href{http://arxiv.org/abs/1502.02114}{{\ttfamily arXiv:1502.02114
  [astro-ph.CO]}}.

\bibitem{Inomata:2016rbd}
K.~Inomata, M.~Kawasaki, K.~Mukaida, Y.~Tada, and T.~T. Yanagida,
  ``{Inflationary primordial black holes for the LIGO gravitational wave events
  and pulsar timing array experiments},''
  \href{http://dx.doi.org/10.1103/PhysRevD.95.123510}{{\em Phys. Rev. D}
  {\bfseries 95} no.~12, (2017) 123510},
  \href{http://arxiv.org/abs/1611.06130}{{\ttfamily arXiv:1611.06130
  [astro-ph.CO]}}.

\bibitem{Kohri:2018awv}
K.~Kohri and T.~Terada, ``{Semianalytic calculation of gravitational wave
  spectrum nonlinearly induced from primordial curvature perturbations},''
  \href{http://dx.doi.org/10.1103/PhysRevD.97.123532}{{\em Phys. Rev. D}
  {\bfseries 97} no.~12, (2018) 123532},
  \href{http://arxiv.org/abs/1804.08577}{{\ttfamily arXiv:1804.08577 [gr-qc]}}.

\bibitem{Steinhardt:2004rf}
P.~J. Steinhardt, ``{Cosmological perturbations: Myths and facts},''
  \href{http://dx.doi.org/10.1142/S0217732304014252}{{\em Mod. Phys. Lett. A}
  {\bfseries 19} (2004) 967--982}.

\bibitem{Cai:2019cdl}
R.-G. Cai, S.~Pi, and M.~Sasaki, ``{Universal infrared scaling of gravitational
  wave background spectra},''
  \href{http://dx.doi.org/10.1103/PhysRevD.102.083528}{{\em Phys. Rev. D}
  {\bfseries 102} no.~8, (2020) 083528},
  \href{http://arxiv.org/abs/1909.13728}{{\ttfamily arXiv:1909.13728
  [astro-ph.CO]}}.

\bibitem{Choudhury:2024one}
S.~Choudhury, A.~Karde, S.~Panda, and M.~Sami, ``{Realisation of the ultra-slow
  roll phase in Galileon inflation and PBH overproduction},''
  \href{http://arxiv.org/abs/2401.10925}{{\ttfamily arXiv:2401.10925
  [astro-ph.CO]}}.

\end{thebibliography}\endgroup
\bibliographystyle{utphys}

\end{document}